\documentclass{aa}  
\usepackage{graphicx}
\usepackage{enumitem}
\usepackage{txfonts}
\usepackage[colorlinks=true,linkcolor=blue,citecolor=blue]{hyperref}
\usepackage{url}
\usepackage{mathtools}
\usepackage{multirow}

\newcommand{\av}{$\rm A_{V}$}
\newcommand{\afuv}{$\rm A_{FUV}$}
\newcommand{\afuvsalim}{$\rm A_{FUVs}$}
\newcommand{\afuvsalimerr}{$\rm A_{FUVs\mbox{ }err}$}
\newcommand{\afuverr}{$\rm A_{FUV\mbox{ }err}$}
\newcommand{\afuvprior}{$\rm A_{FUVp}$}
\newcommand{\afuvpriorerr}{$\rm A_{FUVp\mbox{ }err}$}
\newcommand{\sbu}{$\rm \mu_{u}$}
\newcommand{\ur}{$\left(u-r\right)$}
\newcommand{\ui}{$\left(u-i\right)$}

\newcommand{\scuba}{\texttt{SCUBA2}}
\newcommand{\spt}{\texttt{SPT}}
\newcommand{\noema}{\texttt{NOEMA}}

\newcommand{\Herschel}{\textit{Hershel}}

\newcommand{\ALMA}{\texttt{ALMA}}

\newcommand{\Spitzer}{\textit{Spitzer}}
\newcommand{\galex}{\texttt{GALEX}}
\newcommand{\sdss}{\texttt{SDSS}}
\newcommand{\wise}{\texttt{WISE}}
\newcommand{\mass}{\texttt{2MASS}}
\newcommand{\Gcat}{\texttt{GSWLC}}
\newcommand{\LSST}{\texttt{LSST}}
\newcommand{\Glike}{\texttt{{FULL--run}}}
\newcommand{\Llike}{\texttt{{LIGHT--run}}}
\newcommand{\mstar}{$\rm M_{star}$}
\newcommand{\ltir}{$\rm L_{TIR}$}
\newcommand{\cigale}{\texttt{CIGALE}}

\usepackage{txfonts}
\begin{document} 

\title{Attenuation proxy hidden in surface brightness  -- colour diagrams.}
\subtitle{A new strategy for the LSST era. }

\author{
    K. Ma\l{}ek\inst{1,2} 
    \and
    Junais\inst{1} \and 
    A.~Pollo\inst{1,3} \and  
    M.~Boquien\inst{4} \and 
    V.~Buat\inst{2} \and 
    {S.~Salim}\inst{5} \and 
    S.~Brough\inst{6} \and  
    {R.~Demarco}\inst{7} \and  
    A.~W.~Graham\inst{8} 
    M.~Hamed\inst{1} \and 
    {J.~R.~Mullaney}\inst{9} \and 
    {M.~Romano}\inst{1,10} \and 
    {C.~Sif\'on}\inst{11} \and 
    {M.~Aravena}\inst{12} \and 
    {J.~A.~Benavides}\inst{13} \and 
    {I.~Bus\`{a}}\inst{14} \and  
    D.~Donevski\inst{1,15} \and 
    O.~Dorey\inst{16} \and 
    {H.~M.~Hernandez-Toledo}\inst{17} 
    {A.~Nanni}\inst{1,18} \and 
    {W.~J.~Pearson}\inst{1} \and 
    {F.~Pistis}\inst{1} \and 
    {R.~Ragusa}\inst{19} \and 
    {G.~Riccio}\inst{1} \and  
    {J.~Rom\'{a}n}\inst{20,21} 
}

\institute{
    National Centre for Nuclear Research, Pasteura 7, 02-093 Warsaw, Poland
    \email{katarzyna.malek@ncbj.gov.pl}
    \and
    Aix Marseille Univ. CNRS, CNES, LAM, Marseille, France
    \and
    Astronomical Observatory of the Jagiellonian University, Orla 171, PL-30-244 Cracow,  Poland
    \and
    Instituto de Alta Investigación, Universidad de Tarapacá, Casilla 7D, Arica, Chile 
    \and
    Department of Astronomy, Indiana University, Bloomington, Indiana 47405, USA 
    \and
    School of Physics, University of New South Wales, NSW 2052, Australia 
    \and 
    Institute of Astrophysics, Facultad de Ciencias Exactas, Universidad Andr\'es Bello, Sede Concepci\'on, Talcahuano, Chile
    \and
    Centre for Astrophysics and Supercomputing, Swinburne University of Technology, Hawthorn, VIC 3122, Australia 
    \and 
    Department of Physics and Astronomy, The University of Sheffield, Sheffield, S3 7RH, U.K 
    \and 
    INAF  Osservatorio Astronomico di Padova, Vicolo dell'Osservatorio 5, 35122, Padova, Italy 
    \and
    Instituto de F\'isica, Pontificia Universidad Cat\'olica de Valpara\'iso, Casilla 4059, Valpara\'iso, Chile 
    \and 
    Instituto de Estudios Astrof\'{\i}sicos, Facultad de Ingenier\'{\i}a y Ciencias, Universidad Diego Portales, Av. Ej\'ercito 441, Santiago, Chile 
    \and 
    University of California, Riverside, 900 University Ave, Riverside, CA 92521, USA 
    \and
    INAF, Osservatorio Astrofisico di Catania, Via S. Sofia 78, I-95123 Catania, Italy 
    \and 
    SISSA, Via Bonomea 265, 34136, Trieste, Italy 
    \and
    Centro de Astronomía (CITEVA), Universidad de Antofagasta, Avenida Angamos 601, Antofagasta, Chile 
    \and
    Instituto de Astronom\'{i}a, Universidad Nacional Aut\'{o}noma de Mexico, A.P. 70-264, 04510, M\'{e}xico, D.F., Mexico 
    \and
    INAF Osservatorio Astronomico d’Abruzzo, Via Maggini SNC 64100 Teramo, Italy 
     \and
    INAF Osservatorio Astronomico di Capodimonte, Salita Moiariello 16, 80131 Napoli, Italy 
    \and
    Kapteyn Astronomical Institute, University of Groningen, PO Box 800, 9700 AV Groningen, The Netherlands 
    \and 
    Departamento de Astrof\'{\i}sica, Universidad de La Laguna, E-38206, La Laguna, Tenerife, Spain 
}
\date{}

 
  \abstract
  {}
{Large future sky surveys, such as the Legacy Survey of Space and Time (\LSST{}), will provide optical photometry for billions of objects. Reliable estimation of the physical properties of galaxies requires information about dust attenuation, which is usually derived from ultraviolet (UV) and infrared (IR) data. This paper aims to construct a proxy for the far{-UV (FUV)} attenuation (\afuvprior{}) from the optical data alone, enabling the rapid estimation of the star formation rate (SFR) for galaxies that lack UV or IR data. This will accelerate and improve the estimation of key physical properties of billions of \LSST--like observed galaxies (observed in the optical bands only).} 
{To mimic \LSST{} observations, we use{d} the deep panchromatic optical coverage of the Sloan Digital Sky Survey (\sdss{}) Photometric Catalogue, Data Release 12, complemented by the estimated physical properties for the SDSS galaxies from the GALEX-SDSS-WISE Legacy Catalog (\Gcat{}) and inclination information obtained from the SDSS Data Release 7. We restricted our sample to the 0.025-0.1 spectroscopic redshift range and investigated relations among surface brightness, colours, and dust attenuation in the FUV range  {for star-forming galaxies} obtained from the spectral energy distribution (SED).}
{Dust attenuation is best correlated with colour measured between  \textit{u} and \textit{r} bands \ur{} and the surface brightness in the \textit{u} band (\sbu{}). We provide a dust attenuation proxy for galaxies on the star-forming main sequence. This relation can be used for the \LSST{} or any other type of broadband optical survey. The mean ratio between the catalogue values of SFRs and those estimated using optical-only \sdss{} data with  {the \afuvprior{}} prior calculated as $\Delta$SFR=log(SFR$_{\tiny{\mbox{this work}}}$/SFR$_{\tiny{}\texttt{GSWLC}}$) is found to be less than 0.1~dex, while runs without priors result in an SFR overestimation larger than 0.3~dex. The presence or absence of the \afuvprior{} has a negligible influence on the stellar mass (\mstar{}) estimation (with $\Delta$M$_{star}$ in the range from 0 to $-0.15$ dex). 
}
{{We note that} \afuvprior{} is reliable for low-redshift main sequence galaxies. Forthcoming deep optical observations of the \LSST{} Deep Drilling Fields, which also have multi-wavelength data, will enable one to calibrate the obtained relation for higher redshift galaxies and, possibly, extend the study towards other types of galaxies, such as early-type galaxies off the main sequence.}


   \maketitle
%

\section{Introduction}
\label{sec:introduction}

The modelling of galaxies' spectral energy distribution (SED) is a well-established method for measuring key physical properties of galaxies, such as their stellar mass (\mstar{}) and star formation rate (SFR). 
Both are used as the primary building blocks to classify galaxies as quiescent, star-forming, or starbursting and to reconstruct the evolutionary pathways of galaxies \citep{Brinchmann2004, Noeske2007, Elbaz2007, Speagle2014,Pearson2018,Pearson2023, Graham2024}.
The complex nature of the baryonic components of galaxies, including stars, gas, dust, and active galactic nuclei, and how they interact add considerable complexity to modelling the SED.

To link, via the SED fitting process, \mstar{} and SFR in a galaxy, the star formation history (SFH) must be considered.  
Moreover, galaxy merger events also have an influence on  SFHs as it  {boost} the {SFR} and increase \mstar{}. 
A complex interplay between evolved and newborn stars and dust inevitably accompanying star formation makes both measurements surprisingly challenging  \cite[{for example}][]{Walcher2011,Conroy2013}, as dust strongly affects the shape of the SED. 
Astrophysical dust originates from stellar evolution and is one of the key components of the interstellar medium (ISM) of galaxies. 
The presence of dust particles is wide-ranging: dust plays a fundamental role in star and planet formations, molecule production, and galaxy evolution \cite[{for example}][]{Galliano2018}.  
Small dust particles, called grains, typically ranging in size from 5 to 250 nm \citep{Weingartner2001}, are highly influential. 
Dust grains impact the observations of stars and gas by absorbing and scattering short-wavelength photons and then re-emitting energy in much longer wavelengths. 
Moreover, the star-to-dust geometry can change the effect of dust in a non-negligible way \citep[e.g.][]{buat2019, Hamed2023}.

Since star-forming regions are dust-enshrouded in the dense cores of molecular clouds, the earliest stages of star formation can be observed at {millimetre} wavelengths. 
When the clouds collapse and the proto-stars form, the dust near them starts emitting in the near-- and mid-infrared {(IR)} range. 
In the next step in the formation of stars, the warmest regions of the cloud around the newly formed stars are heated by stars' ultraviolet (UV) emission, and this energy is re-radiated in the {IR} domain. 
This process makes dust emission a powerful indicator of star formation if IR-{sub-millimetre} detections are accessible. 
With the advent of IR and sub-{millimetre} facilities {such as} \Spitzer{}, \Herschel{}, \wise{}, \ALMA{}, \scuba{}, \spt{}{,} and \noema{}, the galaxy's dust content (dust mass and dust emission) is measured routinely at low and high redshift \citep[e.g.][]{Dunne2011, Cortese2012, Santini2014,Shirley2019,Harikane2020,Hamed2023,Zavala2023}.

Modified by dust grains, photons hold information about young and evolved stellar populations, active galactic nuclei, or even interactions with other galaxies, for example, merger events. 
Unfortunately, the primary information from the UV--optical spectra is distorted by dust grains and diffused along different wavelengths. 
This process can be described by the dust attenuation curve, which refers to the total effect of dust absorption and scattering on a galaxy SED. 
Though the issue is very complex, detailed studies of the attenuation curves in galaxies are numerous, and various strategies are used.

\cite{Calzetti1994,Calzetti2000} used observational spectra of local UV-bright star-forming galaxies to derive an empirical law for the dust attenuation. 
Another method is to estimate the attenuation in a galaxy and to calculate the SFR in the modelling of its SED. 
This method has also been used on larger samples of galaxies both at low and high redshifts \citep{Wild2011,Batisti2016}. 
In the literature, two prominent attenuation curves, with some additional modifications, are those of \cite{Calzetti2000} and \cite{CharlotFall2000}. 
The \cite{Calzetti2000} attenuation law is described by a single curve for the continuum and a differential reddening with respect to emission lines while \cite{CharlotFall2000} 
assume{d} a different attenuation to the ISM and to the birth cloud regions. 
Moreover, {the }\cite{CharlotFall2000} attenuation law is not a universal curve since it depends on the metallicity of the ISM, as it was shown in \cite{Shivaei2020} and also on the relative distribution of dust between star-forming regions and the {ISM} \citep[e.g.][ and others]{Boquien2022}, and thus, on the SFH.  
In addition to these two, \cite{LoFaro2017} introduced yet another attenuation recipe for $z\sim2$ ultra luminous IR galaxies, although this recipe is similar in concept to the two-component attenuation curve of \cite{CharlotFall2000}. 

It has been shown, however, that attenuation laws are not universal, and a single attenuation law cannot reproduce the physical properties of a large, varied sample of galaxies \citep[e.g.][]{Buat12,Buat14,Malek2018,Salim2018,Hamed2023}. 
Even galaxies with optical -- far--IR (FIR) observations are best modelled with different attenuation laws, resulting in slightly different estimated SFRs \citep{buat2019,Hamed2021}. 
There are different ways to check which attenuation curve is the closest to the physical one. 
Among these methods we can list the comparison of the reduced $\chi^2$ of the SED modelled assuming different attenuation laws \citep{Malek2018,buat2019, Hamed2021}, calculation of Bayesian information criterion (BIC) between different models \citep[used for example in works of][]{Ciesla2018, buat2019, Buat2021}, or the comparison with  {radiative} transfer on a library of hydrodynamic simulations for isolated disk and mergers \citep[i.e.][ and checked with SED models in \citealp{Buat2018}]{Chevallard2013, Roebuck2019}. 
Yet another method is based on the IRX-$\beta$ diagram \citep{Meurer1999,Takeuchi2012,Salim2019, Hamed2023b} which relates the slope of the {UV} continuum ($\beta$) and the ratio between the {IR} and {FIR} luminosities (the IR excess, IRX).

In the case of limited IR measurements, this topic becomes even more complex, as galaxies with different dust properties can appear similar in the optical wavelength range \citep[e.g. both young dusty galaxies and old dust-free galaxies look red in the optical part of the spectrum, more detailed description of classification problems related to the limited wavelength spectrum; a more detailed description can be found, for example in ][]{Siudek2018}. 
Galaxies with full SED coverage, from UV to FIR, are rarely available, creating obstacles to studying them at a significant statistical level. 
This problem will become even more urgent and important in the upcoming era of the Legacy Survey of Space and Time \citep[\LSST{},][]{Ivezic2019} from the Vera C.~Rubin Observatory, where types of galaxies still poorly understood and difficult to observe, such as faint low-surface brightness (LSB) galaxies, or even ultra-diffuse galaxies \citep[e.g.][]{Sandage1984,vanDokkum2015} are expected to be routinely discovered. 
While LSB galaxies were usually assumed to be dust-free, \cite{Junais2023} found that a non-negligible fraction of them (4\% of their sample, namely 23 LSB galaxies {from their sample})  can actually contain enough dust to affect the shapes of their SEDs, with attenuation in the V band, \av{}$\sim$0.8 mag.

The 10~year \LSST{} observations will provide high-quality optical data in the \textit{ugrizy} bands for $\sim$20 billion galaxies \citep{Ivezic2019,Robertson2019}. 
However, most of these galaxies will have no counterparts in existing (or forthcoming) IR catalogues. 
Another issue is that with a large number of galaxies observed by \LSST{}, the traditional SED fitting method will be very computationally expensive. 
Planned joint observations of \LSST{} and near-{IR} satellites, including Euclid \citep{Euclid} for Deep Drilling Fields and the Nancy Grace Roman Space Telescope \citep[formerly the Wide Field Infrared Survey Telescope, WFIRST,][]{NancyRoman} for follow-up observations, will shed light on the near-IR properties of the observed \LSST{} galaxies but will not be sufficient to {analyse} the entire \LSST{} sample. 
Moreover, planned FIR missions like The Far-IR Spectroscopy Space Telescope (FIRSST), The SPace Interferometric Cosmology Explorer (SPICE), The Single Aperture Large Telescope for Universe Studies (SALTUS) or The PRobe far-Infrared Mission for Astrophysics (PRIMA) can help to obtain dust measurements for \LSST{} galaxies in the future, although non of these future projects will match the area-depth combination of the LSST. 
Furthermore, with such deep data, the existing IR maps may suffer from source blending \citep[e.g.][]{Hurley2017, Pearson2017}, resulting in flux inaccuracy, further complicating the SED fitting processes \citep{Pearson2018}. 
As a result, an extremely valuable data set from the \LSST{} observations will suffer from a  poor understanding of the dust attenuation and, consequently, mis-estimated SFR. 
As shown by \cite{Riccio2021}, the estimation of the LSST SFR for normal star-forming galaxies up to $z\sim1$ can be greatly overestimated, with a strong redshift-dependent bias. 
The issue can be even more problematic for hitherto poorly known populations of faint galaxies, including LSB galaxies.  \cite{Graham2001} and \cite{Graham2001b} reported on dust and opacity in {LSB} galaxies and provided simple dust corrections for the surface brightness. 
Those faint LSB galaxies are not that different from known and well-studied brighter galaxies --- they are also a mixture of stars, gas, and dust \citep[even though only recently we have found IR counterparts for those unfamiliar objects; see][for the first statistical analysis of the dust properties in LSB galaxies]{Junais2023}. 

LSB galaxies undergo similar processes, such as dust attenuation and emission, essential to explain their physical properties. 
Considering the depth of the forthcoming \LSST{} observations \citep[$\sim$27.5~mag in the \textit{r}  in the 10-years observations, and $\sim$28.5~mag band for Deep Drilling Fields,  equivalent to $\rm \mu_r\sim30-33$~mag/arcsec$^2$,][]{Robertson2019,Brough2020arXiv}, 
it is expected to detect a significant number of LSB galaxies and other types of faint galaxies that have remained undetected in current surveys. 
However, this vast dataset presents a significant challenge: how to account for attenuation when calculating, for instance, SFR. 
The LSST catalogue will require additional IR and spectroscopic observations to address this issue. 
The study of the Deep Drilling Fields holds the promise of providing valuable knowledge that can be harnessed by, for example, machine learning techniques to calculate the physical properties (e.g. \mstar{}, SFR, bolometric and IR luminosities) of these faint sources. 
On the other hand, \LSST{} will deliver unprecedented high-quality flux and morphology data for observed galaxies. 
In this study, we aim to investigate if optical LSST data can suffice --- at least to some extent --- to construct a prior for dust attenuation of young stellar populations. 
Such a prior can then be used as a preliminary input for SED modelling \citep{Bogdanoska2020,Riccio2021}. 
After obtaining a first estimate of the main physical parameters, it can be replaced with more refined priors derived from other \LSST{} pipelines {and} ancillary data.

More precisely, in this paper, we study the possibility of using \LSST{}--like observables to estimate the prior of the dust attenuation in the far UV regime. 
We do not aim to estimate the actual \afuv{} but only the prior value for each optically detected galaxy that can be further used in SED modelling. 
Additionally, we check to what extent we can reduce the number of parameters in the fit without a significant decrease in the estimate of the main physical properties of the \LSST{}-like sources in order to reduce the computing time.

This paper is structured as follows: in Sect.~\ref{data} we describe the data used for our study. Section~\ref{sec:sample_selection} presents the sample selection and all additional calculations of parameters needed for the next steps of the analysis. The main analysis of \LSST{}-like observables and the resultant attenuation proxy is presented in Sect.~\ref{Sec:color_sb_analysis}. The reliability of the obtained dust attenuation prior is checked in Sect.~\ref{sec:comparison_with_salim}. The results are discussed in Sect.~\ref{sec:discussion}, and the summary and future perspectives conclude this paper in Sect.~\ref{sec:summary}. 
Throughout this paper, we adopt the stellar IMF of \citet{Chabrier2003} and $\Lambda$CDM cosmology parameters \citealp[WMAP7,][]{Komatsu2011}): H$_0$ = 70.4 km s$^{-1}$ Mpc$^{-1}$, $\Omega_{M}$ = 0.272, and $\Omega_{\Lambda}$ = 0.728, the default from the \cigale{} SED fitting tool. 

\section{Data}
\label{data}

To construct a prior for the dust attenuation in the FUV from the observational data, we used three catalogues: (1) The \galex{}-\sdss{}-\wise{} Legacy Catalog \citep[\Gcat{}-X2,][]{Salim2016ApJS,Salim2018}\footnote{\url{https://salims.pages.iu.edu/gswlc/}}, (2)  the Sloan Digital Sky Survey Photometric Catalogue, Data Release~12 \citep[\sdss{},][]{Alam2015ApJS}, and the \sdss{} Data Release~7  spectroscopic main galaxy sample with morphological parameters \citep{Meert2015MNRAS.446.3943M}. 

\subsection{Key physical properties: \mstar, SFR, {and }\afuv{}}
\label{physical_properties}

\Gcat{} is a catalog of local galaxies based on the 10$^{{\rm{th}}}$ \sdss{} Data Release \citep{ahn14}, which covers $\sim$8\,000~deg$^2$. 
Three different catalogues were produced depending on the \galex{} exposure time (\Gcat{}-A, -M and -D for all-sky shallow, medium and deep surveys {, respectively}){,} providing a total of 659\,229 objects ($\sim$90\% of \sdss{}  DR10 objects) at $0.01<z<0.3$, with additional selection on the brightness of \sdss{} objects: $r_{petro}<$18~$[\rm mag]$, which is the magnitude limit for \sdss{} galaxies in the \textit{r} band. 
All three primary catalogues listed above, \Gcat{}-A, M, and D, yield reliable SFRs for main-sequence galaxies,  as the SFR were obtained through a simple conversion factor between IR and SFR and then calibrated using mid-IR luminosity and H$\alpha$ line. 
Galaxies on which the calibration was performed were selected via BPT diagram \citep{BPT1981}.
For quiescent or nearly quiescent galaxies, the simple conversions of IR luminosity do SFR produce overestimation of SFR \citep[specific SFR reaches the overestimation up to 2~dex,][]{Salim2016ApJS}. 
\Gcat{}-M and D are recommended for galaxies off the main sequence. For \Gcat{}, the photometry was taken from \textit{(i)} the \galex{} GR6/7 final release (\citealt{Bianchi2014}; \textit{(ii)} the \mass{} Extended Source Catalog (XSC, \citealt{Jarrett2000}); \textit{(iii)} the \sdss{} DR10; \textit{(iv)} and \wise{} from the All\wise{} Source Catalog and u\wise{} \citep{Lang2016}. 
The \sdss{} and \galex{} photometry were corrected for galactic extinction based on \citet{Peek2013} and \citet{Yuan2013} coefficients. 
Moreover, additional corrections are used for \galex{} data {, when the most significant one is correction due to blending.
This correction is a function of the difference in \sdss{}~g magnitude and the range of separations between sources in the \sdss{}  catalogue. 
This correction is the same for \galex{} FUV and NUV bands. 
The other two corrections deal with (1) edge-of-detector correction required for NUV band when the distance from the centre of the tile to the location of the galaxy is larger than 0.47 degrees; (2) and the centroid shift between optical and UV positions due to lower accuracy of the \galex{} astrometry, applied when the shift between \sdss{} and \galex{} position is larger than 0.7 arcsecond. 
All those corrections are described in detail } in   \citealt{Salim2016ApJS}. 
In our analysis, we used the second version of the catalogue, namely \Gcat{}--X2, which is the master catalogue taking the deepest of GSWLC-A, M and D (659,229 galaxies). 
All the details concerning the construction of the catalogue can be found in \citet{Salim2016ApJS} and \cite{Salim2018}.

We selected \Gcat{}  in order to have a homogeneous associated catalogue of physical parameters, which was obtained with the Code Investigating GALaxy Emission \citep[\cigale{},][]{Burgarella2005, Noll2009,Boquien2019}. 
To model the stellar population of galaxies, \cite{Salim2016ApJS} used \cite{BC03} models with four different metallicities from 0.2 to 2.5 $Z_{\odot}$ (according to \citealp{Gallazzi2005}, these values are in the proper range for a majority of \sdss{} galaxies). 
A two-component exponential model of the star formation history was used, and the modified, using a variable slope $\delta$,  \cite{Calzetti2000} dust attenuation curve with an additional burst was used to model physical parameters for \Gcat{}. 
The total dust luminosity $\rm L_{TIR}$, \citep[8--1000~$u$m,][]{Sanders1996} was estimated by interpolating the \cite{CharyElbaz2001ApJ} {IR} templates. 
The final \Gcat{} catalogue contains a list of estimated parameters (\mstar{}, SFR, \afuv{}) and flags which we used in our analysis as described in Sect.~\ref{sec:GSWLC_cleaning}.

\subsection{Photometric measurements and radii}

We cross-matched \Gcat{} with the \sdss{} DR~12 catalogue \citep{Alam2015ApJS}, which contains not only spectroscopic redshifts but also Petrosian magnitudes and Petrosian radii in \textit{u}, \textit{g}, \textit{r}, and \textit{i} bands\footnote{There are no Petrosian radii and magnitudes for the \textit{z} band in this catalogue.}. 
Petrosian magnitudes  {and }radii are a good first-order proxy  {for} more precise magnitude {and }radii from the \LSST{} pipeline. 
This makes the SDSS DR~12 catalogue a perfect sample to study possible changes in the magnitudes and sizes calculated based on the Petrosian measurements as a function of dust attenuation. 
The \sdss{} DR~12 contains 469\,053\,874 primary sources plus 324\,960\,094 secondary sources. 
More than 3\,500\,000 objects have spectroscopic data. 
It is the final release of the \sdss{}~III, and, at the same time, a  perfect  \LSST{}-like sample to study. 
The main difference between DR~10 used by \cite{Salim2016ApJS} and DR~12 used in our analysis are additional dedicated surveys included in the catalogue (BOSS, APOGEE, and MARVELS) as well as the publication of Petrosian data for all four bands.  

\subsection{Inclination}
The \sdss{} DR~12 catalogue does not include information about the angular sizes of galaxies. 
The minor-to-major axis ratio is crucial for our analysis, as it indicates the galaxy's inclination, which can strongly influence attenuation due to non-spherically symmetric dust distribution. 
To obtain morphological information for our sample of galaxies, we use the catalogue of two-dimensional photometric decomposition from the \sdss{}~DR7 spectroscopic main galaxy sample \citep{Meert2015MNRAS.446.3943M}. 
This catalogue provides a robust set of morphological parameters obtained for the \sdss{} \textit{r} band, using the \texttt{GALFIT} \citep{GALFIT2002AJ....124..266P} and \texttt{PYMORPH} \citep{PYMORPH2010MNRAS.409.1379V}  {software}.
The \cite{Meert2015MNRAS.446.3943M} catalogue includes measurements for 607\,722 galaxies.
We cross-matched \Gcat{} catalogue with axis ratio measurements for \textit{r}-band detections from \cite{Meert2015MNRAS.446.3943M}. 


\section{Sample selection}
\label{sec:sample_selection}

\subsection{Cleaning of the \Gcat{} catalogue}
\label{sec:GSWLC_cleaning}

As recommended by \cite{Salim2018}, for statistical studies of the main sequence galaxies, we used galaxies from the \sdss{} Main Galaxy Survey (\texttt{flag\_mgs = 1}) catalogue, known as \Gcat{}-X2. 
We focus on the main sequence galaxies, as the SFR estimated for \Gcat{} are shown to be reliable \citep{Salim2016ApJS,Salim2018}. 
For galaxies in this catalogue, the accuracy of estimated SFRs is similar in three versions of \Gcat{} (A: shallow all-sky catalogue containing 640\,659 galaxies, corresponding to 88\% of DR10 targets; M: medium-deep catalogue with 361\,328 galaxies, 49\% of the SDSS DR10; D: deep catalogue, which contains 48\,401 galaxies, 7\% of the SDSS DR10). 
Selection based on the \texttt{flag\_mgs = 1} results in 610\,518 galaxies. 
Additionally, we only keep galaxies with a good fit to their {SED} (\texttt{FLAG\_SED}=0, also recommended by  \citealp{Salim2016ApJS} and \citealp{Salim2018}). 
This selection gives us an initial catalogue of 603\,615 main sequence galaxies in the redshift range $0.01-0.30$. 
In the next step, we perform further cleaning.

The analysis presented in the following sections aims to construct and test a possible prior for attenuation of the young stellar population (\afuv{}). 
As the accuracy of the \afuv{} and the SFR depends on the depth of the \galex{} observation, we remove all shallow UV detections (all-sky, \Gcat{}A). 
Therefore{,} we perform the analysis based on the medium-deep and deep \galex{} observations (we use \texttt{UV\_SURVEY} flags~2 and 3). 
After this selection, we are left with a sample of 404\,830 galaxies. 
We also remove all objects that belong to the shallow, all-sky, \galex{} catalogue.  
Thus cutting down the selection by a 152\,385 galaxies.

The \Gcat{}-X2 catalogue contains galaxies whose total {IR} luminosity ($\rm L_{TIR}$) was calculated based on the 12~$\mu$m or 22 $\mu$m detection {and} corrected for mid-{IR} AGN emission. 
To homogenise the data used for the analysis, we decided to use galaxies with $\rm L_{TIR}$ estimated based on the 22~$\mu$m \wise{} detection. 
This cut removes all AGN-corrected galaxies, which means that the sample should not contain any AGNs. 
As the IR counterpart is necessary in a standard SED fitting process to estimate reliable attenuation and SFR, we have to limit our analysis to galaxies bright enough to be visible in the \wise{} bands. 
It creates a bias by removing a large fraction of LSB galaxies but not all of them. 
Moreover, in the future, we are planning the next calibration based on more sensitive Euclid measurements. 
After that selection, the catalogue contains 82\,116 galaxies. 
We also reject all objects with \texttt{REDCHISQ} flag, which stands for the reduced goodness-of-fit value ($\chi_{red}^2$) for the SED fitting, larger than five (following \citealp{Salim2016ApJS} and \citealp{Salim2018}).
After all these steps, the final subsample of the \Gcat{}-X2 catalogue used in this analysis contains 78\,725 galaxies.

\subsection{Cleaning of the \sdss{} catalogue}
\label{sec:SDSS_cleaning}

Based on the flags used for the photometric measurements of the \cite{Alam2015ApJS} catalogue, we remove all  {objects} with \texttt{class} flag equal to six (stars from the \sdss{} catalogue), and objects that do not have magnitude and Petriosian radii measurements in all four bands (\textit{u}, \textit{g}, \textit{r}, and \textit{i}). 
This criterion allows us to check all possible relations between \afuv{} and future \LSST{}  data. 
This initial cleaning resulted in 78\,723 galaxies (i.e. only two galaxies from the above sample were removed). 

We also use flags describing the quality of the estimation of the radii. 
We remove all galaxies where no valid Petrosian radius was found (\texttt{NONPETRO}) or with multiple Petrosian radii (\texttt{MANYPETRO}). 
We also remove measurements with  Petrosian radius larger than the radial profile (\texttt{NOPETRO\_BIG)} or with more than one radius including 50\% or 90\% of the light (\texttt{MANYR50}) and \texttt{MANYR90}). 
We do not use radii that were measured at the edge of the frame (\texttt{INCOMPLETE\_PROFILE)}, those rejected because of low surface brightness level (\texttt{PETROFAINT}) or objects larger than 4~arcmin (\texttt{TOO\_LARGE}). 
Additionally, we remove all measurements with possible saturation deception as the centre of the radii is close to the saturated pixel (\texttt{SATUR\_CENTER}) or interpolated pixel (\texttt{INTERP\_CENTER}). 
On top of those flags, we also remove galaxies detected with a very low sky level, which results in the centre pixel of the galaxy being negative (\texttt{BADSKY}) or at the edge of the frame  (\texttt{EDGE}). 
Yet another flag which indicates a possible problem with the image is \texttt{CANONICAL\_CENTER}. 
This flag is set for objects for which it is impossible to measure the centre in the \textit{r} band. 
We also remove all possibly moving objects (\texttt{MOVED}) or galaxies detected at a level larger than 200$\sigma$ in the~\textit{r} band (\texttt{BRIGHT})\footnote{The detailed description of all \sdss{} flags can be found on the \url{https://www.sdss.org/dr12/algorithms/flags_detail/} webpage}. 

 \begin{figure*}[ht!]
    \centering
        \includegraphics[width=1\textwidth]{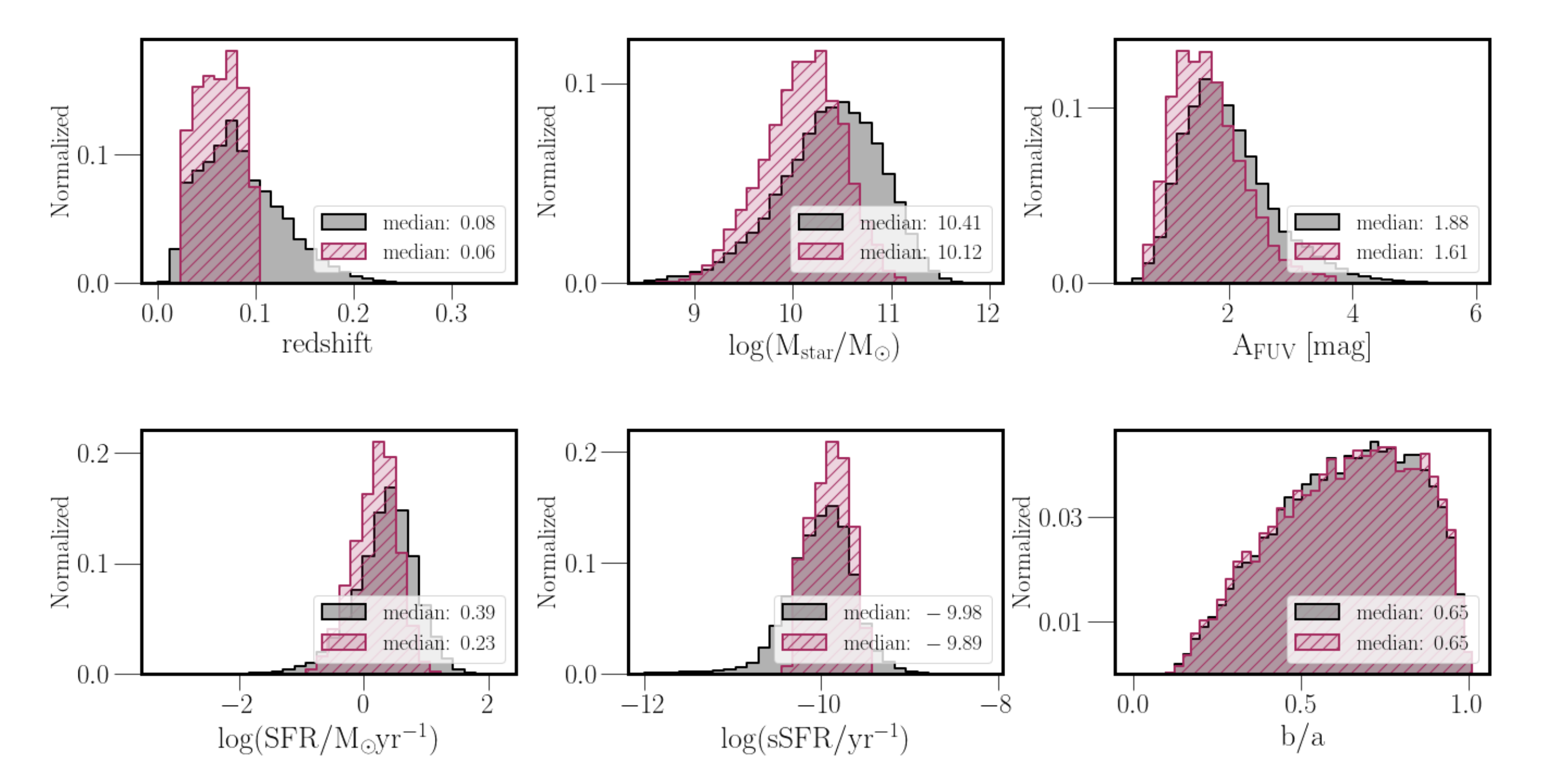}
        \caption{Main physical properties used in our analysis{. Panels above show} spectroscopic redshift, stellar mass ($\rm log(M_{star}/M_{\odot}$)), attenuation in the FUV band ($\rm A_{FUV}$), SFR and specific SFR (sSFR), both in logarithmic scale. The last bottom right panel shows the axis ratio (semi-minor/semi-major) from \cite{Meert2015MNRAS.446.3943M}. Black histograms represent distributions obtained from the whole sample of 29\,487 galaxies ($0.01<z<0.30$), while maroon hatched histograms show distributions for the final sample used for the analysis (7\,986 galaxies). Legends show median values for all parameters calculated for the initial (black histograms) and final (maroon hatched histograms) samples.}
        \label{fig:mpp}
\end{figure*}

We use the same quality condition for the \textit{u},\textit{g},\textit{r}, and \textit{i} bands. 
In total, we remove 34\,676 galaxies from the initial sample created based on the \cite{Salim2018}. 
This selection allows us to create a catalogue of 44\,047 galaxies with good photometric measurements in all four \sdss{} bands, with UV and mid-infrared detections, and reliably estimated key physical parameters from \cite{Salim2018}, namely SFR, \mstar{}, \afuv{}, and A$_{V}$. 

\subsection{Cross match with the \sdss{}~DR7 2D decomposition catalogue}

Next, we cross-match the catalogue with the \cite{Meert2015MNRAS.446.3943M} \sdss{}~DR7 catalogue to obtain information about angular sizes and the inclination of galaxies. 
This cross-matching reduces the sample significantly to  29\,487 galaxies,{ that is,} removing one third of the sample. 
From this sample, we remove all galaxies with an axis ratio (semi-minor/semi-major) lower than zero. This cut removes an additional 106 galaxies. 

\subsection{Final sample}
\label{sec:final_sample}
The selection described above yields our final sample of  29\,487 normal star-forming galaxies in the redshift range 0.01$<z<$0.3. 
In Table~\ref{tab:selection}, we list all steps performed to obtain the final sample. 
This final sample provides by reliable measurements of magnitudes and radii in all four \sdss{} bands, sets of morphological parameters for the \sdss{}~\textit{ugri} bands, and proper estimation of the main physical parameters (\mstar{}, SFR, attenuation in the FUV band, etc) from the \Gcat{}-X2 catalogue. 

\begin{table*}[]
\centering
\caption{Sample selection discussed in Section~\ref{sec:sample_selection}. } 
\label{tab:selection}
\begin{tabular}{l r r}
\hline
 {Selection criteria} &  {Number of selected sources} &  {\% of the initial sample} \\ \hline
\hline
\multicolumn{3}{c}{ catalogue of physical properties \citep{Salim2016ApJS,Salim2018} }\\
\hline
\Gcat{}-X2  & 659\,229   & 100,0\%\\
Objects with \afuv{} estimation & 650\,597 & 98.69\%\\
Main Galaxy Survey \texttt{flag\_msg=1} & 610\,518  & 92.61\%\\
SED fitting flag=0 (all \sdss{} photometry, no broad-line spectrum)  & 603\,615 & 91.56\% \\
At least one \galex{} detection (FUV or NUV) & 404\,830 & 61.41\% \\
Medium and deep UV exposure time  (\Gcat{}-A and D) & 252\,445 & 38.29\% \\
$\rm L_{TIR}$ estimated based on the \wise{} 22~$\mu$m & 82\,116 & 12.45\% \\
\texttt{REDCHISQ}$<$5 \cite[goodness of the fit, following][]{Salim2016ApJS, Salim2018} & 78\,725 & 11.94\%\\
\hline
\multicolumn{3}{c}{photometric  catalogue \text{\citep{Alam2015ApJS}} }\\
\hline
Cross-matching with \sdss{} \cite{Alam2015ApJS} catalogue & 78\,725 & 11.94\%\\
Cleaning based on the \sdss{} flags (Sec.~\ref{sec:SDSS_cleaning}) & 44\,047 &  6.68\% \\ 
\hline
\multicolumn{3}{c}{2D photometric decompositions catalogue \citep{Meert2015MNRAS.446.3943M} }\\
\hline
Cross-matching with \cite{Meert2015MNRAS.446.3943M} catalogue&  29\,593   & 4.49\%\\
The axis ratio (\textit{b/a}) of the total fit $>0$  & 29\,487 & 4.47\% \\
\hline 
\multicolumn{3}{c}{\text{Selection based on Sect.~\ref{sec:final_sample}:}} \\
\hline
Cut for the \afuvsalimerr{}$<$0.25 [mag] & 15\,004 & 2.28\% \\
Redshift range 0.025--0.1 & 9\,837 & 1.49\% \\
Cut between 1$^{st}$ and the 99$^{th}$ percentile of the $\rm A_{FUV}$ & 9\,641 &  1.46\% \\

Main sequence galaxies & 7\,986 & 1.21\% \\
\hline
\end{tabular}
\end{table*}

Black histograms presented in Fig.~\ref{fig:mpp} show distributions of the main physical parameters for the whole sample of 29\,487 galaxies ($0.001<z<0.3$). 
Six panels of this figure show the distribution of the spectroscopic redshift, as well as the main physical properties: stellar mass ($\rm log(M_{star}/M_{\odot}$)), attenuation in the FUV band ($\rm A_{FUV}$), SFR, and specific SFR (sSFR), both in logarithmic scale. 
The bottom right panel shows the axis ratio (semi-minor/semi-major) from \cite{Meert2015MNRAS.446.3943M}.

 \begin{figure}[ht!]
    \centering
        \includegraphics[width=0.5\textwidth]{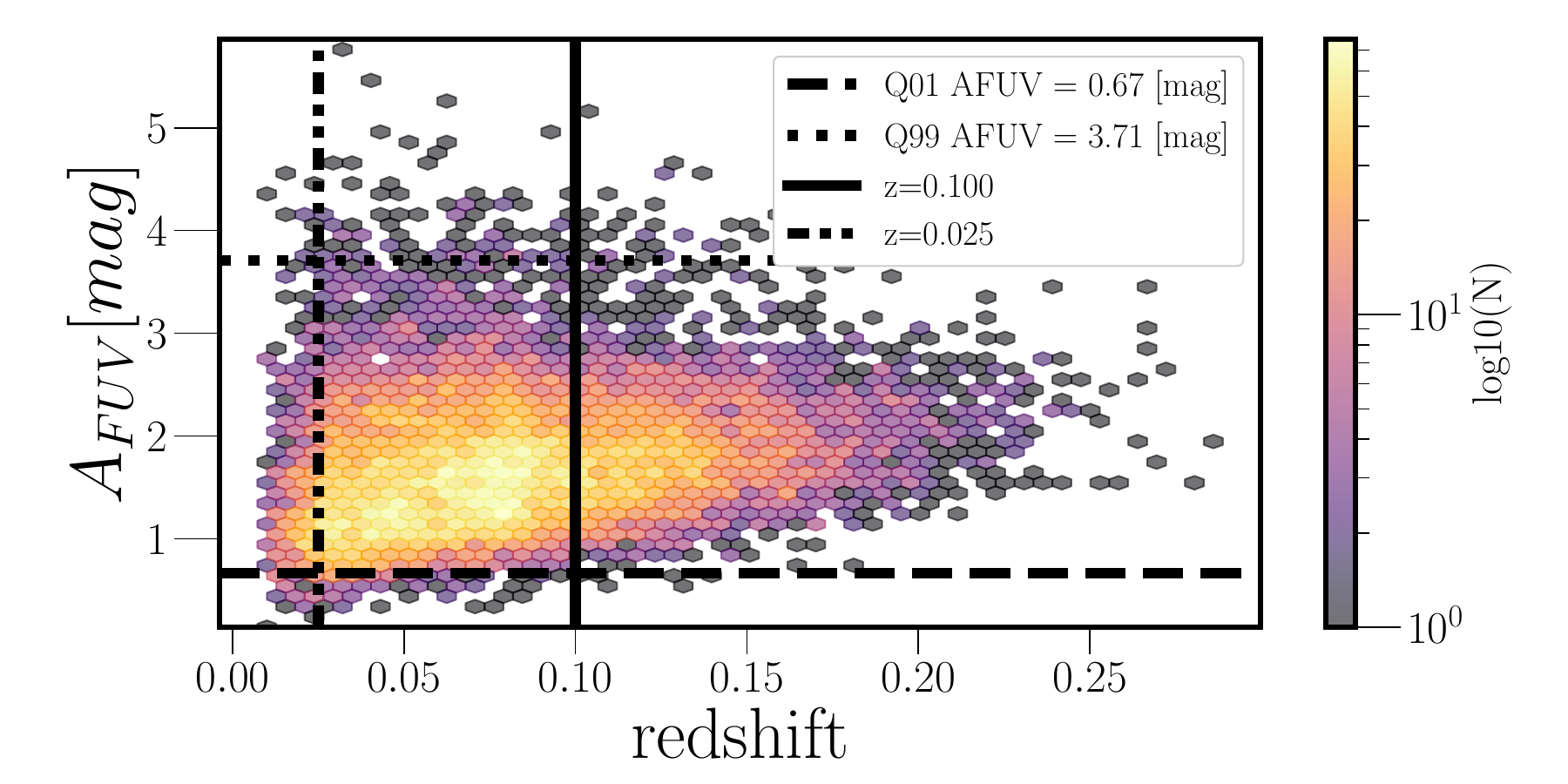}
        \caption{Attenuation in the FUV band ($\rm A_{FUV}$) as a function of redshift. Two horizontal lines, dashed and dotted, represent the 1$^{st}$ and the 99$^{th}$ percentile of the $\rm A_{FUV}$ distribution, respectively. The solid vertical line indicates redshift equal to 0.1, while the dashed double-dotted line represents a redshift cut at 0.025. Above that redshift line, the sample cannot probe the most extreme $\rm A_{FUV}$ values.}
        \label{fig:AFUV_z_cut}
\end{figure}

From the sample, we remove all galaxies with \afuverr{} larger than 0.25 [mag], as in the next step of our analysis, we want to bin galaxies regarding the \afuv{} value, and too large errorbars can influence our binning. 
The final sample, after this cut, contains 15\,004 galaxies. 
We stress here that this cut has negligible influence on the main physical properties of the final sample, as shown in Fig.~\ref{fig:mpp}.

The number of objects in the sample drops above redshift $z\sim0.1$, which can be seen in the upper left panel in Fig.~\ref{fig:mpp}. 
A similar drop (but much steeper) can be seen below redshift 0.025. 
In Fig.~\ref{fig:AFUV_z_cut}, we check the \afuv{} distribution as a function of redshift. 
This figure shows that below redshift 0.025 and above redshift 0.1, the values of $\rm A_{FUV}$ are not spread across the full range of this parameter. 
To obtain a representative sample of galaxies across the attenuation and redshift space range, we remove all galaxies with \afuv{} below the 1$^{st}$ (0.67~mag) and above the 99$^{th}$ (3.71~mag) percentile of the distribution. 
Moreover, we introduce additional cuts in redshift, removing galaxies outside the 0.025 and 0.1 redshift bin. 
Those two cuts allow us to keep a statistically significant galaxy sample characterised by an almost complete distribution of \afuv. 
Therefore, we decide to use in the following analysis only galaxies within the redshift range $0.025<z<0.1$ (which reduces the sample to 9\,837 galaxies), and with $0.67<\rm A_{FUV}<3.71$~mag, further reducing the sample to 9\,641 galaxies.

\begin{figure}[ht!]
    \centering
        \includegraphics[width=0.5\textwidth]{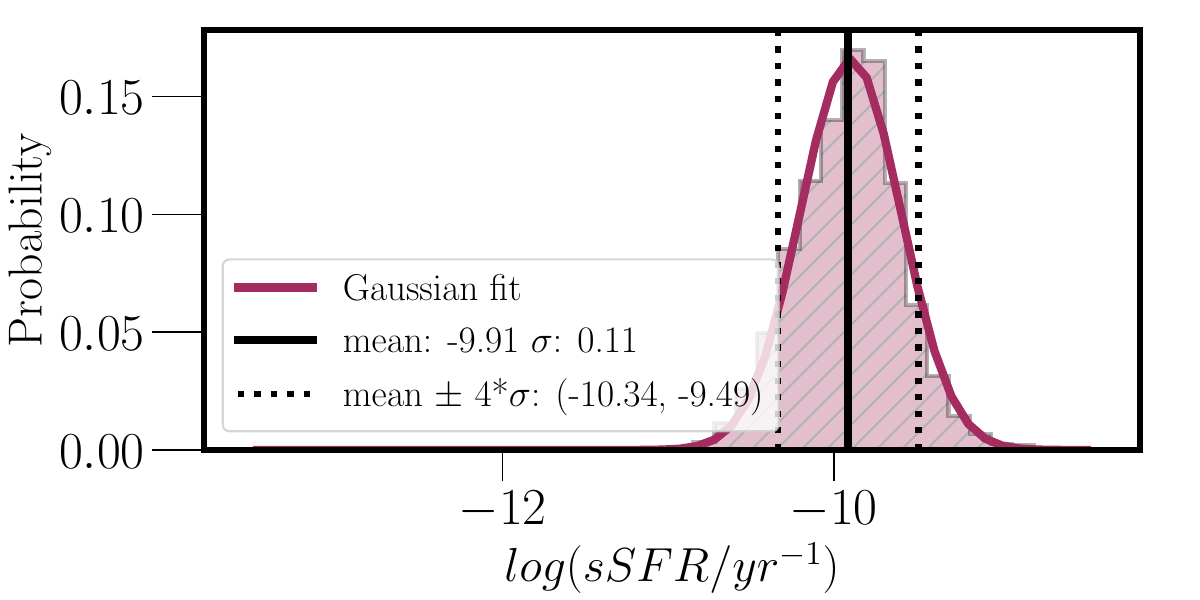}
        \caption{Distribution of log(sSFR/yr$-1$) for the selected 7\,986 galaxies in the redshift range 0.025$<z<$0.1. A vertical solid line and two vertical dotted lines represent the mean value of sSFR of the sample, and the mean value of the sSFR decreased/increased by  $\rm 4\sigma$ of the distribution, respectively.}
        \label{fig:SSFR_cut}
\end{figure}

As the last step of the sample selection, we remove all galaxies in the tail of the sSFR distribution.   
From the \sdss{} distribution, we remove the tail of the main sequence distribution by selecting only galaxies located within $\rm 4\sigma$ of the sSFR distribution (i.e. $-10.34<\mathrm{log(sSFR/yr^{-1})}<-9.49$, as illustrated in Fig.~\ref{fig:SSFR_cut}). 
This cut removes 1\,655 objects. 

Distributions of the main parameters used in the analysis are shown in Fig.~\ref{fig:mpp}. 
The full sample of 29\,487 galaxies, without our internal cuts, is shown in black histograms, while the final sample of 7\,986 galaxies is presented as maroon-hatched histograms.  
The \afuv{} distribution, together with the \mstar{} distribution, show that cuts based on the \afuverr{}, redshift, \afuv{} and sSFR, do not change the main properties of the key physical properties, but only remove the most massive and in the same time the most active in the star formation processes, and the most attenuated galaxies. 

\subsection{Surface brightness calculation}
\label{sec:SB_calculation}

For the final sample of 7\,986 galaxies, we calculate the surface brightness in each of the SDSS bands using the equation:
\begin{equation}
\label{eq:sb}
   \rm  \mu_x = \texttt{mag}_x +2.5\cdot log_{10}(2 \pi   \texttt{r}_x^2), 
\end{equation}
where \textit{x} stands for \textit{u}, \textit{g}, \textit{r} and \textit{i} band, \texttt{mag} for Petrosian magnitude and \texttt{r} for circular Petrosian radius. 
{We n}ote that we use the Petrosian radius for the surface brightness measurements, unlike the half-light radius, which is generally used in the literature \citep[e.g.][]{Paudel2017,perez-montano2022}. 
As we do not have a half-light radius measurement for all the photometric bands used in this work, we chose to adopt the Petrosian radius in all the bands for consistency (\citealt{Meert2015MNRAS.446.3943M,Meert2016MNRAS.455.2440M} provides half-light radius measurements only for the \textit{g}, \textit{r} and \textit{i} bands, without the \textit{u}-band). 
\citet{Graham2005} shows that the Petrosian radius for a galaxy with S\'ersic index $n=1$ (which is a reasonable assumption for the main-sequence galaxies studied in this work) is about twice larger than its S\'ersic half-light radius. 
This will result in our surface brightness measurements $\sim$1.5 mag/arcsec$^2$ fainter than those estimated using the half-light radii. 
However, such a systematic offset does not affect any of the trends studied in this work. 
Therefore, from here upon, we adopt the surface brightness measurements obtained using the Petrosian quantities. 

We apply the correction for inclination \citep[following][]{Zhong2008,Pahwa2018}:
\begin{equation}
\label{eq:inclination}
   \rm  \mu_{x, corrected} = \mu_x +2.5\cdot log_{10}\cdot (b/a)-10\cdot log_{10}(1+z), 
\end{equation}
where \textit{ b/a } represents the ratio of a galaxy's minor and major axis. 
From now on, we always use only 'corrected' surface brightness in the analysis, and thus we drop the subscript \texttt{corrected} from the definition of $\mu_{\rm x,corr}$. 
Figure~\ref{fig:mag_sb} shows the distribution of the magnitudes and calculated surface brightness based on the Eqs.~\ref{eq:sb} and~\ref{eq:inclination} for the final sample used in the analysis.

\subsection{$\rm A_{FUV}$-- LSST-like observables relations}
\label{sec:AFUV_prior_estimation}

We look for possible relations between observed LSST-like data (fluxes, magnitudes, colours, surface brightness in different bands, as well as the ratio between different surface brightness) in four SDSS bands (\textit{u}, \textit{g}, \textit{r}, and \textit{i}) and the attenuation in the FUV band estimated via fitting the UV to IR SED. 
The \afuv{} relation with colours and surface brightness calculated in different bands are presented in App.~\ref{app:mix_colours}. 
We also tried to use colours, but the relations are too narrow to separate different attenuation levels. 
\Gcat{}-X2 \afuv{} values were obtained using \galex{}, \sdss{}, and \wise{} photometry calibrated on the \Herschel{} ATLAS \citep{Salim2018}, ensuring proper dust attenuation estimation. 
The method used by \cite{Salim2018} combined SED constrained with CIGALE \citep{Burgarella2005,Noll2009,Boquien2019} fitting code with infrared luminosity (SED+\ltir{} fitting;  more details can be found in  \citealp{Salim2018}). 
Our main goal is to find a simple proxy for \afuv{} based on the observational \LSST{}-like data.

We checked all possible relations between colours, magnitudes, surface brightnesses, and their ratios. 
As a result, we find a promising relation between \ur{} colour, the surface brightness calculated in the \textit{u} band, and the \afuv{}, which is characterised by a monotonic rise of the ratio of \ur{} colour and the \textit{u} surface brightness with the \afuv{} values, and an extensive parameter locus (more than 0.6 magnitude in colour; for example, the \textit{(g$-$i)} colour gives only $\sim$0.3 magnitude width, which makes the \afuv{} analysis more complicated taking into account the uncertainties of photometric measurements, {and so forth}). 
We also find a very similar relation using \ui{} instead of \ur. 
The two main changes between both relations (the chosen one \ur{}--\sbu{}--\afuv{} and the second best one \ui{}--\sbu{}--\afuv{}) are the larger global slope uncertainty for \ui{} shown in Fig.~\ref{App:A_global}, and larger uncertainties for the final \afuvprior{} equation based on larger errors for the intercept equation (Eq.~\ref{eq:slop_inetr}). 
The \ur{} or \ui{} colour is a natural indicator of dust attenuation since dust affects the slope of the galaxy SED. 
Both colours also cover the Balmer break, so they are very good indicators of the age of the stellar population. 
We are aware that we have a degeneracy between age and dust attenuation; since we have no information on the ages of the \Gcat{} and a very narrow redshift range, {we subsequently} analyse this degeneracy in the forthcoming analysis using much smaller but more informative, reference catalogues. 
The surface brightness in the band closest to UV \sdss{} is an indicator of the {SFR} as it traces light from the young stellar populations.
The combination of these two parameters can be thus expected to be sensitive to dust attenuation for young stellar populations. 
However, this is the first time, to our knowledge, that these parameters have been combined to derive the proxy for \afuv{}.
In Sections~\ref{Sec:color_sb_analysis} and \ref{sec:comparison_with_salim}, we present and {analyse} this relation in detail.

\section{\afuv{} prior: \ur{} colour versus surface brightness in the \textit{u} band}
\label{Sec:color_sb_analysis}

 \begin{figure}[ht!]
    \centering
        \includegraphics[width=0.5\textwidth]{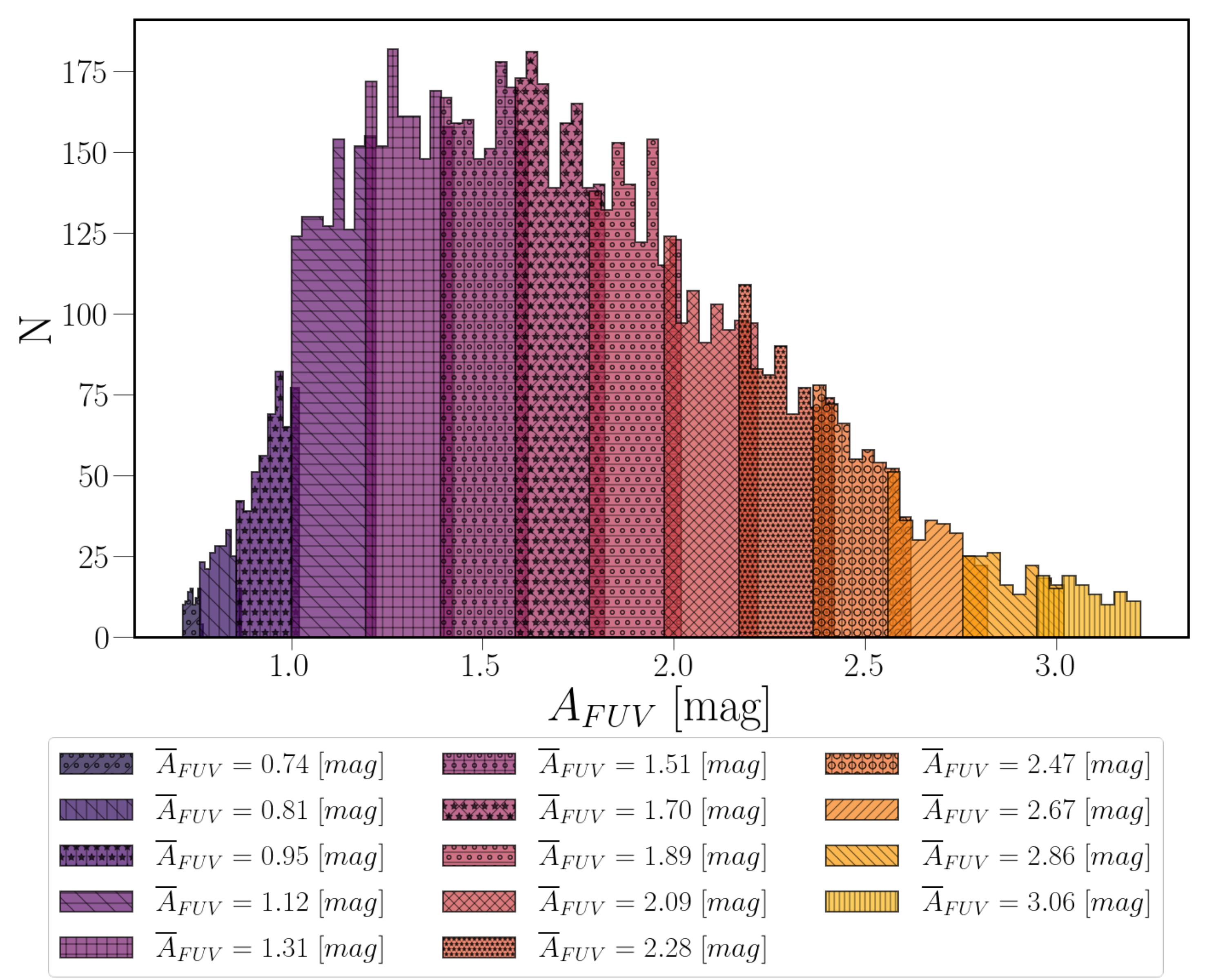}
        \caption{Distribution of \afuv{} for the final sample of 7\,814 main sequence galaxies in the redshift range $0.025-0.1$. The palette of colours represents the \afuv{}  bins used in our analysis. The mean \afuv{} value for each bin ($\rm \overline{A}_{FUV}$) is denoted in the legend.}
        \label{fig:AFUV_bins}
\end{figure}

To study the relation between the colour and the surface brightness, we divide our sample into 14 \afuv{} bins to check if different attenuation values follow different relations in the colour~$-$~surface brightness space and if they can be separated. 
These bins are presented in Tab.~\ref{tab:AFUVbins} and graphically shown in Fig.~\ref{fig:AFUV_bins}. 
Each bin contains at least one percent of the final sample (at least 70 galaxies). 
Due to small numbers of galaxies in the \Gcat{}-X2 having \afuv{} lower than~1~mag, we use a variable bin width to probe as densely as possible the lowest attenuation range, which is underrepresented in the catalogue. 
Thus, the first bin has a width of 0.05~mag, the second is 0.10~mag wide, and the third is 0.15~mag wide. 
Starting from the fourth, the width is greater at 0.20~mag. 
Additionally, to increase the number of galaxies in each bin, as well as to include possible uncertainties arising from the redshift estimation or physical properties, we add overlaps between bins.
These overlaps increase with increasing \afuv{} according to the relation: \texttt{bin$_n \cdot$0.005}, where \texttt{bin$_n$} refers to the bin number. 
This helps us to gather enough galaxies in the less populated bins of high \afuv{} (larger than 2~mag), but also to take into account the uncertainties of the \afuv{} estimated by \cite{Salim2018}, as for larger \afuv{} we also have larger \afuverr{} (it can be seen later in the right bottom panel of Fig.~\ref{fig:AFUV_prior_distribution}, blue circles).

\begin{table}[]
\small
\begin{center}

\caption{\afuv{} bins discussed in Section~\ref{Sec:color_sb_analysis} and used in our analysis.  } 
\label{tab:AFUVbins}

\begin{tabular}{c| c | c |  r | r}

{\afuv{} bin}  & {bin width} & $\rm \overline{A}_{FUV}$ &
{\# gal.} & {\% sample}  \\ 

\hline
\hline
0.72 -- 0.77 &  0.06 & 0.74 & 91 & 1.14 \\
0.76 -- 0.87 &  0.11 & 0.81 & 199 & 2.49 \\
0.86 -- 1.02 &  0.16 & 0.95 & 481 & 6.02 \\
1.00 -- 1.22 &  0.22 & 1.12 & 1098 & 13.75 \\
1.20 -- 1.42 &  0.23 & 1.31 & 1303 & 16.32 \\
1.39 -- 1.62 &  0.23 & 1.51 & 1290 & 16.15 \\
1.59 -- 1.82 &  0.24 & 1.70 & 1267 & 15.87 \\
1.78 -- 2.02 &  0.24 & 1.89 & 1077 & 13.49 \\
1.98 -- 2.22 &  0.25 & 2.09 & 812 & 10.17 \\
2.17 -- 2.42 &  0.25 & 2.28 & 654 & 8.19 \\
2.37 -- 2.62 &  0.25 & 2.47 & 472 & 5.91 \\
2.56 -- 2.82 &  0.26 & 2.67 & 267 & 3.34 \\
2.76 -- 3.02 &  0.27 & 2.86 & 161 & 2.02 \\
2.95 -- 3.22 &  0.27 & 3.06 & 117 & 1.47 \\
\hline

\end{tabular}
\end{center}
\vspace{1mm}
\small{
 {\it  {Notes}:} The first column represents the minimal and the maximal value of \afuv{} in each bin, the second -- the bin width, and the third is the mean value of \afuv{} in each bin. The fourth column presents the number of galaxies in each bin, while the fifth column is the percentage of the full sample of 7\,934 galaxies.}
\end{table}

\begin{figure}[t!]
    \centering
        \includegraphics[width=0.5\textwidth]{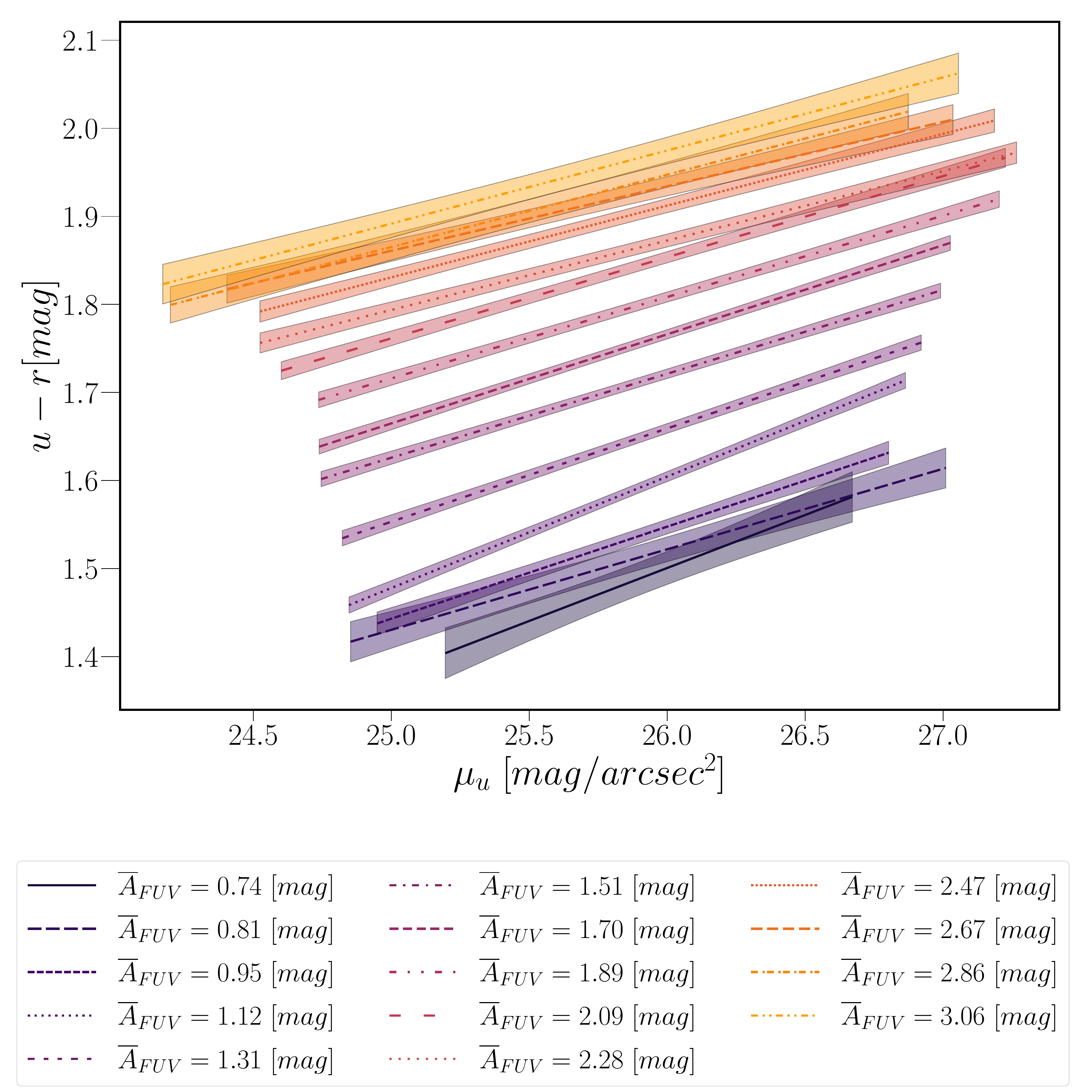}
        \caption{Relations fitted between observed \ur{} colours and \sbu{} for 14 \afuv{} bins. The sequence of colours represents the one used for \afuv{} bin in Fig.~\ref{fig:AFUV_bins}. Filled areas display the $\pm 1\sigma$ uncertainty around estimated lines.}
        \label{fig:AFUV_all_fits}
\end{figure}

We perform a linear fit in each \afuv{} bin in the \ur{}--\sbu{} parameter space, as this relation shows the most prominent slope of the general relation (0.0796 $\pm$ 0.0024)\footnote{The second most prominent relation is \ui{}--\sbu{} with slope 0.0799 $\pm$ 0.0047, however, the slope uncertainty is almost twice larger than for the slope uncertainty of the \ur{}--\sbu{} plane.}, Fig.~\ref{app:mix_colours}. 
Moreover, the \ur{} colour space is wide enough to separate different attenuation levels, taking into account measurement errors for future \LSST{} observations. 
Since the \sbu{} range is much wider than the range of \ur{}, and more often is contaminated by outliers caused by uncertainties in calculating Petrosian radii and Pertosian magnitudes, we perform the fit only between the 10$^{th}$ and the 90$^{th}$ percentile of the  \sbu{} distribution in each bin. 
Figure~\ref{fig:AFUVbins_linear_fitting} shows fits of the relation \ur{} versus \sbu{} fits for all \afuv{} bins, while Fig.~\ref{fig:AFUV_all_fits} presents combined linear fits for all 14 bins. 
In Figure~\ref{fig:AFUV_all_fits}, it is evident that for all \afuv{} bins, the \ur{}-\sbu{} relation maintains a consistent slope, with the intercept gradually increasing as \afuv{} rises. 
We have interpreted the flattening observed in the lowest and highest \afuv{} values as being related to the much less densely populated part of our sample. 
This can be seen in Table~\ref{tab:AFUVbins} and Figure~\ref{fig:AFUV_bins}, where \afuv{} values lower than $\sim$0.8~mag and greater than 2.7~mag constitute only about 8\% of the total sample analysed in our manuscript.  
Figure~\ref{fig:AFUV_all_bins_bdg} shows the same relation between \ur$-$\sbu{} and \afuv{}, but with an additional background of galaxies used in our analysis  {colour~$-$~coded} according to the value of \afuv{}, and with the interpolated linear fit with \sbu{} in a range of $22.5-27.5$ [mag/arcsec$^2$]. 
For the simplicity in the main manuscript, we show only fits with the $\rm \pm 1\sigma$ uncertainty around estimated lines.

\subsection{Global \ur{} -- \sbu{} -- \afuv{} relation}

\begin{figure}[t!]
    \centering
        \includegraphics[width=0.48\textwidth]{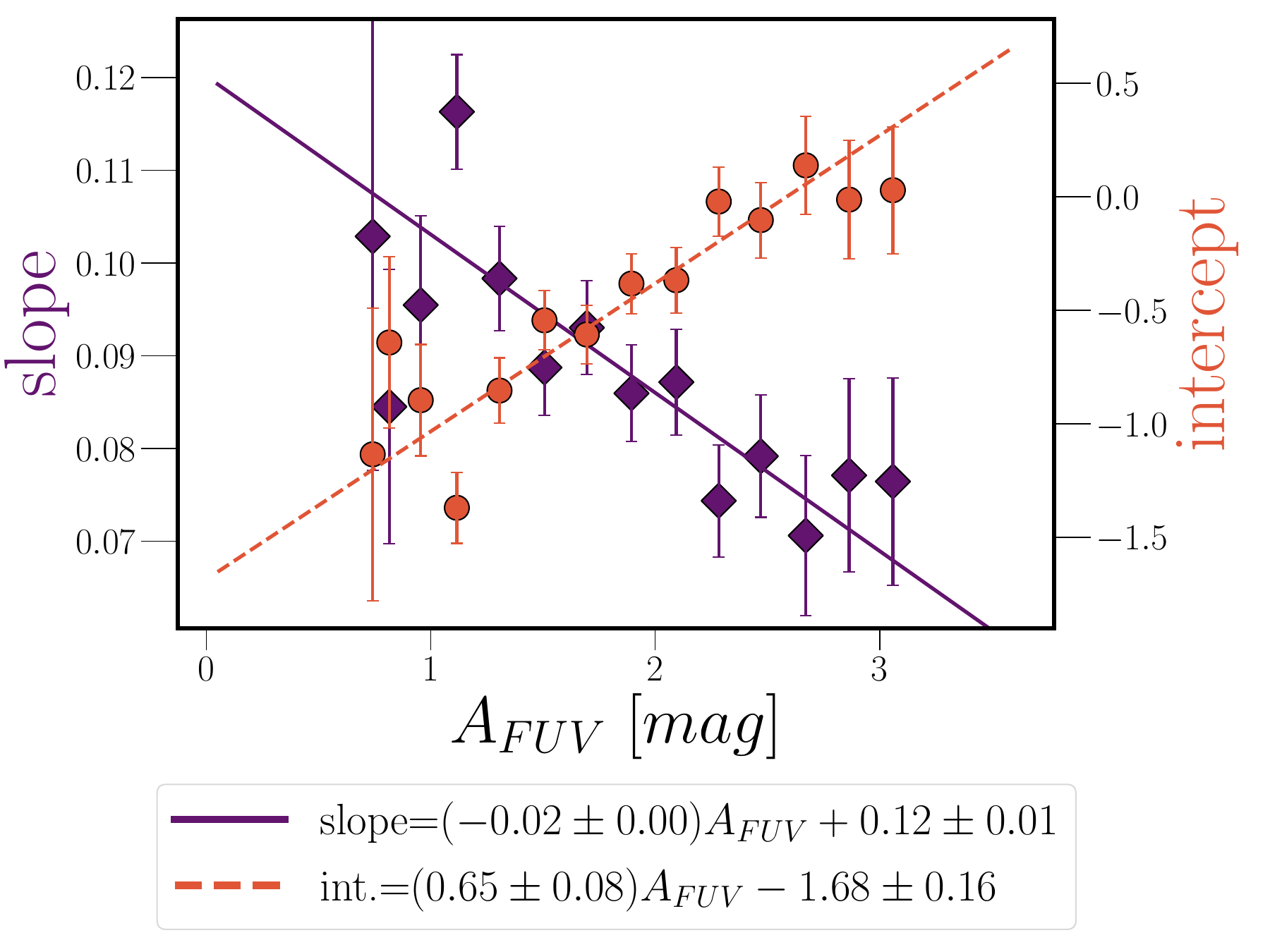}
        \caption{{Slopes and intercept from \ur{}-\sbu{} fits.} Interpolated slopes (left axis) and intercepts (right axis) obtained from linear fits for all 14 bins of \afuv{} (Fig.~\ref{fig:AFUVbins_linear_fitting}), as a function of \afuv{}. The left axis represents slopes (purple full diamonds), and the right axis represents intercepts (red full circles). Linear fits and associated fitted parameters are presented using dashed-dotted lines for slopes and dashed lines for intercepts.}
        \label{fig:slopes_intercepts}
\end{figure}

Figures~\ref{fig:AFUV_all_fits} and \ref{fig:AFUV_all_bins_bdg} indicate that there is a possibility to create a global relation between observed \ur{} colour, surface brightness in the \textit{u} band and the attenuation in the FUV band. 
To find a relation, we examine the slopes and the intercepts of the relation between \ur{} and \sbu{} for each \afuv{} bin. 
We show in Fig.~\ref{fig:slopes_intercepts},  
a relation between both slopes and intercepts as a function of \afuv, together with
two linear fits: one for the slopes and one for the intercepts, both as a function of \afuv{}:
\begin{equation}
\begin{aligned}
\label{eq:slop_inetr}
  \rm  slope= (-0.02\pm0.00)\cdot A_{FUV}+0.12\pm0.01{,} \\
\rm  intercept= (0.65\pm0.08)\cdot{} A_{FUV}-1.68\pm0.16{.}
\end{aligned}  
\end{equation}
The increase in scatter for slope ad intercept with decreasing \afuv{} can be explained by smaller bin sizes and less representative samples in the global distribution of \afuv{}. 

We derived a solution for this set of two equations based on the fitted relations (slopes and intercepts with respect to \afuv{}  shown in Eq.~\ref{eq:slop_inetr}), resulting in a linear expression that characterises \afuv{} through the combination of \ur{} and \sbu{}. 
This final relation (Eq.~\ref{eq:AFUV_relation}) incorporates all three values: two entirely observational (\ur{} and \sbu{}) and one physical property (\afuv{}) obtained from the SED fitting from \Gcat{}:
\begin{equation}
\label{eq:AFUV_relation}
\rm A_{FUVp }= \frac{(u-r)-(0.12\cdot \mu_{u})+1.68}{(-0.02 \cdot \mu_{u})+0.65}.
\end{equation}
This equation provides a proxy for the \afuv{} when only optical measurements (fluxes and radii) are available, as will be the case for the majority of the \LSST{} survey\footnote{As discussed above, we obtained a very similar result for the \ui{}--\sbu{} plane:\begin{equation}\label{eq:ui_usb}
 \rm A_{FUVp }= \frac{(u-i)-(0.13\cdot \mu_{u})+1.68}{(-0.02 \cdot \mu_{u})+0.65}. 
\end{equation} The main difference between equations are the larger uncertainties for slopes and intercepts in Eq.~\ref{eq:slop_inetr}.} 
This proxy can significantly shorten the time needed to estimate all physical parameters through the SED fitting, as the grid for the dust attenuation properties can be much narrower and more specific. 

We have also checked that using half-light instead of Petrosian radii will not change{ our relation significantly}.
Using approximately $\sim$1.5 mag/arcs$^2$ brighter values of surface brightness (see Sec.~\ref{sec:SB_calculation}) results in changes of two values from Eq.~\ref{eq:AFUV_relation}: from 1.68 to 2.50 and from 0.65 to 0.74. 
This change results in the mean difference between\afuvprior{} obtained with Petrosian end effective values equal to 0.05 [mag], with $\sigma$=0.45 [mag].

Future \LSST{} observations will provide more precise magnitude and morphology measurements than the data employed in this manuscript, where we adopt Petrosian radii and magnitudes from \sdss{} DR12. 
We plan to perform a similar test for the data acquired from \LSST{} Deep Drilling Fields to better calibrate Eq.~\ref{eq:AFUV_relation} as soon as the observations, both optical from \LSST{} and from other ground-based and satellite observatories are collected. 
For those fields, near-IR data will also be available, for example, VISTA-NIR, which will further help us to constrain reliable dust properties, including attenuation in the FUV band in the broader \afuv{} range. 
\LSST{} will enable the investigation of the lower \afuv{} range, since a substantial percentage of galaxies observed by \LSST{} will be LSB~galaxies. 
While low in comparison to many other types of galaxy, FUV attenuation levels among LSB galaxies are still non-negligible (i.e. \afuv{}<0.4 mag), as shown recently by \cite{Junais2023}.  
While waiting for observations and estimates of the main physical properties of galaxies from the Deep Drilling Field, we can use Eq.~\ref{eq:AFUV_relation} to calculate a proxy representing dust attenuation for star-forming galaxies. 
This relation will be used to prepare a software pipeline to estimate physical properties from future \LSST{} data by members of the \LSST{} Galaxy Science Collaboration \citep{Robertson2017}.  

\subsection{Final \afuv{} prior}
\label{Sec:AFUV_proxy}

\begin{figure*}[h!]
    \centering
        \includegraphics[width=0.99\textwidth]{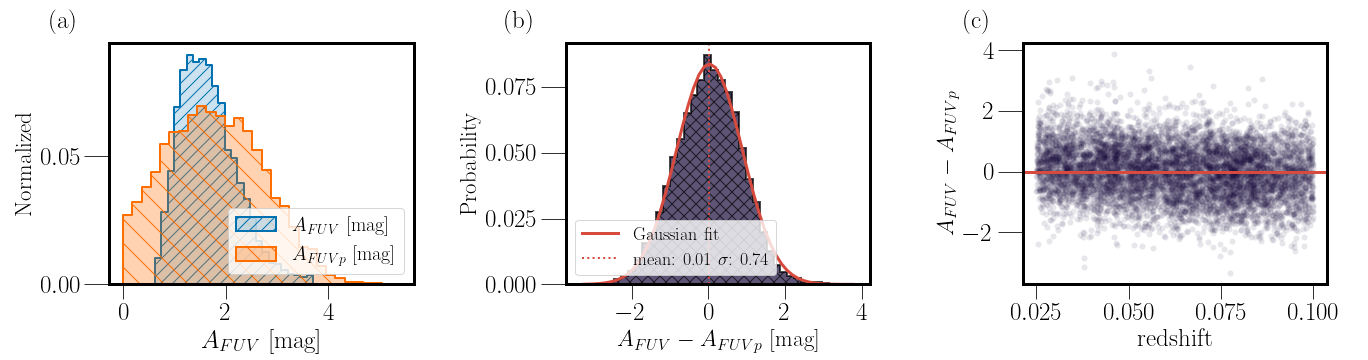}
        \caption{Main properties of obtained \afuv{} priors. Panel \texttt{(a)} shows the distributions of original \afuv{} from the \cite{Salim2018} work obtained from careful SED fitting based on measurements from UV to mid-IR (\afuv{}, blue left hatched histogram), and \afuvprior{} calculated based on Eq.~\ref{eq:AFUV_relation} (orange right hatched histogram).  Panel \texttt{(b)} presents the distribution of the difference between \afuv{} and \afuvprior{} and denotes its median value. In panel \texttt{(c)}, the difference between \afuv{} and \afuvprior{} is shown as a function of redshift. We removed for clarity from panel (a) 426 galaxies, for which calculated \afuvprior{} was lower than~0.   
        }
        \label{fig:AFUV_prior_distribution}
\end{figure*}

The final distribution of the obtained \afuvprior{}, as well as the comparison with \afuv{} from the original work of \cite{Salim2018}, is shown in Fig.~\ref{fig:AFUV_prior_distribution}. 
Panel (a) in this Figure shows the \afuv{} and \afuvprior{} distributions. 
It is clearly seen that the distributions are different and that the prior obtained from \ur{} and \sbu{} can reach both much lower (to the \afuvprior{}=0 mag) and larger (up to \afuvprior{}$>4$ mag) values than the original \afuv{} from \cite{Salim2018}. 
The Kolmogorov–Smirnov test, which is a commonly used non-parametric test based on the distance between two cumulative distributions, confirms that  \afuv{} and \afuvprior{} do not come from the same distribution ($p_{KS}=1.11\cdot 10^{-128}$).
For this plot, we have removed 426 galaxies for which the value of \afuvprior{} from Eq.~\ref{eq:AFUV_relation} is less than 0. 
All galaxies from the removed sample occupy \ur{}-\sbu{} loci not included in our analysis (just below the bin with the lowest value of mean \afuv{} ($\overline{A}_{FUV}=0.74$ [mag]). 
Fig.~\ref{app:AFUV_0} shows the position of all 426 galaxies in the \ur{}-\sbu{} plane. 

We notice here the sharp cut in the low-end of the \afuv{} distribution (visible also in Fig.~\ref{fig:mpp}, right upper panel). 
It can be related to the specific sets of parameters used in \cite{Salim2018} or the data set for which the SED fitting was performed. 
As shown in \cite{Osborne2023}, the \galex{} data, which give direct insight into the young stellar population and were used to create \Gcat{}, are partially affected by blending. 
The new catalogue built with a new software pipeline \texttt{EMphot}, which uses forced photometry from the \sdss{} catalogue, presented by \cite{Osborne2023}, revealed that magnitudes used in \Gcat{} were systematically fainter (up to 0.5 mag) due to insufficient background subtraction for faint sources. 
The new, deblended, \galex{} catalogue of \cite{Osborne2023} shows that $\sim$15\% of galaxies in the \Gcat{} catalogue were moderately affected by blending (contamination $>$0.2 mag), and 2.4\% of galaxies were contaminated at the level of more than 1~magnitude. 
To summarise, the NUV and FUV \galex{} magnitudes originally used to estimate \afuv{} by \cite{Salim2018} were fainter than the corrected deblended magnitudes, but only a small percentage of galaxies in our sample can be affected by this effect. 
It means that \galex{} data used by \cite{Salim2016ApJS} has negligible influence on the lack of low \afuv{} values in the original \Gcat{}-X2 catalogue for the main sequence galaxies\footnote{Based on private communication with S.~Salim, we have found that the \afuv{} distributions based on the previous \galex{} data used for the \Gcat{} catalogue and the \afuv{} obtained with the new, deblended \galex{} measurements from \cite{Osborne2023} have a statistically negligible change.}.

The mean difference between \afuv{} and \afuvprior{} equals $0.01$~mag, with a $\sigma=0.74$ {magnitude} (see middle panel of Fig.~\ref{fig:AFUV_prior_distribution}). 
From hereupon, we use this $\sigma$, which is the scatter in the difference between fiducial values of \afuv{} and \afuvprior{}, as a constant uncertainty for our estimated \afuvprior{} (hereafter: \afuvpriorerr{}). 
We want to stress that, in the future, with the \LSST{}-like observations, it will also be possible to calculate the \afuvpriorerr{} directly for individual sources based on the uncertainties in the observed colour, surface brightness, and the fit coefficients shown in Eq.~\ref{eq:AFUV_relation}. 
However, we are currently restricted to a limited number of galaxies, which affects the uncertainties of our fits from Eq~\ref{eq:slop_inetr}. 
Additionally, we have significant photometric errors (for both radii and magnitudes).  
Furthermore, the fiducial \afuv{} is not free of uncertainties (limited to \afuverr{} $<0.25$ {magnitude}, based on our selection in Table~\ref{table:parameters}). 
Therefore, for the simplicity of this work and to avoid significant overestimations of errors, we decided to use a constant uncertainty of \afuvpriorerr{}=0.74 {magnitude}.

We do not observe any redshift dependence (panel (c), Fig.~\ref{fig:AFUV_prior_distribution}); however, the redshift range used in this analysis is very narrow (0.025--0.100). 
We can reasonably expect that the much deeper \LSST{} data will require adding a redshift-dependent calibration. 

The obtained priors do not follow a 1:1 relation with the \afuv{} values calculated directly from fits to the UV-IR range SED. 
This is due to many reasons, where the most important ones are 
(1) uncertainties of the original \afuv{}, \ur{} and \sbu{}, which were not taken into account when constructing the \ur{} - \sbu{} - \afuv{} relation; 
(2) the quality of the SED fits obtained by \cite{Salim2018} (our only selection is based on  \texttt{REDCHISQ}$<$5); 
(3) the quality of the data used for full fitting --- there is no information about the {signal-to-noise ratio} for specific bands or the goodness of the measurement; 
(4) but even more importantly, the \ur{}-\sbu{} relation is not tight as the \ur{} colour depends both on the age of the stellar population and on the dust. 
Nevertheless, the median difference between \afuv{} obtained based on the careful fitting of broadband photometry from UV to IR and \afuvprior{} determined from \ur{} and \sbu{} observed quantities is only 0.10 {magnitude} larger than the median \afuvpriorerr{} (0.12 {magnitude}). 

\section{{Reliability of \afuvprior{} derived from LSST-like observations}}
\label{sec:comparison_with_salim}

Even if, as mentioned above, the agreement between \afuvprior{}~$\pm$~\afuvpriorerr{} and \afuv{} estimated via SED fitting by \cite{Salim2016ApJS,Salim2018} is not perfect, the correlation is evident, and the advantages of such an approach are numerous. 
An \afuv{} prior obtained from an optical-only dataset (without any information about dust emission or proxy from UV observations) can help to reduce the number of parameters needed to estimate the main physical properties of studied galaxies. 
It can also reduce the risk of overfitting by decreasing the number of free parameters used for SED fitting based on five optical broadbands only. 

To check whether the obtained \afuvprior{} values can lead to reliable estimates of the main physical properties of galaxies, we perform a set of tests using only optical broadband data with and without two priors: the original \afuv{} from \Gcat{} obtained by \cite{Salim2018}, which we call now \afuvsalim{} (and \afuvsalimerr{}) to distinguish between both \afuv{} values used in the test, and the \afuvprior{} (and \afuvpriorerr{}) from Eq.~\ref{eq:AFUV_relation}.

To test using \afuvprior{} obtained via Eq.~\ref{eq:AFUV_relation} as a prior, we perform six CIGALE runs to fit the SED of our sample. 
We use \sdss{} DR 12 \textit{ugriz} measurements from \cite{Alam2015ApJS}  \cite[in case of ][they used \sdss{} DR10]{Salim2016ApJS,Salim2018}.  
\cite{Salim2016ApJS} and \cite{Salim2018} made use of additional data from \galex{} and \wise{} which are not taken into account in our analysis. 
A simple run based on five optical bands only is intended to reproduce the future \LSST{}-like observations (without the \LSST{}~\textit{y} band). 
We employ the same SED fitting code as used for the \Gcat{} data set. 
Parameters and modules used are described in Table~\ref{table:parameters}. 
As our SED coverage consists of only five optical SDSS data points, we do not include any dust emission module. 

We categorise the SED fitting parameters into two main groups: \Glike{} (based on the description given by \citealp{Salim2016ApJS,Salim2018}) and \Llike{}, with a significantly reduced number of parameters describing dust attenuation  {and the age of the late burst for the star formation history module (details are listed in Table~\ref{table:parameters})}. 
Runs based on \Glike{} parameters produce 332\,640 templates per redshift bin. 
In contrast, {utilising} \Llike{} parameters reduces the number of templates to only 5\,540 per redshift bin. 
Thus, the number of generated templates decreases by 98\%. 
We performed our runs for 7\,934 galaxies using Intel Core i9--9900K CPU @ 3.60~GHz processor with 64~GB memory and 8 cores (16 threads).  
\Glike{}  required 188~seconds to compute models while the \Llike{}  did the same in 11~seconds. 
The Bayesian estimates of the physical properties for these templates for the \Glike{}  \cigale{} took 203~seconds, while for the \Llike{}, only one second. 
Thus, both runs used the same time to estimate best-fit properties for all galaxies (85~seconds). 
Thus, the \Llike{}  required only 3\% of the time spent on the \Glike{}. 
This reduction in the number of templates is of particular significance for `big data' galaxy samples like the \LSST{}. 
For \LSST{}-like surveys, with billions of observed galaxies, running full, detailed SED fitting will be impossible due to CPU and memory limitations. 

We used the \LSST{}-like data set, and we performed the SED fitting using \Glike{} and \Llike{} parameter sets.  For \Glike{} and \Llike{} parameters we further divide runs into three groups:  (
1) without any priors (tagged as NO prior),  {without any indication of a preferable \afuv{} value,} (
2) with the \afuvsalimerr{} and \afuvsalimerr{}{} used as priors for the CIGALE run, and 
(3) with \afuvprior{} and \afuvpriorerr{}. 
To add priors, we used the \texttt{properties} option from the original \cigale{} tool\footnote{\label{fn_properties} To run \cigale{} with the \texttt{properties} option, one has to add to the initial input file additional columns filled with prior values and corresponding errors. 
In the case of this work, we added two columns: attenuation.AFUV and attenuation.AFUV\_err, and we filled them with \afuvprior{} from Eq.~\ref{eq:AFUV_relation} and \afuvpriorerr{}  equal to 0.74 {magnitude}, respectively. },  while for runs without priors, we left the \texttt{properties} option empty. 
Runs with and without priors still need to be supported by the input parameters for the dust attenuation module. In all six cases, we used input parameters as listed in Table~\ref{table:parameters}.  

In Fig.~\ref{fig:comparison_Mstar_SFR}, we present the difference between estimated \mstar{} and SFRs obtained via our six runs and the original (fiducial) values from the \Gcat{} catalogue. 
We stress that for all six runs, only \textit{u}, \textit{g}, \textit{r},  \textit{i}, and \textit{z} SDSS broadband photometry was used. 
The resultant loci of the obtained six main sequences are shown in Fig.~\ref{fig:comparison_six_runs}. 

It is important to note that processing \Llike{} data with \cigale{} without \afuv{} prior leads to large overestimates of the SFR \cite[see also][]{Riccio2021}. 
Priors, even if based only on optical measurements, can reproduce the physical parameters of the fiducials well enough. 
Moreover, priors can also help to reduce the number of parameters, and hence the CPU time, required to analyse large numbers of galaxies.  

\begin{table*}
\small
\caption{Input parameters for the code CIGALE.}
\label{table:parameters}  
\centering       
\renewcommand{\arraystretch}{1.1} 
\begin{tabular}{p{0.57\textwidth}|p{0.37\textwidth}}       
\hline             
Parameters & Values \\   
\hline
\hline
\multicolumn{2}{c}{Star formation history:} \\\hline\hline
\multicolumn{2}{l}{double exponential (delayed with additional burst)}\\ \hline
e-folding time of the main stellar population model (Myr) & 500, 1000, 3000, 5000, 8000, 10000, 15000, 20000 \\
e-folding time of the late starburst population model (Myr) & 20000\\
Mass fraction of the late burst population & 0, 0.05, 0.1, 0.15, 0.2,  0.25, 0.3, 0.35,0.4, 0.45, 0.5 \\
Age of the main stellar population (Myr) &  6500\\
Age of the late burst (Myr) & 10,  {30}$^*$,  {100}$^*$, 300, 1000, 3000, 5000\\ \hline \hline                                   
\multicolumn{2}{l}{ Single stellar population  \cite{BC03}} \\\hline\hline
Initial mass function & \cite{Chabrier2003}\\
Metallicities (solar metallicity) & $0.02$\\
Age of the separation between the young and the old star population (Myr) & $10$\\
\hline\hline
\multicolumn{2}{l}{ Nebular} \\\hline\hline
Ionisation parameter & -3.0\\
\hline\hline
\multicolumn{2}{l}{ \Glike{}: Dust attenuation law  \cite{Calzetti2000}} \\\hline\hline
$E(B-V)$:  the colour excess of the stellar continuum light for the young population &  0, 0.01, 0.02, 0.05, 0.1, 0.15, 0.2, 0.25, 0.3, 0.35, 0.4, 0.5, 0.6, 0.7, 0.8\\
 Amplitude of the UV bump at 217.5~nm & 0, 2.0, 4.0, 6.0\\
Slope of the power law modifying the attenuation curve&  0.4, 0.2, 0, -0.2, -0.4, -0.6, -0.8, -1., -1.2\\
\hline\hline
\multicolumn{2}{l}{  {\Llike{}: Dust attenuation law}  \cite{Calzetti2000}  } \\\hline\hline
 {$E(B-V)$:  the colour excess of the stellar continuum light for the young population} &  {0,  0.05, 0.15, 0.3, 0.4, 0.5, 0.6, 0.7, 0.8} \\
 { Amplitude of the UV bump at 217.5~nm} &  {0}\\
 {Slope of the power law modifying the attenuation curve} &  {0}\\
\hline\hline

 \multicolumn{2}{l}{{ \scriptsize{$^{*}$   {\textit{Notes: }values 30 and 100 was removed from the \Llike{}. }  }}}

\end{tabular}
\vspace{1mm}
\end{table*}

\section{Discussion}
\label{sec:discussion}
 \begin{figure*}[t!]
    \centering
        \includegraphics[width=0.99\textwidth]{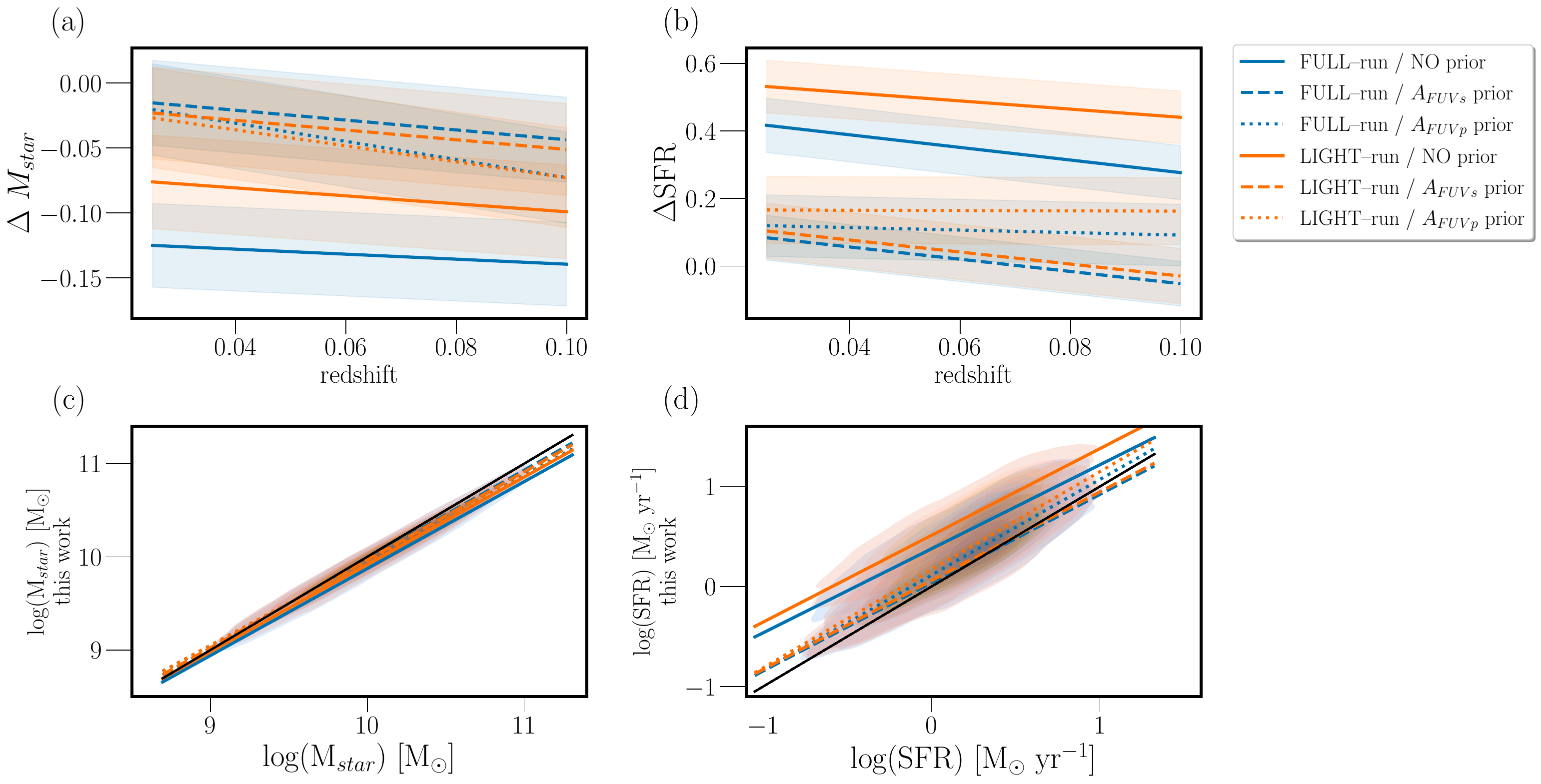}
        \caption{Difference between SFR and M$_{star}$ from the original \Gcat{} catalogue \citep{Salim2018}, and those obtained in this work. The difference $\Delta M_{star}$  was calculated as log(M$_{star\mbox{ \tiny{this work}}}$/M$_{star }$); and $\Delta SFR$ was obtained in an analogous way. Panels \texttt{(a)} and \texttt{(b)} show the difference in the stellar masses and SFRs, correspondingly, as a function of redshift,  {with shaded areas represent standard deviation of the scatter; } panels \texttt{(c)} and \texttt{(d)} present the relation between the fiducial values and the estimates obtained in this work {, with contours showing the distribution of the data}. Runs based on \texttt{GSWLC} set of parameters, \Glike{}, are shown in blue-coloured lines, while those obtained from \Llike{} -- in an orange-coloured set of lines.  
        Black solid lines visible in panels \texttt{(c)} and \texttt{(d)} represent 1:1 relations. 
        }
        \label{fig:comparison_Mstar_SFR}
\end{figure*}

Figures~\ref{fig:comparison_Mstar_SFR}~and~\ref{fig:comparison_six_runs} show the differences between the original (fiducial) physical parameters from the \Gcat{} catalogue and the physical parameters obtained with two sets of parameters: \Glike{}, and highly reduced \Llike{}. 
Additionally, different \afuv{} priors (and also no prior for even more detailed comparison) were applied in these runs. 
Table~\ref{tab:meanSFR_Mstar} shows the mean SFR and the mean \mstar{} accompanied by their uncertainties for fiducial parameters from the \Gcat{} and those obtained from six additional runs described in Sec.~\ref{sec:comparison_with_salim}.

\begin{table}
    \centering
    \begin{tabular}{l| c | c}
    \multirow{2}{*}{data set} &  \multirow{2}{*}{log(SFR/M$_{\odot}$yr$^{-1}$)} & \multirow{2}{*}{log(\mstar/M$_{\odot}$)}\\
     & & \\
    \hline \hline
    fiducial values                & 0.19 $\pm$ 0.07 & 10.09 $\pm$ 0.05 \\
    \Glike{}/ NO prior          & 0.54 $\pm$ 0.41 & 9.96 $\pm$ 0.17\\
    \Glike{}/ \afuvsalim{} prior & 0.20 $\pm$ 0.14 & 10.06 $\pm$ 0.09\\
    \Glike{}/ \afuvprior{} prior & 0.30 $\pm$ 0.21  & 10.04 $\pm$ 0.11\\
    \Llike{}/ NO prior           & 0.68 $\pm$ 0.36 &  10.01 $\pm$ 0.11\\
    \Llike{} / \afuvsalim{} prior & 0.23 $\pm$ 0.07 & 10.06 $\pm$ 0.08\\
    \Llike{}/ \afuvprior{} prior & 0.28 $\pm$ 0.10 & 10.05 $\pm$ 0.08\\
    \hline \hline
    \end{tabular}
    \caption{Mean values of log(SFR) and log(\mstar{}) for the fiducial catalogue of 7\,986 galaxies and those estimated from all six runs described in Sec.~\ref{sec:discussion}, and presented in Figs.~ \ref{fig:comparison_Mstar_SFR} and \ref{fig:comparison_six_runs}. }
    \label{tab:meanSFR_Mstar}
\end{table}

\subsection{Influence of the \afuvprior{} or lack of it on the \mstar{} estimation for the \LSST{}-like observations}
\label{sec:stellar_mass}

Results presented in Fig.~\ref{fig:comparison_Mstar_SFR}, panels \texttt{(a)} and \texttt{(c)}, show a negligible, $\leq0.1$~dex,  difference in the estimated \mstar{} between all six runs and the original \mstar{} from the \Gcat{} catalogue. 
The consistency in the estimations is actually expected given the complete set of the \sdss{} DR~12 optical broadband data used by \cite{Salim2018} and in our work. 
Detailed coverage of the optical spectrum, from \textit{u} to \textit{z} bands, allows reconstruction of the old stellar population in the galaxy (especially at low redshift), which is the main ingredient of the total stellar mass of both quiescent and normal star-forming galaxies. 

However, it is worth mentioning that only for runs without prior $\rm \Delta M_{star}$, defined as log(M$_{\rm star\mbox{ \tiny{this work}} }$/\mstar), is lower than $-0.07$, and monotonically decreases with redshift. 
It implies that for optical data only, without any proxy for dust emission or \afuv{}, the Bayesian method of estimating physical parameters prefers to choose templates corresponding to somewhat less massive galaxies. 
It is clearly visible when comparing runs without priors: for the run with a larger number of templates (marked as \Glike{}) $\rm \Delta$\mstar{} is systematically shifted towards lower values of \mstar{} when compared to the \Llike{}. 
We want to stress here that this effect can be caused by overfitting \citep[{it can be seen} also Fig.~7 in][ where the \mstar{} is very slightly, but still overestimated due to the number of used templates]{Riccio2021}. 
The number of parameters used in \cite{Salim2018} was allowed thanks to a larger number of measurements available, which is not the case in the \LSST{}-like dataset. 
We stress that possible overfitting should be avoided for the \LSST{}-like data analysis, as it may cause the choice of templates of systematically less massive galaxies, which is another reason why introducing a prior to reduce the size of the parameter grid is needed.

\subsection{Influence of the \afuvprior{} {and }lack of the prior on the SFR estimation for the \LSST{}-like observations}
\label{sec:sfr}

In this section, we check how the use of priors influences the estimations of the SFR. 
Again, we calculate the ratio between the fiducial physical value from \Gcat{} catalogue and those obtained in our runs. 
We define this ratio, $\Delta$SFR, as log(SFR$_{\tiny{\mbox{this work}}}$/SFR). 
The results are shown in  Fig.~\ref{fig:comparison_Mstar_SFR}, panels \texttt{(b)} and \texttt{(d)}.

\subsubsection{{No \afuv{} prior involved}}

The first conclusion from this test, presented in the right panels of Fig.~\ref{fig:comparison_Mstar_SFR}, is that using optical data only without any prior 
(that is, without giving any values for the \texttt{properties} option, see note~\ref{fn_properties}) for the dust attenuation results in a significant overestimation of the SFR, which decreases with redshift. 
The same overestimation and its redshift dependence were found by \cite{Riccio2021} for a sample of $\sim$50\,000 main sequence galaxies. 
They used a sample of observed galaxies to estimate the expected \LSST{} fluxes in the \textit{ugrizy} bands and then performed SED fitting. 
They found that the $\rm M_{star}$ remains well estimated (similarly to our result shown in Sect.~\ref{sec:stellar_mass}) by the \LSST{}--like data set. 
However, at the same time, the SFR and the dust luminosity are overestimated when the \LSST{}--like sample alone is used. 
The SFR overestimation found by \cite{Riccio2021} is redshift dependent and clearly decreases with redshift, disappearing at about redshift $\sim$1. 
This effect can be explained by the wavelength range of the \LSST{} observations. 
At redshift $\sim$0 \LSST{} probes mainly old stellar population, without any band probing young stellar population or dust properties. 
As redshift increases, the \LSST{} $ugriz$ filters start to cover the UV rest frame, and the estimates of the SFR significantly improve. 

The lack of information about the UV and MIR rest-frame wavelengths for the \LSST{} low redshift sample causes a large overestimation of the attenuation during the fitting \citep{Riccio2021}, which then translates into an overestimation of the dust luminosity and, finally, the SFR. 
As shown in  Fig.~\ref{fig:comparison_Mstar_SFR}, both \Glike{} and \Llike{} sets of parameters, applied without any \afuv{} prior (blue and orange solid lines, respectively), result in significant SFR overestimation: mean $\Delta$SFR for \Glike{} and \Llike{} runs without priors equal to 0.34 and 0.48, respectively. 
Based on Eq.~1 from \cite{Riccio2021}, {that is,} $\Delta\mathrm{SFR}=\rm SFR_{LSST}/SFR_{UV\mbox{-}FIR}$, the expected $\Delta$SFR at redshift 0.062 (which is the mean redshift of our galaxy sample) is equal to $\sim$0.5. 
As seen from panels \texttt{(b)} and \texttt{(d)} of Fig.~\ref{fig:comparison_Mstar_SFR}, our results, although limited to a much narrower redshift range, are in agreement with predictions of \cite{Riccio2021}.

We stress that enlarging the parameter space in the SED fitting process, with only a stellar population as a proxy, cannot solve the problem of the overestimation of SFR. 
This conclusion is illustrated in Fig.~\ref{fig:comparison_six_runs}. 
In this figure, the black contours are based on the original catalogue of \cite{Salim2018},  while the orange contours {show} the estimates used in this work. 
It is clearly seen that runs based on the optical data only without any priors result in a significant systematic overestimation of SFR (panels \texttt{(a)} and \texttt{(b)}). 
The overestimation occurs both for the \Glike{} and the \Llike{} parameter grid. 
The overestimation of the SFR results in a shift of the main sequence locus and can affect not only directly the physical analysis but also the classification of normal star-forming and starbursting galaxies. 
Such classification can propagate through the statistical analysis of evolutionary paths and many other science cases.
  
\subsubsection{{Overcoming SFR overestimation with \afuv{} and \afuvprior{}} }

In \cite{Riccio2021}, it was also found that the $\rm M_{star}$--\afuv{} prior can solve the SFR overestimation for objects without UV or IR observation. 
As seen in the panel~\texttt{(b)} of Fig.~\ref{fig:comparison_Mstar_SFR}, SFR estimated with \afuv{} priors (both \Glike{} and \Llike{}) are much closer to the fiducial  {values} than the SFR obtained without any priors. 
The $\Delta$SFR for runs with priors are lower than 0.1. 

\begin{figure*}[t!]
    \centering
        \includegraphics[width=0.98\textwidth]{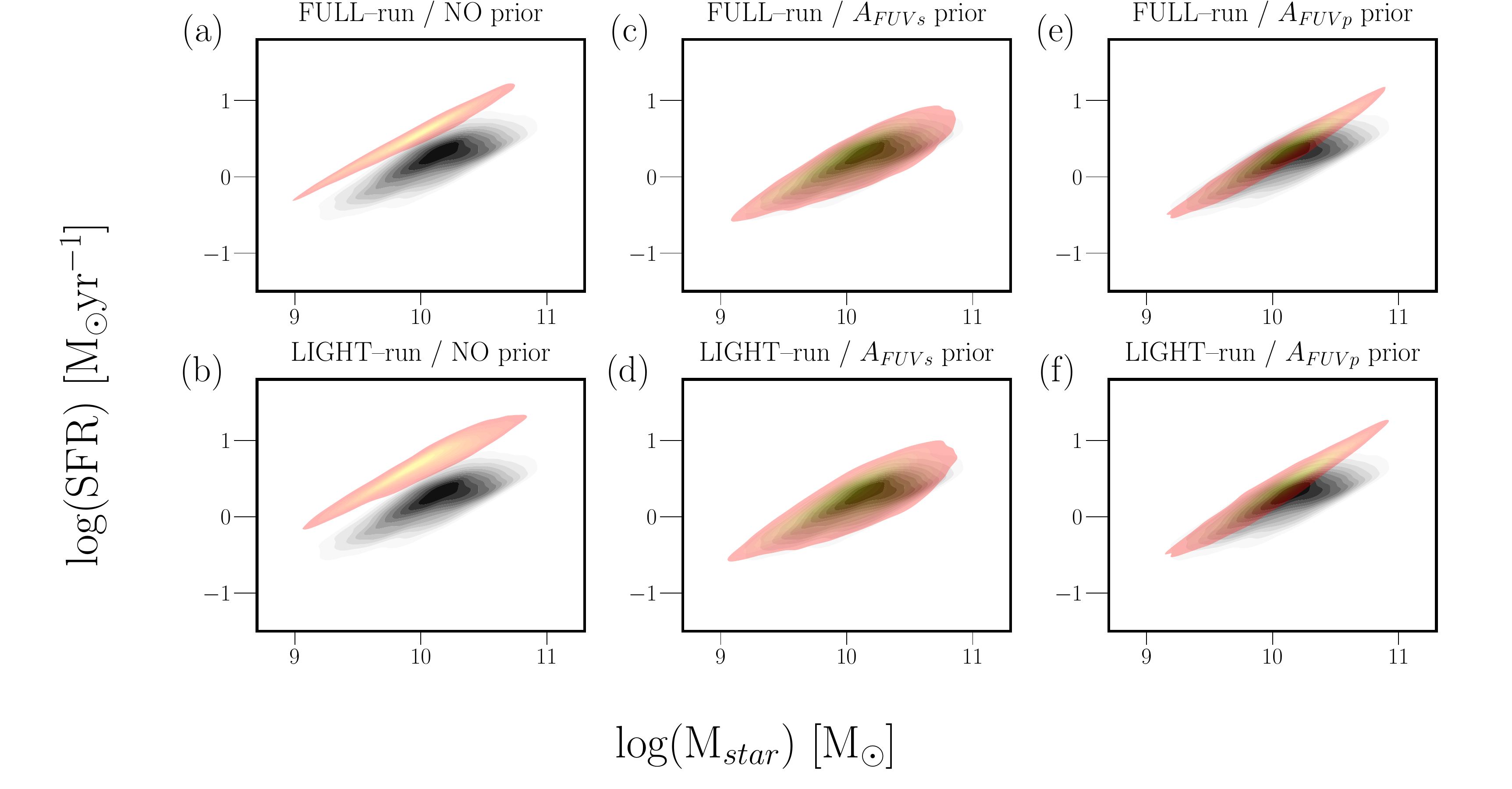}
        \caption{{Comparison of physical properties obtained in \Glike{}, and \Llike{} runs.} Upper panels present results obtained based on \Glike{} parameter sets (332\,640 templates per redshift bin), while the results shown in bottom panels were estimated based on the \Llike{} sets of parameters (5\,544 templates per redshift bin).  {Black} contours shown in each panel correspond to fiducial values from the \Gcat{}-X2 catalogue (SFR, $\rm M_{star}$), while {orange} contours illustrate physical parameters estimated in this work. The left column presents results obtained without priors, the middle column -- with \afuvsalim{} from the \Gcat{} catalogue, and the right column -- with  \afuvprior{} calculated from Eq.~\ref{eq:AFUV_relation}.   
        }
        \label{fig:comparison_six_runs}
\end{figure*}

The quality of SFR reconstruction can be seen in the central and the right columns in Fig.~\ref{fig:comparison_six_runs}. 
Adding \afuv{} prior (original \afuvsalim{} from \Gcat{} catalogue or \afuvprior{} from Eq.~\ref{eq:AFUV_relation}) clearly reduces the overestimation. 
The calculated SFR ratio between mean fiducial values from \cite{Salim2018} and those obtained in this work with \afuvsalim{} prior is less than 0.05, and with \afuvprior{} -- less than 0.1. \\  

\paragraph{\underline{Original \afuv{} prior.}}
Results presented as blue and orange dashed lines (\Glike{} and \Llike{}, respectively) in Fig.~\ref{fig:comparison_Mstar_SFR}  are based on the \afuvsalim{} prior. 
Although the difference with respect to the fiducial SFR values is almost negligible ($\Delta\mathrm{SFR}=0.1$), the relation is not 1:1 (which would correspond to $\Delta\mathrm{SFR}=0$). 
In both cases, \Glike{} and \Llike{} sets of parameters, the difference with respect to the fiducial value of SFR, $\Delta$SFR, decreases with redshift.

This small difference results from a narrower wavelength coverage than used by \citet{Salim2018}. 
At the same time, these two runs demonstrate that \afuvsalim{} prior works almost equally well for the dense (\Glike{}) and significantly reduced (\Llike{}) grids of parameters.

The central panels in Fig.~\ref{fig:comparison_six_runs} (\texttt{c} and \texttt{d}) show the main sequence built using the original \afuvsalim{} prior for \Glike{} (panel \texttt{c}) and \Llike{} (panel \texttt{d}) runs. 
The original main sequence from the \Gcat{} catalogue is below the orange contours. 
These two panels show that the use of the original \afuvsalim{} prior reproduces the fiducial values of the SFR equally well with fewer parameters (we remind that the number of templates in the \Llike{} case is 98\% lower than in the \Glike{} case). 

Of course, without ancillary data from other facilities, the actual value of \afuv{} will be unknown for most of \LSST{} galaxies. 
The purpose of this test was to demonstrate that 
(1) providing an \afuv{} prior is instrumental in the recovery of SFR and 
(2) that an \afuv{} prior can be used along with a small number of free parameters in the fitting process, which will reduce the risks of overfitting in the case of {a} low number of data points. 
However, galaxies at redshifts higher than $\sim 2$ will benefit from the restframe FUV observations with the $u$ band of LSST. 
This will give unprecedented power for probing the attenuation in this band specifically. 
Therefore, attenuation in FUV can be used as a prior for high redshift galaxies, correcting their SFR and \mstar{} estimations when IR data are missing.\\ 

\paragraph{\underline{Results based on the \afuvprior{} from \ur{}--\sbu{} relation.}}

To test the effect of the prior obtained from Eq.~\ref{eq:AFUV_relation}, we run \Glike{} and \Llike{} setup of parameters with \afuvprior{}. 
SFRs obtained from this test are shown with blue and orange dotted lines (\Glike{}, and \Llike{}, respectively) in panels~\texttt{(b) and (d)} of Fig.~\ref{fig:comparison_Mstar_SFR}. 
Both results are very close to the fiducial values \citep[$\Delta\mathrm{SFR}<0.1$, very similar to the results obtained with an original prior estimated from full UV--IR SED fitting,][]{Salim2018}. 
The \Glike{} parameter setup with an \afuvprior{} provides values of SFR which have a small bias almost uniform along all the probed ranges of the SFR, with both mean and median $\Delta$SFR equal to 0.11). 
In the run with a smaller grid of parameters (\Llike{}+\afuvprior{}), a similar small bias is observed (the median and the mean $\Delta\mathrm{SFR}=0.09$. 
This small overestimation rises somewhat with redshift.

The right column of Fig.~\ref{fig:comparison_six_runs} (panels \texttt{e} and \texttt{f}) shows the comparison of the main sequences of 7\,934 galaxies based on physical properties obtained from the original \Gcat{} catalogue and those calculated using \afuvprior{} from Eq.~\ref{eq:AFUV_relation}. 
It can be seen that for the case of a large number of parameters (\Glike{}, panel \texttt{e}) and a much smaller number of templates (\Llike{}, panel \texttt{f}), the obtained main sequence is narrower than the original one, and the overestimation can be seen for more massive galaxies. 
The obtained main sequence is, however, much closer to the fiducial one than in the case of fitting with no priors. 
We emphasize that this may be related to a lower fraction of massive galaxies in our sample (see Fig.~\ref{fig:mpp}) but also to the representative original \afuvsalim{} estimates, which have not been taken into account in our fit of the slopes--intercepts. 

The linear relations found between \sbu{} and the \ur{} colours for all the\afuv{} bins show that the morphological aspect of galaxies might be an indicator of attenuation. 
Surface brightness encodes in itself the spatial extent of galaxies, and indications of a correlation between the amount of attenuation on the one hand and the compactness of galaxies on the other hand were found in \citet{buat2019,Hamed2023, Hamed2023b}. 
The \sbu{} can also be correlated with the age of stellar populations, and so it helps to break the degeneracy of the \ur{} colour (which depends both on the dust attenuation and the stellar population age).

We conclude that the \Llike{}+\afuvprior{} run results in only the small effect of SFR overestimation, visible mainly for massive galaxies.  
For the case of the optical \sdss{} data only, the reduced number of templates and \afuvprior{} created based on \ur{} colour and \sbu{} are thus optimal, although not perfect, solution{s} to reproduce the main physical properties of galaxies even without IR or UV data.  

\section{Summary}
\label{sec:summary}

In this work, we aimed at constructing the proxy of dust attenuation in the FUV wavelength range from the optical photometric \LSST-like data only. 
To imitate the \LSST{} optical detections of low redshift main-sequence galaxies, we used \sdss{} observations. 
Furthermore, for the analysis of the dust attenuation, we selected only galaxies with IR auxiliary observations, with homogeneous data analysis and estimated main physical parameters, such as \mstar{}, SFR and \afuv{}.  
We selected a sample of 7\,934 local ($0.025<z<0.1$) main-sequence star-forming \sdss{} galaxies with properties previously measured based on multiwavelength IR-to-UV photometric data. 
We used the \Gcat{} catalogue of physical properties  citep{Salim2018} together with their \sdss{} DR~12 \textit{ugriz} measurements \citep{Alam2015ApJS} and inclination from cite{Meert2015MNRAS.446.3943M}. 
We calculated their surface brightness in the \textit{ugri} bands and corrected for inclination and cosmological surface brightness dimming. 

We tested different setups of the \cigale{} SED fitting code in order to find an optimal method to estimate the main physical properties of galaxies when only the $ugriz$ photometric data are available. 
Our results can be {summarised} as follows:

\begin{enumerate}
    \item We find that the proxy for dust attenuation, \afuvprior{}, can be constructed based on the combination of the \ur{} colour and the galaxy surface brightness in the $u$ band (\sbu). 
    The formula is given by Eq.~\ref{eq:AFUV_relation}. 
    The formula remains almost the same if the colour \ui{} is used instead of \ur{}, Eq.~\ref{eq:ui_usb} in the footnote.  
    \item Surface brightness measured in bands other than $u$ do not provide an equally good estimate for \afuv{}, as it can be seen in Fig.~\ref{App:A_global}. 
    \item The \mstar{} can be well recovered by the SED fitting in the optical range only, with no necessity to use additional priors. 
    \item The SFR cannot be well measured based on the SED fitting method in the optical range only due to missing information about dust attenuation. 
    In the case of \cigale{}, it results in systematic and redshift-dependent overestimation of SFR at the level of $\Delta\mathrm{SFR}>0.3$~dex, which is in agreement with the findings of \cite{Riccio2021}.
    \item The SFR can be recovered almost unbiased, with $\Delta\mathrm{SFR} <0.05$~dex, if the fitting is performed with an \afuv{} proxy. 
    However, such a proxy will not be available for all future \LSST{} data without auxiliary data.
    \item The \afuvprior{} given by Eq.~\ref{eq:AFUV_relation}, used as a proxy for the SED fitting, allows for very good recovery of SFR with only a small bias $\Delta\mathrm{SFR}\sim 0.1$~dex.
    \item This bias is both \mstar{} and redshift dependent, which implies that the \afuvprior{} proposed in Eq.~\ref{eq:AFUV_relation} in the future will have to be refined and {generalised} based on the calibration with deeper data and larger galaxy samples.
\end{enumerate}
Additionally, we have found that when a prior for dust attenuation is used, the parameter grid used by the fitting code can be significantly reduced in order to avoid overfitting, which, in particular, tends to lead to overestimation of SFR. 
Also, thanks to the reduction of the number of generated templates by $\sim$98\%, we decrease the computing time needed for fitting by a comparable factor, which will be of great importance for the 'big data' analysis of the \LSST{} data.

Thus, this work proposes a strategy based only on optical photometric measurements to reliably and efficiently measure galaxy physical properties through SED fitting in future large surveys. 
The next steps will include the analyses based on galaxy samples deeper than the \sdss, complemented by the state-of-the-art simulations, which will allow for extensions of the proposed strategy toward{s} higher redshift and lower brightness data sets, as well as cover different types of galaxies, which will likely require a diversified approach.



\begin{acknowledgements}
We acknowledge and thank the referee for a thorough and constructive report, which helped improve this work. \\
This research was done as a part of the NCBJ’s In-kind Contributions to the Vera C. Rubin Observatory Legacy Survey of Space and Time POL-NCB-S3.  
This research was supported by the Polish National Science Centre grant UMO-2018/30/M/ST9/00757 and the Polish Ministry of Science and Higher Education grant DIR/WK/2018/12. 
K.~M and J. have been supported by the National Science Centre (UMO-2018/30/E/ST9/00082). 
This research was also partially supported by the 'PHC POLONIUM' programme (project number: 49136QB), funded by the French Ministry for Europe and Foreign Affairs, the French Ministry for Higher Education and Research and the Polish NAWA. 
The project is co-financed by the Polish National Agency for Academic Exchange (BPN/BFR/2022/1/00005).
{ This research was supported by} from COST Action CA21136 – 'Addressing observational tensions in cosmology with systematics and fundamental physics (CosmoVerse)', supported by COST (European Cooperation in Science and Technology). 
R.~D. gratefully acknowledges support by the ANID BASAL project FB210003. M.~B., C.~S., and M.~A. acknowledge support from the  ANID through FONDECYT grants (no. 1211000, 11191125 and 1211951) and Agencia Nacional de Investigaci\'on y Desarrollo (ANID) BASAL project FB210003.  
D.~D. acknowledges support from the National Science Center (NCN) grant SONATA (UMO-2020/39/D/ST9/00720). 
J.~R.~M. is supported by STFC funding for UK participation in LSST, through grant ST/Y00292X/1. 
A.~N. and M.~R. acknowledge support from the Narodowe Centrum Nauki (UMO-
2020/38/E/ST9/00077), and M.~R. also acknowledge support from the Foundation for Polish Science (FNP) under the program START 063.2023.
O.~D. acknowledges support from the Programa de doctorado en Astrofísica y Astroinformática of Universidad de Antofagasta. 
W.~J.~P. has been supported by the Polish National Science Center project UMO-2020/37/B/ST9/00466. 
JR acknowledges funding from the University of La Laguna through the Margarita Salas Program from the Spanish Ministry of Universities ref. UNI/551/2021-May 26, and under the EU Next Generation.

This work made use of Astropy:\footnote{http://www.astropy.org} a community-developed core Python package and an ecosystem of tools and resources for astronomy \citep{astropy:2013, astropy:2018, astropy:2022}

\end{acknowledgements}

\bibliographystyle{aa}
\bibliography{aa48432.bib}

\begin{thebibliography}{88}
\expandafter\ifx\csname natexlab\endcsname\relax\def\natexlab#1{#1}\fi

\bibitem[{{Ahn} {et~al.}(2014){Ahn}, {Alexandroff}, {Allende Prieto}, {Anders},
  {Anderson}, {Anderton}, {Andrews}, {Aubourg}, {Bailey}, {Bastien},
  {Bautista}, {Beers}, {Beifiori}, {Bender}, {Berlind}, {Beutler}, {Bhardwaj},
  {Bird}, {Bizyaev}, {Blake}, {Blanton}, {Blomqvist}, {Bochanski}, {Bolton},
  {Borde}, {Bovy}, {Shelden Bradley}, {Brandt}, {Brauer}, {Brinkmann},
  {Brownstein}, {Busca}, {Carithers}, {Carlberg}, {Carnero}, {Carr},
  {Chiappini}, {Chojnowski}, {Chuang}, {Comparat}, {Crepp}, {Cristiani},
  {Croft}, {Cuesta}, {Cunha}, {da Costa}, {Dawson}, {De Lee}, {Dean},
  {Delubac}, {Deshpande}, {Dhital}, {Ealet}, {Ebelke}, {Edmondson},
  {Eisenstein}, {Epstein}, {Escoffier}, {Esposito}, {Evans}, {Fabbian}, {Fan},
  {Favole}, {Femen{\'\i}a Castell{\'a}}, {Fern{\'a}ndez Alvar}, {Feuillet},
  {Filiz Ak}, {Finley}, {Fleming}, {Font-Ribera}, {Frinchaboy},
  {Galbraith-Frew}, {Garc{\'\i}a-Hern{\'a}ndez}, {Garc{\'\i}a P{\'e}rez}, {Ge},
  {G{\'e}nova-Santos}, {Gillespie}, {Girardi}, {Gonz{\'a}lez Hern{\'a}ndez},
  {Gott}, {Gunn}, {Guo}, {Halverson}, {Harding}, {Harris}, {Hasselquist},
  {Hawley}, {Hayden}, {Hearty}, {Herrero Dav{\'o}}, {Ho}, {Hogg}, {Holtzman},
  {Honscheid}, {Huehnerhoff}, {Ivans}, {Jackson}, {Jiang}, {Johnson},
  {Kinemuchi}, {Kirkby}, {Klaene}, {Kneib}, {Koesterke}, {Lan}, {Lang}, {Le
  Goff}, {Leauthaud}, {Lee}, {Lee}, {Long}, {Loomis}, {Lucatello}, {Lupton},
  {Ma}, {Mack}, {Mahadevan}, {Maia}, {Majewski}, {Malanushenko},
  {Malanushenko}, {Manchado}, {Manera}, {Maraston}, {Margala}, {Martell},
  {Masters}, {McBride}, {McGreer}, {McMahon}, {M{\'e}nard}, {M{\'e}sz{\'a}ros},
  {Miralda-Escud{\'e}}, {Miyatake}, {Montero-Dorta}, {Montesano}, {More},
  {Morrison}, {Muna}, {Munn}, {Myers}, {Nguyen}, {Nichol}, {Nidever},
  {Noterdaeme}, {Nuza}, {O'Connell}, {O'Connell}, {O'Connell}, {Olmstead},
  {Oravetz}, {Owen}, {Padmanabhan}, {Palanque-Delabrouille}, {Pan}, {Parejko},
  {Parihar}, {P{\^a}ris}, {Pepper}, {Percival}, {P{\'e}rez-R{\`a}fols}, {Dotto
  Perottoni}, {Petitjean}, {Pieri}, {Pinsonneault}, {Prada}, {Price-Whelan},
  {Raddick}, {Rahman}, {Rebolo}, {Reid}, {Richards}, {Riffel}, {Robin},
  {Rocha-Pinto}, {Rockosi}, {Roe}, {Ross}, {Ross}, {Rossi}, {Roy},
  {Rubi{\~n}o-Martin}, {Sabiu}, {S{\'a}nchez}, {Santiago}, {Sayres},
  {Schiavon}, {Schlegel}, {Schlesinger}, {Schmidt}, {Schneider}, {Schultheis},
  {Sellgren}, {Seo}, {Shen}, {Shetrone}, {Shu}, {Simmons}, {Skrutskie},
  {Slosar}, {Smith}, {Snedden}, {Sobeck}, {Sobreira}, {Stassun}, {Steinmetz},
  {Strauss}, {Streblyanska}, {Suzuki}, {Swanson}, {Terrien}, {Thakar},
  {Thomas}, {Thompson}, {Tinker}, {Tojeiro}, {Troup}, {Vandenberg}, {Vargas
  Maga{\~n}a}, {Viel}, {Vogt}, {Wake}, {Weaver}, {Weinberg}, {Weiner}, {White},
  {White}, {Wilson}, {Wisniewski}, {Wood-Vasey}, {Y{\`e}che}, {York}, {Zamora},
  {Zasowski}, {Zehavi}, {Zhao}, {Zheng}, \& {Zhu}}]{ahn14}
{Ahn}, C.~P., {Alexandroff}, R., {Allende Prieto}, C., {et~al.} 2014, \apjs,
  211, 17

\bibitem[{{Alam} {et~al.}(2015){Alam}, {Albareti}, {Allende Prieto}, {Anders},
  {Anderson}, {Anderton}, {Andrews}, {Armengaud}, {Aubourg}, {Bailey}, {Basu},
  {Bautista}, {Beaton}, {Beers}, {Bender}, {Berlind}, {Beutler}, {Bhardwaj},
  {Bird}, {Bizyaev}, {Blake}, {Blanton}, {Blomqvist}, {Bochanski}, {Bolton},
  {Bovy}, {Shelden Bradley}, {Brandt}, {Brauer}, {Brinkmann}, {Brown},
  {Brownstein}, {Burden}, {Burtin}, {Busca}, {Cai}, {Capozzi}, {Carnero
  Rosell}, {Carr}, {Carrera}, {Chambers}, {Chaplin}, {Chen}, {Chiappini},
  {Chojnowski}, {Chuang}, {Clerc}, {Comparat}, {Covey}, {Croft}, {Cuesta},
  {Cunha}, {da Costa}, {Da Rio}, {Davenport}, {Dawson}, {De Lee}, {Delubac},
  {Deshpande}, {Dhital}, {Dutra-Ferreira}, {Dwelly}, {Ealet}, {Ebelke},
  {Edmondson}, {Eisenstein}, {Ellsworth}, {Elsworth}, {Epstein}, {Eracleous},
  {Escoffier}, {Esposito}, {Evans}, {Fan}, {Fern{\'a}ndez-Alvar}, {Feuillet},
  {Filiz Ak}, {Finley}, {Finoguenov}, {Flaherty}, {Fleming}, {Font-Ribera},
  {Foster}, {Frinchaboy}, {Galbraith-Frew}, {Garc{\'\i}a},
  {Garc{\'\i}a-Hern{\'a}ndez}, {Garc{\'\i}a P{\'e}rez}, {Gaulme}, {Ge},
  {G{\'e}nova-Santos}, {Georgakakis}, {Ghezzi}, {Gillespie}, {Girardi},
  {Goddard}, {Gontcho}, {Gonz{\'a}lez Hern{\'a}ndez}, {Grebel}, {Green},
  {Grieb}, {Grieves}, {Gunn}, {Guo}, {Harding}, {Hasselquist}, {Hawley},
  {Hayden}, {Hearty}, {Hekker}, {Ho}, {Hogg}, {Holley-Bockelmann}, {Holtzman},
  {Honscheid}, {Huber}, {Huehnerhoff}, {Ivans}, {Jiang}, {Johnson},
  {Kinemuchi}, {Kirkby}, {Kitaura}, {Klaene}, {Knapp}, {Kneib}, {Koenig},
  {Lam}, {Lan}, {Lang}, {Laurent}, {Le Goff}, {Leauthaud}, {Lee}, {Lee},
  {Licquia}, {Liu}, {Long}, {L{\'o}pez-Corredoira}, {Lorenzo-Oliveira},
  {Lucatello}, {Lundgren}, {Lupton}, {Mack}, {Mahadevan}, {Maia}, {Majewski},
  {Malanushenko}, {Malanushenko}, {Manchado}, {Manera}, {Mao}, {Maraston},
  {Marchwinski}, {Margala}, {Martell}, {Martig}, {Masters}, {Mathur},
  {McBride}, {McGehee}, {McGreer}, {McMahon}, {M{\'e}nard}, {Menzel},
  {Merloni}, {M{\'e}sz{\'a}ros}, {Miller}, {Miralda-Escud{\'e}}, {Miyatake},
  {Montero-Dorta}, {More}, {Morganson}, {Morice-Atkinson}, {Morrison},
  {Mosser}, {Muna}, {Myers}, {Nandra}, {Newman}, {Neyrinck}, {Nguyen},
  {Nichol}, {Nidever}, {Noterdaeme}, {Nuza}, {O'Connell}, {O'Connell},
  {O'Connell}, {Ogando}, {Olmstead}, {Oravetz}, {Oravetz}, {Osumi}, {Owen},
  {Padgett}, {Padmanabhan}, {Paegert}, {Palanque-Delabrouille}, {Pan},
  {Parejko}, {P{\^a}ris}, {Park}, {Pattarakijwanich}, {Pellejero-Ibanez},
  {Pepper}, {Percival}, {P{\'e}rez-Fournon}, {P{\'e}rez-R{\`a}fols},
  {Petitjean}, {Pieri}, {Pinsonneault}, {Porto de Mello}, {Prada}, {Prakash},
  {Price-Whelan}, {Protopapas}, {Raddick}, {Rahman}, {Reid}, {Rich}, {Rix},
  {Robin}, {Rockosi}, {Rodrigues}, {Rodr{\'\i}guez-Torres}, {Roe}, {Ross},
  {Ross}, {Rossi}, {Ruan}, {Rubi{\~n}o-Mart{\'\i}n}, {Rykoff},
  {Salazar-Albornoz}, {Salvato}, {Samushia}, {S{\'a}nchez}, {Santiago},
  {Sayres}, {Schiavon}, {Schlegel}, {Schmidt}, {Schneider}, {Schultheis},
  {Schwope}, {Sc{\'o}ccola}, {Scott}, {Sellgren}, {Seo}, {Serenelli}, {Shane},
  {Shen}, {Shetrone}, {Shu}, {Silva Aguirre}, {Sivarani}, {Skrutskie},
  {Slosar}, {Smith}, {Sobreira}, {Souto}, {Stassun}, {Steinmetz}, {Stello},
  {Strauss}, {Streblyanska}, {Suzuki}, {Swanson}, {Tan}, {Tayar}, {Terrien},
  {Thakar}, {Thomas}, {Thomas}, {Thompson}, {Tinker}, {Tojeiro}, {Troup},
  {Vargas-Maga{\~n}a}, {Vazquez}, {Verde}, {Viel}, {Vogt}, {Wake}, {Wang},
  {Weaver}, {Weinberg}, {Weiner}, {White}, {Wilson}, {Wisniewski},
  {Wood-Vasey}, {Ye`che}, {York}, {Zakamska}, {Zamora}, {Zasowski}, {Zehavi},
  {Zhao}, {Zheng}, {Zhou}, {Zhou}, {Zou}, \& {Zhu}}]{Alam2015ApJS}
{Alam}, S., {Albareti}, F.~D., {Allende Prieto}, C., {et~al.} 2015, \apjs, 219,
  12

\bibitem[{{Astropy Collaboration} {et~al.}(2022){Astropy Collaboration},
  {Price-Whelan}, {Lim}, {Earl}, {Starkman}, {Bradley}, {Shupe}, {Patil},
  {Corrales}, {Brasseur}, {N{"o}the}, {Donath}, {Tollerud}, {Morris},
  {Ginsburg}, {Vaher}, {Weaver}, {Tocknell}, {Jamieson}, {van Kerkwijk},
  {Robitaille}, {Merry}, {Bachetti}, {G{"u}nther}, {Aldcroft},
  {Alvarado-Montes}, {Archibald}, {B{'o}di}, {Bapat}, {Barentsen}, {Baz{'a}n},
  {Biswas}, {Boquien}, {Burke}, {Cara}, {Cara}, {Conroy}, {Conseil}, {Craig},
  {Cross}, {Cruz}, {D'Eugenio}, {Dencheva}, {Devillepoix}, {Dietrich},
  {Eigenbrot}, {Erben}, {Ferreira}, {Foreman-Mackey}, {Fox}, {Freij}, {Garg},
  {Geda}, {Glattly}, {Gondhalekar}, {Gordon}, {Grant}, {Greenfield}, {Groener},
  {Guest}, {Gurovich}, {Handberg}, {Hart}, {Hatfield-Dodds}, {Homeier},
  {Hosseinzadeh}, {Jenness}, {Jones}, {Joseph}, {Kalmbach}, {Karamehmetoglu},
  {Ka{l}uszy{'n}ski}, {Kelley}, {Kern}, {Kerzendorf}, {Koch}, {Kulumani},
  {Lee}, {Ly}, {Ma}, {MacBride}, {Maljaars}, {Muna}, {Murphy}, {Norman},
  {O'Steen}, {Oman}, {Pacifici}, {Pascual}, {Pascual-Granado}, {Patil},
  {Perren}, {Pickering}, {Rastogi}, {Roulston}, {Ryan}, {Rykoff}, {Sabater},
  {Sakurikar}, {Salgado}, {Sanghi}, {Saunders}, {Savchenko}, {Schwardt},
  {Seifert-Eckert}, {Shih}, {Jain}, {Shukla}, {Sick}, {Simpson},
  {Singanamalla}, {Singer}, {Singhal}, {Sinha}, {Sip{H{o}}cz}, {Spitler},
  {Stansby}, {Streicher}, {{{S}}umak}, {Swinbank}, {Taranu}, {Tewary},
  {Tremblay}, {Val-Borro}, {Van Kooten}, {Vasovi{'c}}, {Verma}, {de Miranda
  Cardoso}, {Williams}, {Wilson}, {Winkel}, {Wood-Vasey}, {Xue}, {Yoachim},
  {Zhang}, {Zonca}, \& {Astropy Project Contributors}}]{astropy:2022}
{Astropy Collaboration}, {Price-Whelan}, A.~M., {Lim}, P.~L., {et~al.} 2022,
  apj, 935, 167

\bibitem[{{Astropy Collaboration} {et~al.}(2018){Astropy Collaboration},
  {Price-Whelan}, {Sip{\H{o}}cz}, {G{\"u}nther}, {Lim}, {Crawford}, {Conseil},
  {Shupe}, {Craig}, {Dencheva}, {Ginsburg}, {Vand erPlas}, {Bradley},
  {P{\'e}rez-Su{\'a}rez}, {de Val-Borro}, {Aldcroft}, {Cruz}, {Robitaille},
  {Tollerud}, {Ardelean}, {Babej}, {Bach}, {Bachetti}, {Bakanov}, {Bamford},
  {Barentsen}, {Barmby}, {Baumbach}, {Berry}, {Biscani}, {Boquien}, {Bostroem},
  {Bouma}, {Brammer}, {Bray}, {Breytenbach}, {Buddelmeijer}, {Burke},
  {Calderone}, {Cano Rodr{\'\i}guez}, {Cara}, {Cardoso}, {Cheedella}, {Copin},
  {Corrales}, {Crichton}, {D'Avella}, {Deil}, {Depagne}, {Dietrich}, {Donath},
  {Droettboom}, {Earl}, {Erben}, {Fabbro}, {Ferreira}, {Finethy}, {Fox},
  {Garrison}, {Gibbons}, {Goldstein}, {Gommers}, {Greco}, {Greenfield},
  {Groener}, {Grollier}, {Hagen}, {Hirst}, {Homeier}, {Horton}, {Hosseinzadeh},
  {Hu}, {Hunkeler}, {Ivezi{\'c}}, {Jain}, {Jenness}, {Kanarek}, {Kendrew},
  {Kern}, {Kerzendorf}, {Khvalko}, {King}, {Kirkby}, {Kulkarni}, {Kumar},
  {Lee}, {Lenz}, {Littlefair}, {Ma}, {Macleod}, {Mastropietro}, {McCully},
  {Montagnac}, {Morris}, {Mueller}, {Mumford}, {Muna}, {Murphy}, {Nelson},
  {Nguyen}, {Ninan}, {N{\"o}the}, {Ogaz}, {Oh}, {Parejko}, {Parley}, {Pascual},
  {Patil}, {Patil}, {Plunkett}, {Prochaska}, {Rastogi}, {Reddy Janga},
  {Sabater}, {Sakurikar}, {Seifert}, {Sherbert}, {Sherwood-Taylor}, {Shih},
  {Sick}, {Silbiger}, {Singanamalla}, {Singer}, {Sladen}, {Sooley},
  {Sornarajah}, {Streicher}, {Teuben}, {Thomas}, {Tremblay}, {Turner},
  {Terr{\'o}n}, {van Kerkwijk}, {de la Vega}, {Watkins}, {Weaver}, {Whitmore},
  {Woillez}, {Zabalza}, \& {Astropy Contributors}}]{astropy:2018}
{Astropy Collaboration}, {Price-Whelan}, A.~M., {Sip{\H{o}}cz}, B.~M., {et~al.}
  2018, \aj, 156, 123

\bibitem[{{Astropy Collaboration} {et~al.}(2013){Astropy Collaboration},
  {Robitaille}, {Tollerud}, {Greenfield}, {Droettboom}, {Bray}, {Aldcroft},
  {Davis}, {Ginsburg}, {Price-Whelan}, {Kerzendorf}, {Conley}, {Crighton},
  {Barbary}, {Muna}, {Ferguson}, {Grollier}, {Parikh}, {Nair}, {Unther},
  {Deil}, {Woillez}, {Conseil}, {Kramer}, {Turner}, {Singer}, {Fox}, {Weaver},
  {Zabalza}, {Edwards}, {Azalee Bostroem}, {Burke}, {Casey}, {Crawford},
  {Dencheva}, {Ely}, {Jenness}, {Labrie}, {Lim}, {Pierfederici}, {Pontzen},
  {Ptak}, {Refsdal}, {Servillat}, \& {Streicher}}]{astropy:2013}
{Astropy Collaboration}, {Robitaille}, T.~P., {Tollerud}, E.~J., {et~al.} 2013,
  \aap, 558, A33

\bibitem[{{Baldwin} {et~al.}(1981){Baldwin}, {Phillips}, \&
  {Terlevich}}]{BPT1981}
{Baldwin}, J.~A., {Phillips}, M.~M., \& {Terlevich}, R. 1981, \pasp, 93, 5

\bibitem[{{Battisti} {et~al.}(2016){Battisti}, {Calzetti}, \&
  {Chary}}]{Batisti2016}
{Battisti}, A.~J., {Calzetti}, D., \& {Chary}, R.~R. 2016, \apj, 818, 13

\bibitem[{{Bianchi} {et~al.}(2014){Bianchi}, {Conti}, \& {Shiao}}]{Bianchi2014}
{Bianchi}, L., {Conti}, A., \& {Shiao}, B. 2014, Advances in Space Research,
  53, 900

\bibitem[{{Bogdanoska} \& {Burgarella}(2020)}]{Bogdanoska2020}
{Bogdanoska}, J. \& {Burgarella}, D. 2020, \mnras, 496, 5341

\bibitem[{{Boquien} {et~al.}(2022){Boquien}, {Buat}, {Burgarella}, {Bardelli},
  {B{\'e}thermin}, {Faisst}, {Ginolfi}, {Hathi}, {Jones}, {Koekemoer},
  {Lemaux}, {Narayanan}, {Romano}, {Schaerer}, {Vergani}, {Zamorani}, \&
  {Zucca}}]{Boquien2022}
{Boquien}, M., {Buat}, V., {Burgarella}, D., {et~al.} 2022, \aap, 663, A50

\bibitem[{{Boquien} {et~al.}(2019){Boquien}, {Burgarella}, {Roehlly}, {Buat},
  {Ciesla}, {Corre}, {Inoue}, \& {Salas}}]{Boquien2019}
{Boquien}, M., {Burgarella}, D., {Roehlly}, Y., {et~al.} 2019, \aap, 622, A103

\bibitem[{{Brinchmann} {et~al.}(2004){Brinchmann}, {Charlot}, {White},
  {Tremonti}, {Kauffmann}, {Heckman}, \& {Brinkmann}}]{Brinchmann2004}
{Brinchmann}, J., {Charlot}, S., {White}, S.~D.~M., {et~al.} 2004, \mnras, 351,
  1151

\bibitem[{{Brough} {et~al.}(2020){Brough}, {Collins}, {Demarco}, {Ferguson},
  {Galaz}, {Holwerda}, {Martinez-Lombilla}, {Mihos}, \&
  {Montes}}]{Brough2020arXiv}
{Brough}, S., {Collins}, C., {Demarco}, R., {et~al.} 2020, arXiv e-prints,
  arXiv:2001.11067

\bibitem[{{Bruzual} \& {Charlot}(2003)}]{BC03}
{Bruzual}, G. \& {Charlot}, S. 2003, \mnras, 344, 1000

\bibitem[{{Buat} {et~al.}(2018){Buat}, {Boquien}, {Ma{\l}ek}, {Corre}, {Salas},
  {Roehlly}, {Shirley}, \& {Efstathiou}}]{Buat2018}
{Buat}, V., {Boquien}, M., {Ma{\l}ek}, K., {et~al.} 2018, \aap, 619, A135

\bibitem[{{Buat} {et~al.}(2019){Buat}, {Ciesla}, {Boquien}, {Ma{\l}ek}, \&
  {Burgarella}}]{buat2019}
{Buat}, V., {Ciesla}, L., {Boquien}, M., {Ma{\l}ek}, K., \& {Burgarella}, D.
  2019, \aap, 632, A79

\bibitem[{{Buat} {et~al.}(2014){Buat}, {Heinis}, {Boquien}, {Burgarella},
  {Charmandaris}, {Boissier}, {Boselli}, {Le Borgne}, \& {Morrison}}]{Buat14}
{Buat}, V., {Heinis}, S., {Boquien}, M., {et~al.} 2014, \aap, 561, A39

\bibitem[{{Buat} {et~al.}(2021){Buat}, {Mountrichas}, {Yang}, {Boquien},
  {Roehlly}, {Burgarella}, {Stalevski}, {Ciesla}, \& {Theul{\'e}}}]{Buat2021}
{Buat}, V., {Mountrichas}, G., {Yang}, G., {et~al.} 2021, \aap, 654, A93

\bibitem[{{Buat} {et~al.}(2012){Buat}, {Noll}, {Burgarella}, {Giovannoli},
  {Charmandaris}, {Pannella}, {Hwang}, {Elbaz}, {Dickinson}, {Magdis}, {Reddy},
  \& {Murphy}}]{Buat12}
{Buat}, V., {Noll}, S., {Burgarella}, D., {et~al.} 2012, \aap, 545, A141

\bibitem[{{Burgarella} {et~al.}(2005){Burgarella}, {Buat}, \&
  {Iglesias-P{\'a}ramo}}]{Burgarella2005}
{Burgarella}, D., {Buat}, V., \& {Iglesias-P{\'a}ramo}, J. 2005, \mnras, 360,
  1413

\bibitem[{{Calzetti} {et~al.}(2000){Calzetti}, {Armus}, {Bohlin}, {Kinney},
  {Koornneef}, \& {Storchi-Bergmann}}]{Calzetti2000}
{Calzetti}, D., {Armus}, L., {Bohlin}, R.~C., {et~al.} 2000, \apj, 533, 682

\bibitem[{{Calzetti} {et~al.}(1994){Calzetti}, {Kinney}, \&
  {Storchi-Bergmann}}]{Calzetti1994}
{Calzetti}, D., {Kinney}, A.~L., \& {Storchi-Bergmann}, T. 1994, \apj, 429, 582

\bibitem[{{Chabrier}(2003)}]{Chabrier2003}
{Chabrier}, G. 2003, \pasp, 115, 763

\bibitem[{{Charlot} \& {Fall}(2000)}]{CharlotFall2000}
{Charlot}, S. \& {Fall}, S.~M. 2000, \apj, 539, 718

\bibitem[{{Chary} \& {Elbaz}(2001)}]{CharyElbaz2001ApJ}
{Chary}, R. \& {Elbaz}, D. 2001, \apj, 556, 562

\bibitem[{{Chevallard} {et~al.}(2013){Chevallard}, {Charlot}, {Wandelt}, \&
  {Wild}}]{Chevallard2013}
{Chevallard}, J., {Charlot}, S., {Wandelt}, B., \& {Wild}, V. 2013, \mnras,
  432, 2061

\bibitem[{{Ciesla} {et~al.}(2018){Ciesla}, {Elbaz}, {Schreiber}, {Daddi}, \&
  {Wang}}]{Ciesla2018}
{Ciesla}, L., {Elbaz}, D., {Schreiber}, C., {Daddi}, E., \& {Wang}, T. 2018,
  \aap, 615, A61

\bibitem[{{Conroy}(2013)}]{Conroy2013}
{Conroy}, C. 2013, \araa, 51, 393

\bibitem[{{Cortese} {et~al.}(2012){Cortese}, {Ciesla}, {Boselli}, {Bianchi},
  {Gomez}, {Smith}, {Bendo}, {Eales}, {Pohlen}, {Baes}, {Corbelli}, {Davies},
  {Hughes}, {Hunt}, {Madden}, {Pierini}, {di Serego Alighieri}, {Zibetti},
  {Boquien}, {Clements}, {Cooray}, {Galametz}, {Magrini}, {Pappalardo},
  {Spinoglio}, \& {Vlahakis}}]{Cortese2012}
{Cortese}, L., {Ciesla}, L., {Boselli}, A., {et~al.} 2012, \aap, 540, A52

\bibitem[{{Dunne} {et~al.}(2011){Dunne}, {Gomez}, {da Cunha}, {Charlot}, {Dye},
  {Eales}, {Maddox}, {Rowlands}, {Smith}, {Auld}, {Baes}, {Bonfield}, {Bourne},
  {Buttiglione}, {Cava}, {Clements}, {Coppin}, {Cooray}, {Dariush}, {de Zotti},
  {Driver}, {Fritz}, {Geach}, {Hopwood}, {Ibar}, {Ivison}, {Jarvis}, {Kelvin},
  {Pascale}, {Pohlen}, {Popescu}, {Rigby}, {Robotham}, {Rodighiero}, {Sansom},
  {Serjeant}, {Temi}, {Thompson}, {Tuffs}, {van der Werf}, \&
  {Vlahakis}}]{Dunne2011}
{Dunne}, L., {Gomez}, H.~L., {da Cunha}, E., {et~al.} 2011, \mnras, 417, 1510

\bibitem[{{Elbaz} {et~al.}(2007){Elbaz}, {Daddi}, {Le Borgne}, {Dickinson},
  {Alexander}, {Chary}, {Starck}, {Brandt}, {Kitzbichler}, {MacDonald},
  {Nonino}, {Popesso}, {Stern}, \& {Vanzella}}]{Elbaz2007}
{Elbaz}, D., {Daddi}, E., {Le Borgne}, D., {et~al.} 2007, \aap, 468, 33

\bibitem[{{Gallazzi} {et~al.}(2005){Gallazzi}, {Charlot}, {Brinchmann},
  {White}, \& {Tremonti}}]{Gallazzi2005}
{Gallazzi}, A., {Charlot}, S., {Brinchmann}, J., {White}, S. D.~M., \&
  {Tremonti}, C.~A. 2005, \mnras, 362, 41

\bibitem[{{Galliano} {et~al.}(2018){Galliano}, {Galametz}, \&
  {Jones}}]{Galliano2018}
{Galliano}, F., {Galametz}, M., \& {Jones}, A.~P. 2018, \araa, 56, 673

\bibitem[{{Graham}(2001)}]{Graham2001b}
{Graham}, A.~W. 2001, \mnras, 326, 543

\bibitem[{{Graham} \& {de Blok}(2001)}]{Graham2001}
{Graham}, A.~W. \& {de Blok}, W.~J.~G. 2001, \apj, 556, 177

\bibitem[{{Graham} {et~al.}(2005){Graham}, {Driver}, {Petrosian}, {Conselice},
  {Bershady}, {Crawford}, \& {Goto}}]{Graham2005}
{Graham}, A.~W., {Driver}, S.~P., {Petrosian}, V., {et~al.} 2005, \aj, 130,
  1535

\bibitem[{{Graham} {et~al.}(2024){Graham}, {Jarrett}, \& {Cluver}}]{Graham2024}
{Graham}, A.~W., {Jarrett}, T.~H., \& {Cluver}, M.~E. 2024, \mnras, 527, 10059

\bibitem[{{Hamed} {et~al.}(2021){Hamed}, {Ciesla}, {B{\'e}thermin}, {Ma{\l}ek},
  {Daddi}, {Sargent}, \& {Gobat}}]{Hamed2021}
{Hamed}, M., {Ciesla}, L., {B{\'e}thermin}, M., {et~al.} 2021, \aap, 646, A127

\bibitem[{{Hamed} {et~al.}(2023{\natexlab{a}}){Hamed}, {Ma{\l}ek}, {Buat},
  {Junais}, {Ciesla}, {Donevski}, {Riccio}, \& {Figueira}}]{Hamed2023}
{Hamed}, M., {Ma{\l}ek}, K., {Buat}, V., {et~al.} 2023{\natexlab{a}}, \aap,
  674, A99

\bibitem[{{Hamed} {et~al.}(2023{\natexlab{b}}){Hamed}, {Pistis}, {Figueira},
  {Ma{\l}ek}, {Nanni}, {Buat}, {Pollo}, {Vergani}, {Bolzonella}, {Junais},
  {Krywult}, {Takeuchi}, {Riccio}, \& {Moutard}}]{Hamed2023b}
{Hamed}, M., {Pistis}, F., {Figueira}, M., {et~al.} 2023{\natexlab{b}}, \aap,
  679, A26

\bibitem[{{Harikane} {et~al.}(2020){Harikane}, {Ouchi}, {Inoue}, {Matsuoka},
  {Tamura}, {Bakx}, {Fujimoto}, {Moriwaki}, {Ono}, {Nagao}, {Tadaki}, {Kojima},
  {Shibuya}, {Egami}, {Ferrara}, {Gallerani}, {Hashimoto}, {Kohno}, {Matsuda},
  {Matsuo}, {Pallottini}, {Sugahara}, \& {Vallini}}]{Harikane2020}
{Harikane}, Y., {Ouchi}, M., {Inoue}, A.~K., {et~al.} 2020, \apj, 896, 93

\bibitem[{{Hurley} {et~al.}(2017){Hurley}, {Oliver}, {Betancourt}, {Clarke},
  {Cowley}, {Duivenvoorden}, {Farrah}, {Griffin}, {Lacey}, {Le Floc'h},
  {Papadopoulos}, {Sargent}, {Scudder}, {Vaccari}, {Valtchanov}, \&
  {Wang}}]{Hurley2017}
{Hurley}, P.~D., {Oliver}, S., {Betancourt}, M., {et~al.} 2017, \mnras, 464,
  885

\bibitem[{{Ivezi{\'c}} {et~al.}(2019){Ivezi{\'c}}, {Kahn}, {Tyson}, {Abel},
  {Acosta}, {Allsman}, {Alonso}, {AlSayyad}, {Anderson}, {Andrew}, {Angel},
  {Angeli}, {Ansari}, {Antilogus}, {Araujo}, {Armstrong}, {Arndt}, {Astier},
  {Aubourg}, {Auza}, {Axelrod}, {Bard}, {Barr}, {Barrau}, {Bartlett}, {Bauer},
  {Bauman}, {Baumont}, {Bechtol}, {Bechtol}, {Becker}, {Becla}, {Beldica},
  {Bellavia}, {Bianco}, {Biswas}, {Blanc}, {Blazek}, {Blandford}, {Bloom},
  {Bogart}, {Bond}, {Booth}, {Borgland}, {Borne}, {Bosch}, {Boutigny},
  {Brackett}, {Bradshaw}, {Brandt}, {Brown}, {Bullock}, {Burchat}, {Burke},
  {Cagnoli}, {Calabrese}, {Callahan}, {Callen}, {Carlin}, {Carlson},
  {Chandrasekharan}, {Charles-Emerson}, {Chesley}, {Cheu}, {Chiang}, {Chiang},
  {Chirino}, {Chow}, {Ciardi}, {Claver}, {Cohen-Tanugi}, {Cockrum}, {Coles},
  {Connolly}, {Cook}, {Cooray}, {Covey}, {Cribbs}, {Cui}, {Cutri}, {Daly},
  {Daniel}, {Daruich}, {Daubard}, {Daues}, {Dawson}, {Delgado}, {Dellapenna},
  {de Peyster}, {de Val-Borro}, {Digel}, {Doherty}, {Dubois},
  {Dubois-Felsmann}, {Durech}, {Economou}, {Eifler}, {Eracleous}, {Emmons},
  {Fausti Neto}, {Ferguson}, {Figueroa}, {Fisher-Levine}, {Focke}, {Foss},
  {Frank}, {Freemon}, {Gangler}, {Gawiser}, {Geary}, {Gee}, {Geha}, {Gessner},
  {Gibson}, {Gilmore}, {Glanzman}, {Glick}, {Goldina}, {Goldstein}, {Goodenow},
  {Graham}, {Gressler}, {Gris}, {Guy}, {Guyonnet}, {Haller}, {Harris},
  {Hascall}, {Haupt}, {Hernandez}, {Herrmann}, {Hileman}, {Hoblitt}, {Hodgson},
  {Hogan}, {Howard}, {Huang}, {Huffer}, {Ingraham}, {Innes}, {Jacoby}, {Jain},
  {Jammes}, {Jee}, {Jenness}, {Jernigan}, {Jevremovi{\'c}}, {Johns}, {Johnson},
  {Johnson}, {Jones}, {Juramy-Gilles}, {Juri{\'c}}, {Kalirai}, {Kallivayalil},
  {Kalmbach}, {Kantor}, {Karst}, {Kasliwal}, {Kelly}, {Kessler}, {Kinnison},
  {Kirkby}, {Knox}, {Kotov}, {Krabbendam}, {Krughoff}, {Kub{\'a}nek},
  {Kuczewski}, {Kulkarni}, {Ku}, {Kurita}, {Lage}, {Lambert}, {Lange},
  {Langton}, {Le Guillou}, {Levine}, {Liang}, {Lim}, {Lintott}, {Long},
  {Lopez}, {Lotz}, {Lupton}, {Lust}, {MacArthur}, {Mahabal}, {Mandelbaum},
  {Markiewicz}, {Marsh}, {Marshall}, {Marshall}, {May}, {McKercher}, {McQueen},
  {Meyers}, {Migliore}, {Miller}, {Mills}, {Miraval}, {Moeyens}, {Moolekamp},
  {Monet}, {Moniez}, {Monkewitz}, {Montgomery}, {Morrison}, {Mueller},
  {Muller}, {Mu{\~n}oz Arancibia}, {Neill}, {Newbry}, {Nief}, {Nomerotski},
  {Nordby}, {O'Connor}, {Oliver}, {Olivier}, {Olsen}, {O'Mullane}, {Ortiz},
  {Osier}, {Owen}, {Pain}, {Palecek}, {Parejko}, {Parsons}, {Pease},
  {Peterson}, {Peterson}, {Petravick}, {Libby Petrick}, {Petry},
  {Pierfederici}, {Pietrowicz}, {Pike}, {Pinto}, {Plante}, {Plate}, {Plutchak},
  {Price}, {Prouza}, {Radeka}, {Rajagopal}, {Rasmussen}, {Regnault}, {Reil},
  {Reiss}, {Reuter}, {Ridgway}, {Riot}, {Ritz}, {Robinson}, {Roby}, {Roodman},
  {Rosing}, {Roucelle}, {Rumore}, {Russo}, {Saha}, {Sassolas}, {Schalk},
  {Schellart}, {Schindler}, {Schmidt}, {Schneider}, {Schneider}, {Schoening},
  {Schumacher}, {Schwamb}, {Sebag}, {Selvy}, {Sembroski}, {Seppala}, {Serio},
  {Serrano}, {Shaw}, {Shipsey}, {Sick}, {Silvestri}, {Slater}, {Smith},
  {Smith}, {Sobhani}, {Soldahl}, {Storrie-Lombardi}, {Stover}, {Strauss},
  {Street}, {Stubbs}, {Sullivan}, {Sweeney}, {Swinbank}, {Szalay}, {Takacs},
  {Tether}, {Thaler}, {Thayer}, {Thomas}, {Thornton}, {Thukral}, {Tice},
  {Trilling}, {Turri}, {Van Berg}, {Vanden Berk}, {Vetter}, {Virieux},
  {Vucina}, {Wahl}, {Walkowicz}, {Walsh}, {Walter}, {Wang}, {Wang}, {Warner},
  {Wiecha}, {Willman}, {Winters}, {Wittman}, {Wolff}, {Wood-Vasey}, {Wu},
  {Xin}, {Yoachim}, \& {Zhan}}]{Ivezic2019}
{Ivezi{\'c}}, {\v{Z}}., {Kahn}, S.~M., {Tyson}, J.~A., {et~al.} 2019, \apj,
  873, 111

\bibitem[{{Jarrett} {et~al.}(2000){Jarrett}, {Chester}, {Cutri}, {Schneider},
  {Skrutskie}, \& {Huchra}}]{Jarrett2000}
{Jarrett}, T.~H., {Chester}, T., {Cutri}, R., {et~al.} 2000, \aj, 119, 2498

\bibitem[{{Junais} {et~al.}(2023){Junais}, {Ma{\l}ek}, {Boissier}, {Pearson},
  {Pollo}, {Boselli}, {Boquien}, {Donevski}, {Goto}, {Hamed}, {Kim}, {Koda},
  {Matsuhara}, {Riccio}, \& {Romano}}]{Junais2023}
{Junais}, {Ma{\l}ek}, K., {Boissier}, S., {et~al.} 2023, \aap, 676, A41

\bibitem[{{Komatsu} {et~al.}(2011){Komatsu}, {Smith}, {Dunkley}, {Bennett},
  {Gold}, {Hinshaw}, {Jarosik}, {Larson}, {Nolta}, {Page}, {Spergel},
  {Halpern}, {Hill}, {Kogut}, {Limon}, {Meyer}, {Odegard}, {Tucker}, {Weiland},
  {Wollack}, \& {Wright}}]{Komatsu2011}
{Komatsu}, E., {Smith}, K.~M., {Dunkley}, J., {et~al.} 2011, \apjs, 192, 18

\bibitem[{{Lang} {et~al.}(2016){Lang}, {Hogg}, \& {Schlegel}}]{Lang2016}
{Lang}, D., {Hogg}, D.~W., \& {Schlegel}, D.~J. 2016, \aj, 151, 36

\bibitem[{{Laureijs} {et~al.}(2011){Laureijs}, {Amiaux}, {Arduini},
  {Augu{\`e}res}, {Brinchmann}, {Cole}, {Cropper}, {Dabin}, {Duvet}, {Ealet},
  {Garilli}, {Gondoin}, {Guzzo}, {Hoar}, {Hoekstra}, {Holmes}, {Kitching},
  {Maciaszek}, {Mellier}, {Pasian}, {Percival}, {Rhodes}, {Saavedra Criado},
  {Sauvage}, {Scaramella}, {Valenziano}, {Warren}, {Bender}, {Castander},
  {Cimatti}, {Le F{\`e}vre}, {Kurki-Suonio}, {Levi}, {Lilje}, {Meylan},
  {Nichol}, {Pedersen}, {Popa}, {Rebolo Lopez}, {Rix}, {Rottgering},
  {Zeilinger}, {Grupp}, {Hudelot}, {Massey}, {Meneghetti}, {Miller}, {Paltani},
  {Paulin-Henriksson}, {Pires}, {Saxton}, {Schrabback}, {Seidel}, {Walsh},
  {Aghanim}, {Amendola}, {Bartlett}, {Baccigalupi}, {Beaulieu}, {Benabed},
  {Cuby}, {Elbaz}, {Fosalba}, {Gavazzi}, {Helmi}, {Hook}, {Irwin}, {Kneib},
  {Kunz}, {Mannucci}, {Moscardini}, {Tao}, {Teyssier}, {Weller}, {Zamorani},
  {Zapatero Osorio}, {Boulade}, {Foumond}, {Di Giorgio}, {Guttridge}, {James},
  {Kemp}, {Martignac}, {Spencer}, {Walton}, {Bl{\"u}mchen}, {Bonoli},
  {Bortoletto}, {Cerna}, {Corcione}, {Fabron}, {Jahnke}, {Ligori}, {Madrid},
  {Martin}, {Morgante}, {Pamplona}, {Prieto}, {Riva}, {Toledo}, {Trifoglio},
  {Zerbi}, {Abdalla}, {Douspis}, {Grenet}, {Borgani}, {Bouwens}, {Courbin},
  {Delouis}, {Dubath}, {Fontana}, {Frailis}, {Grazian}, {Koppenh{\"o}fer},
  {Mansutti}, {Melchior}, {Mignoli}, {Mohr}, {Neissner}, {Noddle}, {Poncet},
  {Scodeggio}, {Serrano}, {Shane}, {Starck}, {Surace}, {Taylor},
  {Verdoes-Kleijn}, {Vuerli}, {Williams}, {Zacchei}, {Altieri}, {Escudero
  Sanz}, {Kohley}, {Oosterbroek}, {Astier}, {Bacon}, {Bardelli}, {Baugh},
  {Bellagamba}, {Benoist}, {Bianchi}, {Biviano}, {Branchini}, {Carbone},
  {Cardone}, {Clements}, {Colombi}, {Conselice}, {Cresci}, {Deacon}, {Dunlop},
  {Fedeli}, {Fontanot}, {Franzetti}, {Giocoli}, {Garcia-Bellido}, {Gow},
  {Heavens}, {Hewett}, {Heymans}, {Holland}, {Huang}, {Ilbert}, {Joachimi},
  {Jennins}, {Kerins}, {Kiessling}, {Kirk}, {Kotak}, {Krause}, {Lahav}, {van
  Leeuwen}, {Lesgourgues}, {Lombardi}, {Magliocchetti}, {Maguire}, {Majerotto},
  {Maoli}, {Marulli}, {Maurogordato}, {McCracken}, {McLure}, {Melchiorri},
  {Merson}, {Moresco}, {Nonino}, {Norberg}, {Peacock}, {Pello}, {Penny},
  {Pettorino}, {Di Porto}, {Pozzetti}, {Quercellini}, {Radovich}, {Rassat},
  {Roche}, {Ronayette}, {Rossetti}, {Sartoris}, {Schneider}, {Semboloni},
  {Serjeant}, {Simpson}, {Skordis}, {Smadja}, {Smartt}, {Spano}, {Spiro},
  {Sullivan}, {Tilquin}, {Trotta}, {Verde}, {Wang}, {Williger}, {Zhao},
  {Zoubian}, \& {Zucca}}]{Euclid}
{Laureijs}, R., {Amiaux}, J., {Arduini}, S., {et~al.} 2011, arXiv e-prints,
  arXiv:1110.3193

\bibitem[{{Lo Faro} {et~al.}(2017){Lo Faro}, {Buat}, {Roehlly},
  {Alvarez-Marquez}, {Burgarella}, {Silva}, \& {Efstathiou}}]{LoFaro2017}
{Lo Faro}, B., {Buat}, V., {Roehlly}, Y., {et~al.} 2017, \mnras, 472, 1372

\bibitem[{{Ma{\l}ek} {et~al.}(2018){Ma{\l}ek}, {Buat}, {Roehlly}, {Burgarella},
  {Hurley}, {Shirley}, {Duncan}, {Efstathiou}, {Papadopoulos}, {Vaccari},
  {Farrah}, {Marchetti}, \& {Oliver}}]{Malek2018}
{Ma{\l}ek}, K., {Buat}, V., {Roehlly}, Y., {et~al.} 2018, \aap, 620, A50

\bibitem[{{Meert} {et~al.}(2015){Meert}, {Vikram}, \&
  {Bernardi}}]{Meert2015MNRAS.446.3943M}
{Meert}, A., {Vikram}, V., \& {Bernardi}, M. 2015, \mnras, 446, 3943

\bibitem[{{Meert} {et~al.}(2016){Meert}, {Vikram}, \&
  {Bernardi}}]{Meert2016MNRAS.455.2440M}
{Meert}, A., {Vikram}, V., \& {Bernardi}, M. 2016, \mnras, 455, 2440

\bibitem[{{Meurer} {et~al.}(1999){Meurer}, {Heckman}, \&
  {Calzetti}}]{Meurer1999}
{Meurer}, G.~R., {Heckman}, T.~M., \& {Calzetti}, D. 1999, \apj, 521, 64

\bibitem[{{Noeske} {et~al.}(2007){Noeske}, {Weiner}, {Faber}, {Papovich},
  {Koo}, {Somerville}, {Bundy}, {Conselice}, {Newman}, {Schiminovich}, {Le
  Floc'h}, {Coil}, {Rieke}, {Lotz}, {Primack}, {Barmby}, {Cooper}, {Davis},
  {Ellis}, {Fazio}, {Guhathakurta}, {Huang}, {Kassin}, {Martin}, {Phillips},
  {Rich}, {Small}, {Willmer}, \& {Wilson}}]{Noeske2007}
{Noeske}, K.~G., {Weiner}, B.~J., {Faber}, S.~M., {et~al.} 2007, \apjl, 660,
  L43

\bibitem[{{Noll} {et~al.}(2009){Noll}, {Burgarella}, {Giovannoli}, {Buat},
  {Marcillac}, \& {Mu{\~n}oz-Mateos}}]{Noll2009}
{Noll}, S., {Burgarella}, D., {Giovannoli}, E., {et~al.} 2009, \aap, 507, 1793

\bibitem[{{Osborne} {et~al.}(2023){Osborne}, {Salim}, {Boquien}, {Dickinson},
  \& {Arnouts}}]{Osborne2023}
{Osborne}, C., {Salim}, S., {Boquien}, M., {Dickinson}, M., \& {Arnouts}, S.
  2023, \apjs, 268, 26

\bibitem[{{Pahwa} \& {Saha}(2018)}]{Pahwa2018}
{Pahwa}, I. \& {Saha}, K. 2018, \mnras, 478, 4657

\bibitem[{{Paudel} {et~al.}(2017){Paudel}, {Lisker}, {Huxor}, \&
  {Ree}}]{Paudel2017}
{Paudel}, S., {Lisker}, T., {Huxor}, A.~P., \& {Ree}, C.~H. 2017, \mnras, 465,
  1950

\bibitem[{{Pearson} {et~al.}(2023){Pearson}, {Pistis}, {Figueira}, {Ma{\l}ek},
  {Moutard}, {Vergani}, \& {Pollo}}]{Pearson2023}
{Pearson}, W.~J., {Pistis}, F., {Figueira}, M., {et~al.} 2023, \aap, 679, A35

\bibitem[{{Pearson} {et~al.}(2018){Pearson}, {Wang}, {Hurley}, {Ma{\l}ek},
  {Buat}, {Burgarella}, {Farrah}, {Oliver}, {Smith}, \& {van der
  Tak}}]{Pearson2018}
{Pearson}, W.~J., {Wang}, L., {Hurley}, P.~D., {et~al.} 2018, \aap, 615, A146

\bibitem[{{Pearson} {et~al.}(2017){Pearson}, {Wang}, {van der Tak}, {Hurley},
  {Burgarella}, \& {Oliver}}]{Pearson2017}
{Pearson}, W.~J., {Wang}, L., {van der Tak}, F.~F.~S., {et~al.} 2017, \aap,
  603, A102

\bibitem[{{Peek} \& {Schiminovich}(2013)}]{Peek2013}
{Peek}, J.~E.~G. \& {Schiminovich}, D. 2013, \apj, 771, 68

\bibitem[{{Peng} {et~al.}(2002){Peng}, {Ho}, {Impey}, \&
  {Rix}}]{GALFIT2002AJ....124..266P}
{Peng}, C.~Y., {Ho}, L.~C., {Impey}, C.~D., \& {Rix}, H.-W. 2002, \aj, 124, 266

\bibitem[{{P{\'e}rez-Monta{\~n}o} {et~al.}(2022){P{\'e}rez-Monta{\~n}o},
  {Rodriguez-Gomez}, {Cervantes Sodi}, {Zhu}, {Pillepich}, {Vogelsberger}, \&
  {Hernquist}}]{perez-montano2022}
{P{\'e}rez-Monta{\~n}o}, L.~E., {Rodriguez-Gomez}, V., {Cervantes Sodi}, B.,
  {et~al.} 2022, \mnras, 514, 5840

\bibitem[{{Riccio} {et~al.}(2021){Riccio}, {Ma{\l}ek}, {Nanni}, {Boquien},
  {Buat}, {Burgarella}, {Donevski}, {Hamed}, {Hurley}, {Shirley}, \&
  {Pollo}}]{Riccio2021}
{Riccio}, G., {Ma{\l}ek}, K., {Nanni}, A., {et~al.} 2021, \aap, 653, A107

\bibitem[{{Robertson} {et~al.}(2019){Robertson}, {Banerji}, {Brough}, {Davies},
  {Ferguson}, {Hausen}, {Kaviraj}, {Newman}, {Schmidt}, {Tyson}, \&
  {Wechsler}}]{Robertson2019}
{Robertson}, B.~E., {Banerji}, M., {Brough}, S., {et~al.} 2019, Nature Reviews
  Physics, 1, 450

\bibitem[{{Robertson} {et~al.}(2017){Robertson}, {Banerji}, {Cooper}, {Davies},
  {Driver}, {Ferguson}, {Ferguson}, {Gawiser}, {Kaviraj}, {Knapen}, {Lintott},
  {Lotz}, {Newman}, {Norman}, {Padilla}, {Schmidt}, {Smith}, {Tyson}, {Verma},
  {Zehavi}, {Armus}, {Avestruz}, {Barrientos}, {Bowler}, {Bremer}, {Conselice},
  {Davies}, {Demarco}, {Dickinson}, {Galaz}, {Grazian}, {Holwerda}, {Jarvis},
  {Kasliwal}, {Lacerna}, {Loveday}, {Marshall}, {Merlin}, {Napolitano},
  {Puzia}, {Robotham}, {Salim}, {Sereno}, {Snyder}, {Stott}, {Tissera},
  {Werner}, {Yoachim}, {Borne}, \& {Members of the LSST Galaxies Science
  Collaboration}}]{Robertson2017}
{Robertson}, B.~E., {Banerji}, M., {Cooper}, M.~C., {et~al.} 2017, arXiv
  e-prints, arXiv:1708.01617

\bibitem[{{Roebuck} {et~al.}(2019){Roebuck}, {Sajina}, {Hayward}, {Martis},
  {Marchesini}, {Krefting}, \& {Pope}}]{Roebuck2019}
{Roebuck}, E., {Sajina}, A., {Hayward}, C.~C., {et~al.} 2019, \apj, 881, 18

\bibitem[{{Salim} \& {Boquien}(2019)}]{Salim2019}
{Salim}, S. \& {Boquien}, M. 2019, \apj, 872, 23

\bibitem[{{Salim} {et~al.}(2018){Salim}, {Boquien}, \& {Lee}}]{Salim2018}
{Salim}, S., {Boquien}, M., \& {Lee}, J.~C. 2018, \apj, 859, 11

\bibitem[{{Salim} {et~al.}(2016){Salim}, {Lee}, {Janowiecki}, {da Cunha},
  {Dickinson}, {Boquien}, {Burgarella}, {Salzer}, \& {Charlot}}]{Salim2016ApJS}
{Salim}, S., {Lee}, J.~C., {Janowiecki}, S., {et~al.} 2016, \apjs, 227, 2

\bibitem[{{Sandage} \& {Binggeli}(1984)}]{Sandage1984}
{Sandage}, A. \& {Binggeli}, B. 1984, \aj, 89, 919

\bibitem[{{Sanders} \& {Mirabel}(1996)}]{Sanders1996}
{Sanders}, D.~B. \& {Mirabel}, I.~F. 1996, \araa, 34, 749

\bibitem[{{Santini} {et~al.}(2014){Santini}, {Maiolino}, {Magnelli}, {Lutz},
  {Lamastra}, {Li Causi}, {Eales}, {Andreani}, {Berta}, {Buat}, {Cooray},
  {Cresci}, {Daddi}, {Farrah}, {Fontana}, {Franceschini}, {Genzel}, {Granato},
  {Grazian}, {Le Floc'h}, {Magdis}, {Magliocchetti}, {Mannucci}, {Menci},
  {Nordon}, {Oliver}, {Popesso}, {Pozzi}, {Riguccini}, {Rodighiero}, {Rosario},
  {Salvato}, {Scott}, {Silva}, {Tacconi}, {Viero}, {Wang}, {Wuyts}, \&
  {Xu}}]{Santini2014}
{Santini}, P., {Maiolino}, R., {Magnelli}, B., {et~al.} 2014, \aap, 562, A30

\bibitem[{{Shirley} {et~al.}(2019){Shirley}, {Roehlly}, {Hurley}, {Buat},
  {Campos Varillas}, {Duivenvoorden}, {Duncan}, {Efstathiou}, {Farrah},
  {Gonz{\'a}lez Solares}, {Malek}, {Marchetti}, {McCheyne}, {Papadopoulos},
  {Pons}, {Scipioni}, {Vaccari}, \& {Oliver}}]{Shirley2019}
{Shirley}, R., {Roehlly}, Y., {Hurley}, P.~D., {et~al.} 2019, \mnras, 490, 634

\bibitem[{{Shivaei} {et~al.}(2020){Shivaei}, {Reddy}, {Rieke}, {Shapley},
  {Kriek}, {Battisti}, {Mobasher}, {Sanders}, {Fetherolf}, {Azadi}, {Coil},
  {Freeman}, {de Groot}, {Leung}, {Price}, {Siana}, \& {Zick}}]{Shivaei2020}
{Shivaei}, I., {Reddy}, N., {Rieke}, G., {et~al.} 2020, \apj, 899, 117

\bibitem[{{Siudek} {et~al.}(2018){Siudek}, {Ma{\l}ek}, {Pollo}, {Krakowski},
  {Iovino}, {Scodeggio}, {Moutard}, {Zamorani}, {Guzzo}, {Garilli}, {Granett},
  {Bolzonella}, {de la Torre}, {Abbas}, {Adami}, {Bottini}, {Cappi},
  {Cucciati}, {Davidzon}, {Franzetti}, {Fritz}, {Krywult}, {Le Brun}, {Le
  F{\`e}vre}, {Maccagni}, {Marulli}, {Polletta}, {Tasca}, {Tojeiro}, {Vergani},
  {Zanichelli}, {Arnouts}, {Bel}, {Branchini}, {Coupon}, {De Lucia}, {Ilbert},
  {Haines}, {Moscardini}, \& {Takeuchi}}]{Siudek2018}
{Siudek}, M., {Ma{\l}ek}, K., {Pollo}, A., {et~al.} 2018, \aap, 617, A70

\bibitem[{{Speagle} {et~al.}(2014){Speagle}, {Steinhardt}, {Capak}, \&
  {Silverman}}]{Speagle2014}
{Speagle}, J.~S., {Steinhardt}, C.~L., {Capak}, P.~L., \& {Silverman}, J.~D.
  2014, \apjs, 214, 15

\bibitem[{{Spergel} {et~al.}(2015){Spergel}, {Gehrels}, {Baltay}, {Bennett},
  {Breckinridge}, {Donahue}, {Dressler}, {Gaudi}, {Greene}, {Guyon}, {Hirata},
  {Kalirai}, {Kasdin}, {Macintosh}, {Moos}, {Perlmutter}, {Postman},
  {Rauscher}, {Rhodes}, {Wang}, {Weinberg}, {Benford}, {Hudson}, {Jeong},
  {Mellier}, {Traub}, {Yamada}, {Capak}, {Colbert}, {Masters}, {Penny},
  {Savransky}, {Stern}, {Zimmerman}, {Barry}, {Bartusek}, {Carpenter}, {Cheng},
  {Content}, {Dekens}, {Demers}, {Grady}, {Jackson}, {Kuan}, {Kruk}, {Melton},
  {Nemati}, {Parvin}, {Poberezhskiy}, {Peddie}, {Ruffa}, {Wallace}, {Whipple},
  {Wollack}, \& {Zhao}}]{NancyRoman}
{Spergel}, D., {Gehrels}, N., {Baltay}, C., {et~al.} 2015, arXiv e-prints,
  arXiv:1503.03757

\bibitem[{{Takeuchi} {et~al.}(2012){Takeuchi}, {Yuan}, {Ikeyama}, {Murata}, \&
  {Inoue}}]{Takeuchi2012}
{Takeuchi}, T.~T., {Yuan}, F.-T., {Ikeyama}, A., {Murata}, K.~L., \& {Inoue},
  A.~K. 2012, \apj, 755, 144

\bibitem[{{van Dokkum} {et~al.}(2015){van Dokkum}, {Abraham}, {Merritt},
  {Zhang}, {Geha}, \& {Conroy}}]{vanDokkum2015}
{van Dokkum}, P.~G., {Abraham}, R., {Merritt}, A., {et~al.} 2015, \apjl, 798,
  L45

\bibitem[{{Vikram} {et~al.}(2010){Vikram}, {Wadadekar}, {Kembhavi}, \&
  {Vijayagovindan}}]{PYMORPH2010MNRAS.409.1379V}
{Vikram}, V., {Wadadekar}, Y., {Kembhavi}, A.~K., \& {Vijayagovindan}, G.~V.
  2010, \mnras, 409, 1379

\bibitem[{{Walcher} {et~al.}(2011){Walcher}, {Groves}, {Budav{\'a}ri}, \&
  {Dale}}]{Walcher2011}
{Walcher}, J., {Groves}, B., {Budav{\'a}ri}, T., \& {Dale}, D. 2011, \apss,
  331, 1

\bibitem[{{Weingartner} \& {Draine}(2001)}]{Weingartner2001}
{Weingartner}, J.~C. \& {Draine}, B.~T. 2001, \apj, 548, 296

\bibitem[{{Wild} {et~al.}(2011){Wild}, {Charlot}, {Brinchmann}, {Heckman},
  {Vince}, {Pacifici}, \& {Chevallard}}]{Wild2011}
{Wild}, V., {Charlot}, S., {Brinchmann}, J., {et~al.} 2011, \mnras, 417, 1760

\bibitem[{{Yuan} {et~al.}(2013){Yuan}, {Liu}, \& {Xiang}}]{Yuan2013}
{Yuan}, H.~B., {Liu}, X.~W., \& {Xiang}, M.~S. 2013, \mnras, 430, 2188

\bibitem[{{Zavala} {et~al.}(2023){Zavala}, {Buat}, {Casey}, {Finkelstein},
  {Burgarella}, {Bagley}, {Ciesla}, {Daddi}, {Dickinson}, {Ferguson}, {Franco},
  {Jim{\'e}nez-Andrade}, {Kartaltepe}, {Koekemoer}, {Le Bail}, {Murphy},
  {Papovich}, {Tacchella}, {Wilkins}, {Aretxaga}, {Behroozi}, {Champagne},
  {Fontana}, {Giavalisco}, {Grazian}, {Grogin}, {Kewley}, {Kocevski},
  {Kirkpatrick}, {Lotz}, {Pentericci}, {P{\'e}rez-Gonz{\'a}lez}, {Pirzkal},
  {Ravindranath}, {Somerville}, {Trump}, {Yang}, {Yung}, {Almaini},
  {Amor{\'\i}n}, {Annunziatella}, {Haro}, {Backhaus}, {Barro}, {Bell},
  {Bhatawdekar}, {Bisigello}, {Buitrago}, {Calabr{\`o}}, {Castellano},
  {Ch{\'a}vez Ortiz}, {Chworowsky}, {Cleri}, {Cohen}, {Cole}, {Cooke},
  {Cooper}, {Cooray}, {Costantin}, {Cox}, {Croton}, {Dav{\'e}}, {de La Vega},
  {Dekel}, {Elbaz}, {Estrada-Carpenter}, {Fern{\'a}ndez}, {Finkelstein},
  {Freundlich}, {Fujimoto}, {Garc{\'\i}a-Argum{\'a}nez}, {Gardner}, {Gawiser},
  {G{\'o}mez-Guijarro}, {Guo}, {Hamilton}, {Hathi}, {Holwerda}, {Hirschmann},
  {Huertas-Company}, {Hutchison}, {Iyer}, {Jaskot}, {Jha}, {Jogee}, {Juneau},
  {Jung}, {Kassin}, {Kurczynski}, {Larson}, {Leung}, {Long}, {Lucas},
  {Magnelli}, {Mantha}, {Matharu}, {McGrath}, {McIntosh}, {Medrano}, {Merlin},
  {Mobasher}, {Morales}, {Newman}, {Nicholls}, {Pandya}, {Rafelski}, {Ronayne},
  {Rose}, {Ryan}, {Santini}, {Seill{\'e}}, {Shah}, {Shen}, {Simons}, {Snyder},
  {Stanway}, {Straughn}, {Teplitz}, {Vanderhoof}, {Vega-Ferrero}, {Wang},
  {Weiner}, {Willmer}, {Wuyts}, \& {CEERS Team}}]{Zavala2023}
{Zavala}, J.~A., {Buat}, V., {Casey}, C.~M., {et~al.} 2023, \apjl, 943, L9

\bibitem[{{Zhong} {et~al.}(2008){Zhong}, {Liang}, {Liu}, {Hammer}, {Hu},
  {Chen}, {Deng}, \& {Zhang}}]{Zhong2008}
{Zhong}, G.~H., {Liang}, Y.~C., {Liu}, F.~S., {et~al.} 2008, \mnras, 391, 986

\end{thebibliography}

\begin{appendix}

\section{Distribution of used \sdss{} magnitudes and their corresponding surface brightness}
\begin{figure}[h!]
    \centering
        \includegraphics[width=0.5\textwidth]{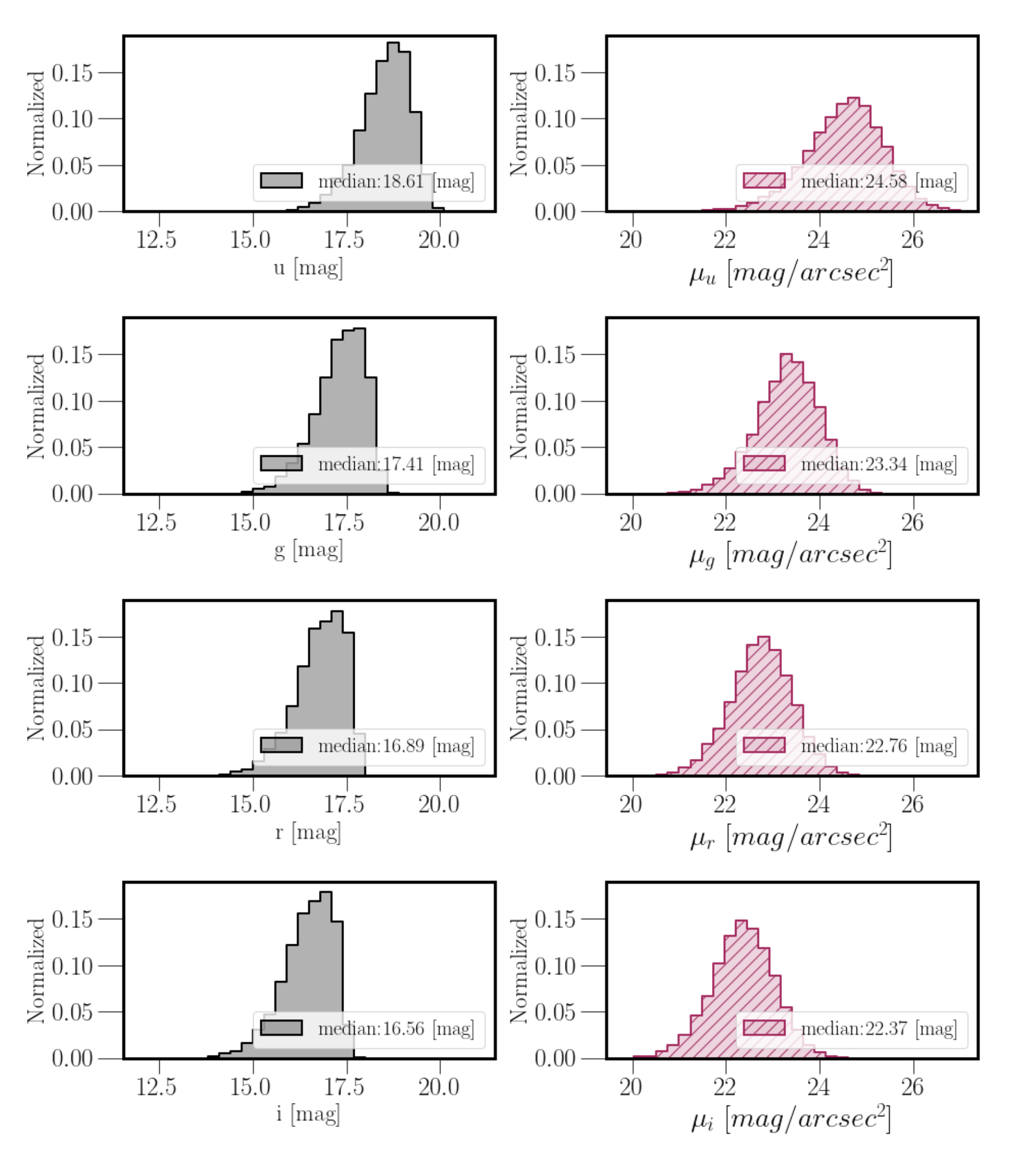}
        \caption{{Distributions of \sdss{} magnitudes and their corresponding surface brightness. The left panel (grey histograms)  show the distribution of  \textit{ugri} magnitudes of \sdss{} 7\,934 galaxies from the final sample used in our analysis, while the right panel (dark red hatched histograms) their corresponding surface brightness corrected for the inclination. } Median values of all distributions are added in each panel.}
        \label{fig:mag_sb}
\end{figure}

Figure~\ref{fig:mag_sb} shows distributions of \sdss{} \textit{ugri} magnitudes and calculated based on Eqs.~\ref{eq:sb} and~\ref{eq:inclination} their corresponding surface brightness. 
 
\begin{figure}[h]
    \centering
        \includegraphics[width=0.5\textwidth]{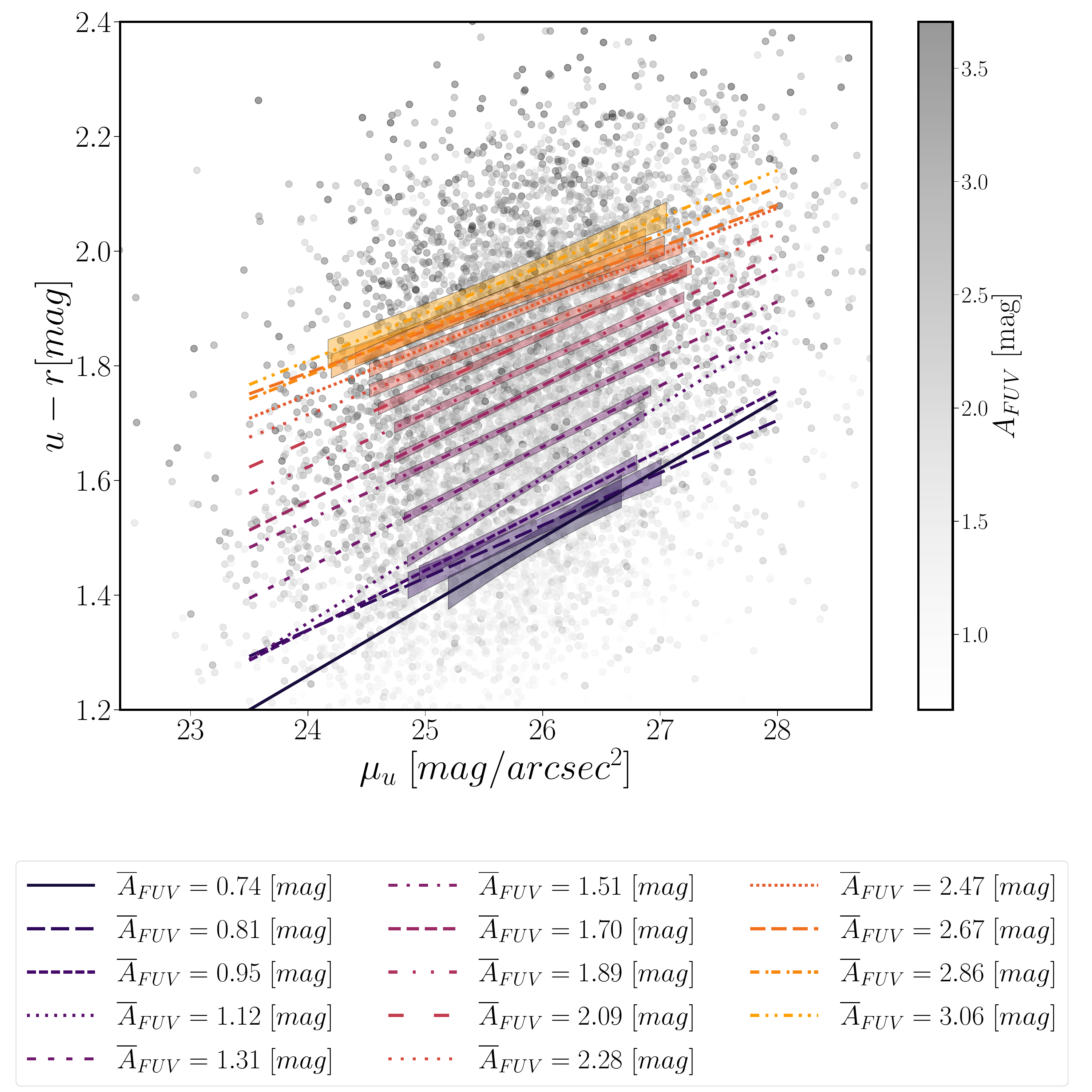}
        \caption{Relations fitted between observed \ur{} colours and \sbu{} for 14 \afuv{} bins with additional background showing all galaxies used in our analysis. The grey scale colour bar axis shows the value of the \afuv{}.} 
        \label{fig:AFUV_all_bins_bdg}
\end{figure}

Figure~\ref{fig:AFUV_all_bins_bdg} shows the same plot as Fig.~\ref{fig:AFUV_all_fits} (fitted \ur{} -- \sbu{} relation for all 14 bins of \afuv{}) but with an additional background of the whole sample, and also interpolated relations between 22.5 and 27.5 \sbu{}. 

\section{Galaxies with \afuvprior{}<0}
\begin{figure}
    \centering    \includegraphics[width=0.5\textwidth]{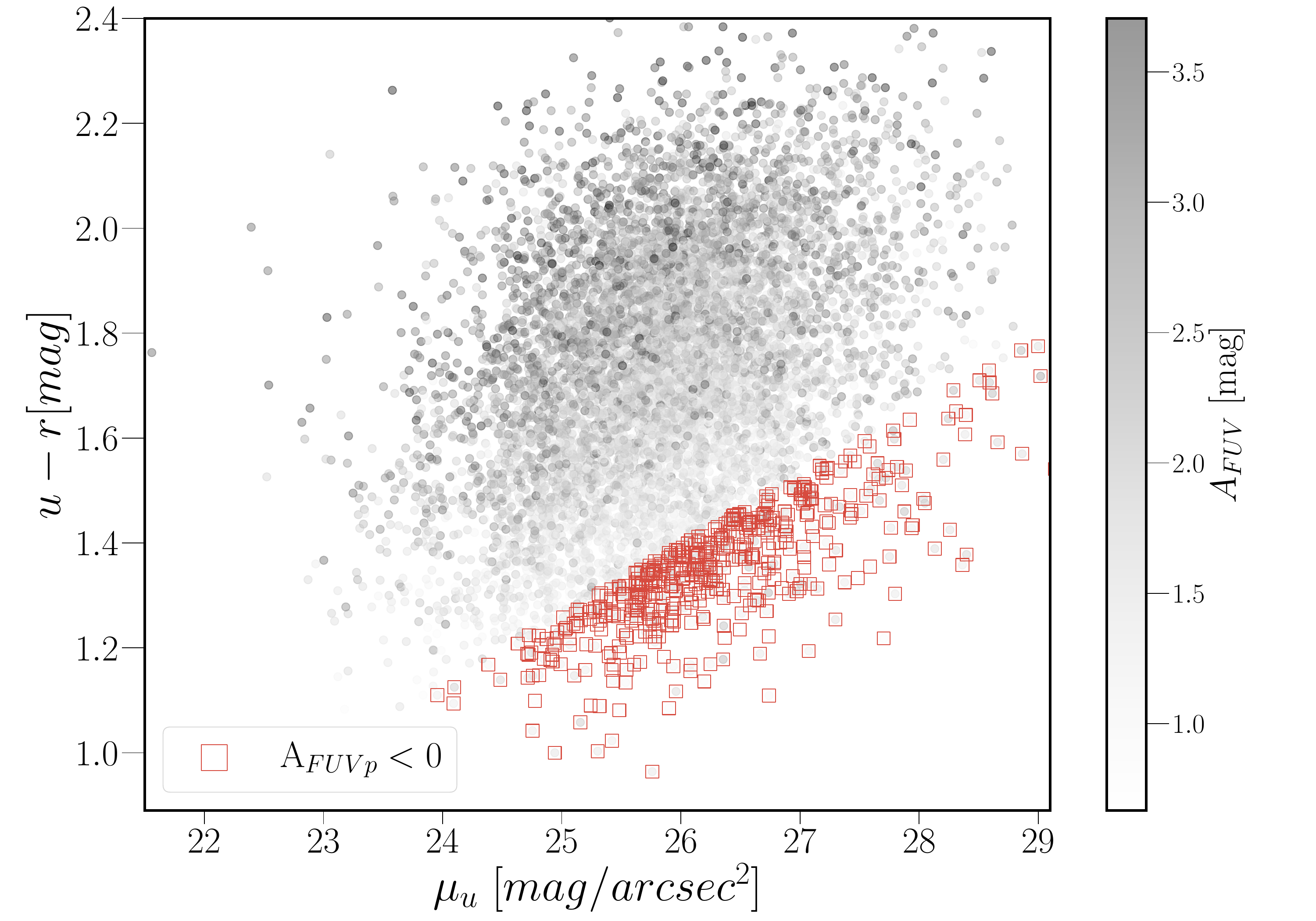}
    \caption{Fiducial \afuv{} values from the \Gcat{} catalogue in the \ur{}-\sbu{} plane. 
    Similarly, as in Fig.~\ref{fig:AFUV_all_bins_bdg}, grey scale colour bar axis shows the value of the \afuv{}. 
    Additionally, galaxies with calculated \afuvprior{}$<0$ are marked as open orange squares.  }
    \label{app:AFUV_0}
\end{figure}

We present here the location in the \ur{}-\sbu{}  plane galaxies for which the calculated based on Eq.~\ref{eq:AFUV_relation} \afuvprior{} is lower than zero. 
Fig.~\ref{app:AFUV_0} shows the location of all 426 galaxies with \afuvprior{}$<0$. 

\section{Other colour$-$surface brightness relatons}
\label{app:mix_colours}
\begin{figure*}
    \centering    \includegraphics[width=0.99\textwidth]{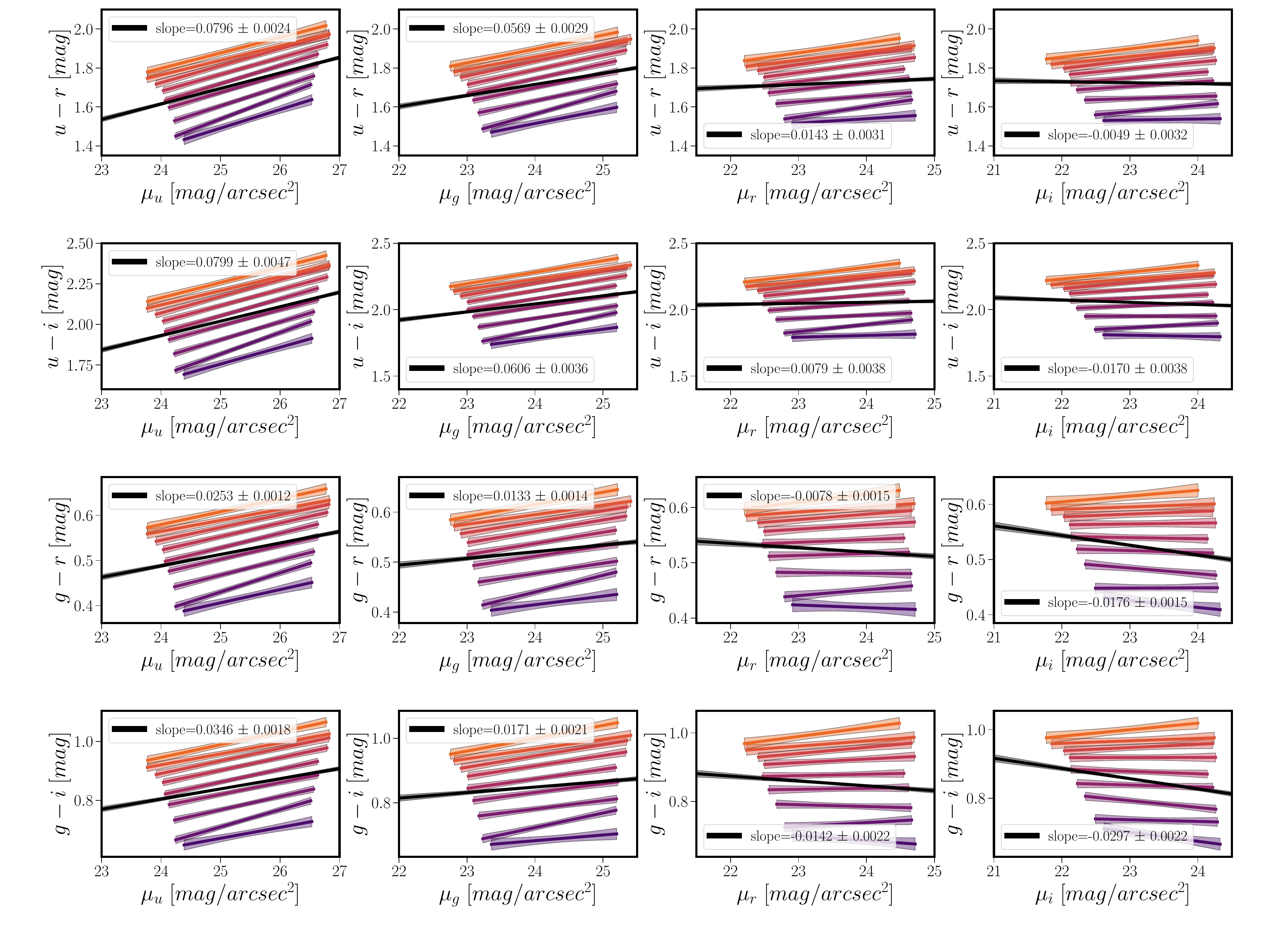}
    \caption{Exemplary colour -- surface brightness-fits for all 14 \afuv{} bins. Colors are the same as in Fig.~\ref{fig:AFUV_all_fits} and Fig.~\ref{fig:AFUV_all_bins_bdg}. The black lines shown in each panel represent the linear slope between colour and surface brightness calculated for the whole sample.  }
    \label{App:A_global}
\end{figure*}

Relations fitted between observed colours and surface brightness in different configurations for 14 \afuv{} bins are presented in Fig.~\ref{App:A_global}. 
The sequence of colours represents the one used for \afuv{} bin in Fig.~\ref{fig:AFUV_bins}. 
Filled areas mimic the $\pm 1\sigma$ uncertainty around estimated lines. 
The black line on each panel represents the linear fit to all bins. The slope of this mean relation is shown on each panel. 

\section{\afuv{} bins fitting }
Here, we present separate fits performed for all 14 \afuv{} bins. 
Each panel represent one bin. 
At the top of each panel, we give the \afuv{} range and the number of galaxies from our sample. 
The fitted linear relation, as well as the deviation of the \ur{} colour from the linear fit ($\Delta_{ur}$) and the variability of \afuv{} within the range of $\pm1\sigma$, are given in each panel. 

  \begin{figure*}[h!]
    \centering
        \includegraphics[width=0.999\textwidth]{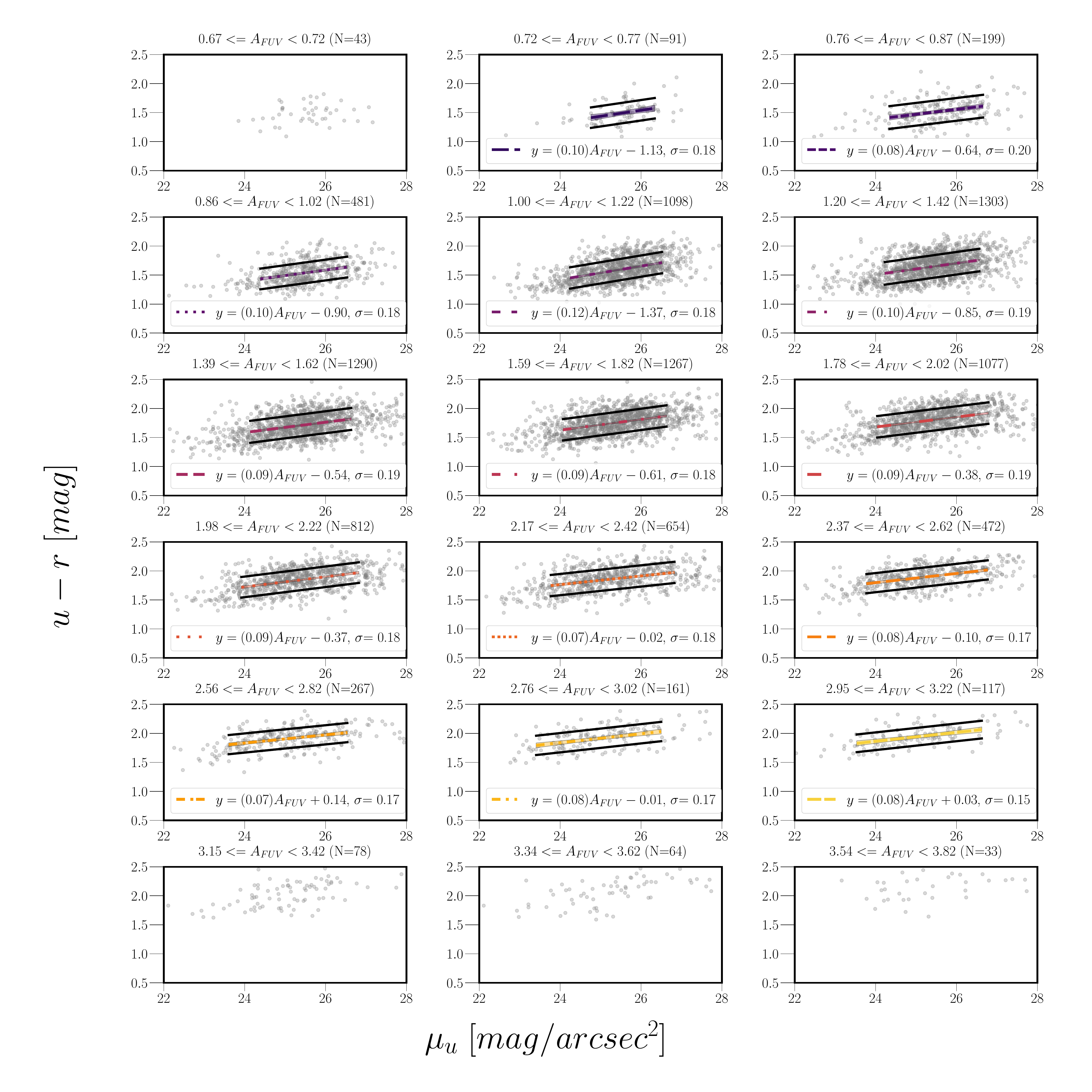}
        \caption{Relation between observed \ur{} and \sbu{} for the sample of 7\,934 galaxies divided into 14 \afuv{} bins. 
        Each panel represents consecutive $\rm A_{FUV}$ bins (Fig.~\ref{fig:AFUV_bins} and Table~\ref{tab:AFUVbins}).
        The resulting slope and intercept of the fit are provided within each corresponding panel. 
        The deviation of the \ur{} colour from the linear fit ($\Delta_{ur}$) and the variability of $\rm A_{FUV}$ within the range of $\pm 1\sigma$ from the linear fit ($\sigma_{AFUV}$) are both indicated in each panel. }
        \label{fig:AFUVbins_linear_fitting}
\end{figure*}

\end{appendix}    

\end{document}